\documentclass[aps,amsmath,prd,showpacs,nofootinbib,amssymb,preprint]{revtex4}

\usepackage{amssymb,graphics,graphicx,color,epstopdf,subfigure,
dcolumn,bm,slashed,tabularx,bm}
\usepackage{afterpage}

\begin{document}
\title{
Complementarity in direct searches for additional Higgs bosons \\
at the LHC and the International Linear Collider
}

\preprint{UT-HET-087}

\author{Shinya Kanemura}
\email{kanemu@sci.u-toyama.ac.jp}
\affiliation{Department of Physics, University of Toyama, Toyama
930-8555, Japan}
\author{Hiroshi Yokoya}
\email{hyokoya@sci.u-toyama.ac.jp}
\affiliation{Department of Physics, University of Toyama, Toyama
930-8555, Japan}
\author{Ya-Juan Zheng}
\email{yjzheng218@gmail.com}
\affiliation{CTS, CASTS and Department of Physics,
National Taiwan University, Taipei 10617, Taiwan}
\date{\today}

\begin{abstract}
We discuss complementarity of discovery reaches of heavier neutral Higgs
 bosons and charged Higgs bosons at the LHC and the International Linear
 Collider (ILC) in two Higgs doublet models (2HDMs).
We perform a comprehensive analysis on their production and decay
 processes for all types of Yukawa interaction under the softly-broken
 discrete symmetry which is introduced to avoid flavour changing neutral
 currents, and we investigate parameter spaces of discovering additional
 Higgs bosons at the ILC beyond the LHC reach. 
We find that the 500~GeV run of the ILC with the integrated luminosity of
 500~fb$^{-1}$ shows an advantage for
 discovering the additional Higgs bosons in the region where the LHC
 cannot discover them with the integrated luminosity of 300~fb$^{-1}$. 
For the 1~TeV run of the ILC with the integrated luminosity of
 1~ab$^{-1}$, production processes of an 
 additional Higgs boson associated with the top quark can be useful as
 discovery channels in some parameter spaces where the LHC with the
 integrated luminosity of 3000~fb$^{-1}$ cannot reach.
It is emphasized that the complementary study at the LHC and the ILC is
 useful not only to survey additional Higgs bosons at the TeV scale, but
 also to discriminate types of Yukawa interaction in the 2HDM. 
\end{abstract}
\maketitle

\section{Introduction}

In July 2012, both the ATLAS and CMS Collaborations announced the
observation of a long-sought new particle with a mass approximately at
126~GeV~\cite{Aad:2012tfa,Chatrchyan:2012ufa}. 
Further measurements of the properties of this new particle manifest
consistency with the Higgs boson in the standard model (SM) within the
errors which are not small up to now~\cite{Aad:2013wqa,Aad:2013xqa,
Chatrchyan:2013iaa,Chatrchyan:2013mxa}. 
It makes the SM much closer to its triumph in explaining electroweak
symmetry breaking. 
However, this does not necessarily mean that the SM is fundamentally
correct. 
There is no theoretical principle to justify the minimal Higgs sector
with only one Higgs doublet in the SM, and many new physics models
beyond the SM predict non-minimal Higgs sectors.
Therefore, it is very important to determine the Higgs sector in order
to understand the structure of the new physics model by future
experiments at the LHC and the International Linear Collider
(ILC)~\cite{Djouadi:2007ik,Behnke:2013lya}.

The two Higgs doublet model (2HDM) is one of the simplest extensions of
the SM Higgs sector, which is useful in both exploring the
phenomenology of extended Higgs sectors and interpreting experimental
results from searches for additional Higgs bosons. 
Some of the new physics models contain two Higgs doublets, such as 
the minimal supersymmetric extension of the SM
(MSSM)~\cite{Haber:1984rc,Gunion:1989we,Djouadi:2005gj}, 
models for extra CP phases, models for electroweak
baryogenesis~\cite{Turok:1990in,Bochkarev:1990gb,Nelson:1991ab}, and
models for radiative neutrino mass generation
mechanism~\cite{Zee:1980ai,Aoki:2008av,Aoki:2011zg}. 
In general, the extension with additional doublet fields causes flavour
changing neutral currents (FCNCs), which are strongly bounded by
experimental data. 
In order to avoid such dangerous FCNCs, different quantum number should
be assigned to each doublet field~\cite{Glashow:1976nt}.
This can be attained by introducing a softly-broken discrete symmetry
under which $\Phi_1\to +\Phi_1$ and $\Phi_2\to-\Phi_2$, where $\Phi_1$
and $\Phi_2$ are the two doublet fields\footnote{%
2HDMs without discrete symmetry have also been considered, such as 
the Type-III 2HDM~\cite{Cheng:1987rs,Atwood:1996vj}, the aligned
2HDM~\cite{Pich:2009sp}, etc.
}. 
In this case, there can be four types of Yukawa interaction, depending
on the assignment of charges of the discrete
symmetry~\cite{Barger:1989fj,Grossman:1994jb}. 
In the 2HDMs, there are two CP-even neutral scalars $h$ and $H$, one
CP-odd neutral scalar $A$, and a pair of charged scalars $H^\pm$. 
We assume that the lighter CP-even neutral scalar $h$ is the discovered
SM-like Higgs boson with the mass of about 126~GeV.
Additional neutral and charged Higgs bosons have rich phenomenology and
serve as a cornerstone for physics beyond the SM. 

In the literature, there have been many discussions on various types of
2HDMs and their signatures at the
LHC~\cite{Barger:2009me,Goh:2009wg,Eriksson:2009ws,Aoki:2009ha}.
For a recent systematic study on the theory and phenomenology of 2HDMs,
we refer to Ref.~\cite{Branco:2011iw} and references therein. 
In light of the recent data collected at the LHC 7-8~TeV run,
many possibilities for explanation of the current data of several
decay channels for the observed Higgs boson are explored in the
framework of the
2HDMs~\cite{Ferreira:2011aa,Burdman:2011ki,Arhrib:2011wc,Arhrib:2012ia, 
Blum:2012kn,Barroso:2012wz,Craig:2012vn,Altmannshofer:2012ar,
Chang:2012ve,Belanger:2012gc,Chen:2013kt,Chiang:2013ixa,
Grinstein:2013npa,Chen:2013rba,Eberhardt:2013uba,Chang:2013ona,
Harlander:2013mla,Celis:2013ixa}. 
Furthermore, the parameter regions in the 2HDMs have been constrained by
direct searches for additional Higgs bosons at the
LHC~\cite{ATLAS:2013zla,CMS:2013eua}. 
For the future run of the LHC with the collision energy of 14~TeV, 
additional Higgs bosons are expected to be detected as long as their
masses are smaller than 350~GeV to 800~GeV, depending on the scenario of
the 2HDMs for the integrated luminosity of
300~fb$^{-1}$~\cite{Asner:2013psa}. 

The ILC is a future electron-positron linear collider with the collision
energies to be from 250~GeV to
1~TeV~\cite{Djouadi:2007ik,Behnke:2013lya}. 
The ILC can be used for precision measurements of the masses and couplings
of the SM particles. 
We can expect that the first run of the ILC with the collision energy at
250~GeV is capable of measuring the properties of the discovered SM-like
Higgs boson with a considerable level. 
By the combination of the results with higher collision energies up to
1~TeV, all the coupling constants with the discovered Higgs boson can be
measured with excellent accuracies. 
For instance, the Higgs couplings with weak gauge bosons can be measured
by better than 1\%, the Yukawa coupling constants can be measured by
percent levels, and the triple Higgs boson coupling can be measured by a
ten percent level~\cite{Asner:2013psa,Dawson:2013bba}. 
Such precision measurements of coupling constants of the discovered
Higgs boson can make it possible to perform fingerprinting of extended
Higgs sectors when deviations from the SM predictions are detected,
because each extended Higgs sector predicts a different pattern in
deviations of coupling
constants~\cite{Asner:2013psa,KTYY,Dawson:2013bba,Kanemura:2014dja,%
Kanemura:2014hja}. 
However, the
deviations in the coupling constants of the SM-like Higgs boson from the
SM predictions can be smaller than those detectable at the ILC, 
even when additional Higgs bosons are not too heavy.

At the ILC, the direct searches can also be well performed for new
particles in the models beyond the SM as long as kinematically
accessible. 
Additional Higgs bosons can be produced mainly in pair if the sum of the
masses is less than the collision energy, via $e^+e^-\to
hA$~\cite{Djouadi:1996ah}, $e^+e^-\to HA$~\cite{Gunion:1988tf} and
$e^+e^-\to H^+H^-$~\cite{Gunion:1988tf}.
For the collision energy below the threshold of the pair production,
single production processes of new additional Higgs bosons can be used
too, although the production cross sections are not large. 
The single charged Higgs boson production has been studied in the
framework of the MSSM~\cite{Kanemura:2000cw,Moretti:2002pa}. 
Preliminary detection possibilities were studied at linear colliders,
and their analysis shows that in the parameter space beyond the
kinematic limit for pair production, single production of $H^\pm$
associated with the top quark turns out to be a useful channel in
studying the charged Higgs boson phenomenology~\cite{Moretti:2002pa}. 
QCD corrections to the process $e^+e^-\to \bar{t}bH^+$ and its charge
conjugate counterpart have been studied in the MSSM in
Ref.~\cite{Kniehl:2002zz}. 
The single production processes of additional neutral Higgs bosons have
been studied in Ref.~\cite{Pocsik:1981bg}, and QCD corrections to the
$e^+e^-\to Q\bar{Q}H$ and $e^+e^-\to Q\bar{Q}A$ processes are calculated
in Refs.~\cite{Dawson:1998qq,Dittmaier:2000tc} where $Q=t$ and $b$. 
The discovery potential for additional Higgs bosons through single and
pair production processes at linear collider are evaluated 
in the MSSM~\cite{Kiyoura:2003tg}, which is useful in distinguishing the
MSSM from the other models. 

In this paper, we perform a comprehensive analysis on the production and
decay processes of additional Higgs bosons for all types of Yukawa
interaction under the discrete symmetry. 
The parameter space of discovering additional Higgs bosons at the LHC is
shown for all types of Yukawa interaction in the 2HDM according to the
analysis given in Ref.~\cite{Asner:2013psa}. 
We then examine detailed signatures of additional Higgs bosons for all
types of Yukawa interaction at the ILC. 
We find that the complementary study at the LHC and the ILC is useful
not only to survey additional Higgs bosons at the TeV scale, but also to
discriminate types of Yukawa interaction in the 2HDM. 

The paper is organized as follows.
In Sec.~\ref{sec:2hdm}, we introduce the 2HDMs and the different types
of Yukawa interaction. 
In Sec.~\ref{sec:constraints}, we present a brief summary of theoretical
and experimental (flavour and collider) constraints on the additional
neutral and charged Higgs bosons. 
Our study on the future prospects of the LHC searches are also presented
in this section. 
Sec.~\ref{sec:ilc} is devoted to our systematic analysis on the ILC
search for the additional Higgs bosons. 
Based on several benchmark scenarios,   
further discussions on the prospects of the direct searches of additional
Higgs bosons at future collider experiments are given in
Sec.~\ref{sec:dis}. 
Finally, we draw a conclusion in Sec.~\ref{sec:sum}.

\section{Two Higgs Doublet Model}\label{sec:2hdm}

\subsection{Basics of the model}

In the 2HDM, two isospin doublet scalar fields, $\Phi_1$ and $\Phi_2$
are introduced with a hypercharge $Y=1$. 
The Higgs potential in the general 2HDM is given as~\cite{Gunion:1989we} 
\begin{align}
 V& = m_1^2|\Phi_1|^2+m_2^2|\Phi_2|^2
 -  (m_3^2 \Phi_1^{\dagger}\Phi_2 + {\rm h.c.} )
 + \frac{\lambda_1}{2}|\Phi_1|^4 +\frac{\lambda_2}{2}|\Phi_2|^4
 \nonumber \\ &
 + \lambda_3|\Phi_1|^2|\Phi_2|^2+\lambda_4|\Phi_1^\dagger\Phi_2|^2
 + \left[\frac{\lambda_5}{2}(\Phi_1^\dagger\Phi_2)^2
 + \left\{\lambda_6(\Phi_1^\dagger\Phi_1)
 + \lambda_7(\Phi_2^\dagger\Phi_2)\right\}\Phi^\dagger_1\Phi_2
 + {\rm h.c.} \right], 
\label{eq:potential}
\end{align}
where $m_1^2$, $m_2^2$, $\lambda_{1-4}$ are real parameters while
$m_3^2$, $\lambda_{5-7}$ are complex in general.

For the most general 2HDM, the presence of Yukawa interactions leads to
the FCNCs via tree-level Higgs-mediated diagrams which is not
phenomenologically acceptable. 
To avoid such FCNCs, we consider 2HDMs with discrete $Z_2$ symmetry,  
under which the two doublets are transformed as
$\Phi_1\to+\Phi_1$ and $\Phi_2\to-\Phi_2$~\cite{Glashow:1976nt,%
Paschos:1976ay,Haber:1978jt,Donoghue:1978cj}. 
For the SM fermions, four sets of parity assignment under the
$Z_2$ transformation are possible~\cite{Barger:1989fj}, which is
summarized in Table~\ref{tab:Z2}.
Because of these types of Yukawa interaction, the 2HDM with $Z_2$
parity contains a variety of phenomenology with quarks and leptons. 

\begin{table}[t]
 \begin{tabular}{c|ccccccc}
  \hline
  & $\Phi_1$ & $\Phi_2$ & $u_R$ & $d_R$ & $\ell_R$ &
  $Q_L$ & $L_L$ \\
  \hline
  \hline
  Type-I & $+$ & $-$ & $-$
              & $-$ & $-$ & $+$
                          & $+$  \\
  Type-II & $+ $& $-$ & $-$
  & $+$ & $+$ & $+$
  & + \\
  Type-X & $+$ & $-$ & $-$
  & $-$ & $+$ & $+$
  & $+$  \\
  Type-Y & $+$ & $-$ & $-$
  & $+$ & $-$ & $+$
  & $+$  \\
  \hline
 \end{tabular}
\caption{Four possible $Z_2$ charge assignments of scalar and fermion
 fields to forbid tree-level Higgs-mediated FCNCs~\cite{Aoki:2009ha}.}
\label{tab:Z2}
\end{table}

To preserve the discrete symmetry, we hereafter restrict ourselves 
with the Higgs potential
in Eq.~(\ref{eq:potential}) with vanishing $\lambda_6$ and $\lambda_7$
which induce the explicit breaking of the symmetry. 
On the other hand, the presence of the $m_3^2$ term induces the soft
breaking of the symmetry characterized by the soft-breaking scale 
$M^2=m_3^2/(\sin\beta\cos\beta)$~\cite{Kanemura:2004mg}.
Therefore, we allow the $m_3^2$ term and the soft breaking of the $Z_2$
symmetry. 
Furthermore, we consider the CP-conserving scenario for simplicity by
taking $m_3^2$ and $\lambda_5$ to be real. 

After the electroweak symmetry breaking and after the three
Nambu-Goldstone bosons are absorbed by the Higgs mechanism, 
five physical states are left; two CP-even neutral Higgs bosons, $h$ and
$H$; one CP-odd neutral Higgs boson, $A$; and charged Higgs bosons,
$H^{\pm}$. 
Masses of these scalars are obtained by solving the stability
conditions of the potential in
Eq.~(\ref{eq:potential})~\cite{Gunion:1989we}. 
In addition to the four kinds of masses $m_h$, $m_H$, $m_A$ and
$m_{H^\pm}$ as well as the soft-breaking parameter $M^2$, 
the remaining two parameters are chosen as follows.
One is $\tan\beta=v_2/v_1$, the ratio of the vacuum expectation values
(VEVs) of the two doublet fields, where $v\equiv \sqrt{v_1^2+v_2^2}
\simeq 246$~GeV is fixed by the Fermi constant $G_F=1/(\sqrt{2}v^2)$. 
The other is $\alpha$, a mixing angle for diagonalizing the
mass matrix for the neutral CP-even component. 
The limit of $\sin(\beta-\alpha)=1$ is called the SM-like limit where
the light CP-even scalar $h$ behaves as the SM Higgs
boson~\cite{Gunion:2002zf}. 
We take $h$ as the observed SM-like Higgs boson with $m_h=126$~GeV.

The input parameters of the model are $v$, $m_h$, $m_H$, $m_A$,
$m_{H^\pm}$, $M$, $\alpha$ and $\beta$.
In terms of these parameters, the quartic coupling constants
are expressed as~\cite{Kanemura:2004mg}
\begin{subequations}
\begin{align}
\lambda_1 &= \frac{1}{v^2\cos^2\beta}
(-M^2\sin^2\beta+m_h^2\sin^2\alpha+m_H^2\cos^2\alpha), \\
\lambda_2 &= \frac{1}{v^2\sin^2\beta}
(-M^2\cos^2\beta+m_h^2\cos^2\alpha+m_H^2\sin^2\alpha), \\
\lambda_3 &= \frac{1}{v^2}\left[-M^2
-\frac{\sin2\alpha}{\sin2\beta}(m_h^2-m_H^2)+2m_{H^{\pm}}^2\right],\\
\lambda_4 &= \frac{1}{v^2}(M^2+m_A^2-2m_{H^{\pm}}^2),\\
\lambda_5 &= \frac{1}{v^2}(M^2-m_A^2).
\end{align}
\end{subequations}

The interactions of the Higgs bosons to weak gauge bosons are common 
among the types of Yukawa interaction. 
Feynman rules for these interactions are read out from the
Lagrangian~\cite{Gunion:1989we,Djouadi:2005gj}; 
\begin{align}
&hZ_\mu Z_\nu:\,2i\frac{m_Z^2}{v}\sin(\beta-\alpha)g_{\mu\nu},\quad
 HZ_\mu Z_\nu:\,2i\frac{m_Z^2}{v}\cos(\beta-\alpha)g_{\mu\nu},\nonumber\\
&hW^+_\mu W^-_\nu:\,2i\frac{m_W^2}{v}\sin(\beta-\alpha)g_{\mu\nu},\quad
 HW^+_\mu W^-_\nu:\,2i\frac{m_W^2}{v}\cos(\beta-\alpha)g_{\mu\nu}
 \label{hVV}
\end{align}
and
\begin{align}
&hAZ_\mu:\,\frac{g_Z^{}}{2}\cos(\beta-\alpha)(p+p')_\mu,\quad
 HAZ_\mu:\,-\frac{g_Z^{}}{2}\sin(\beta-\alpha)(p+p')_\mu,\nonumber\\
&H^+H^-Z_\mu:\,-\frac{g_Z^{}}{2}\cos2\theta_W(p+p')_\mu,\quad
 H^+H^-\gamma_\mu:\,-ie(p+p')_\mu,\nonumber\\
&H^\pm hW^\mp_\mu:\,\mp i\frac{g_W^{}}{2^{}}\cos(\beta-\alpha)(p+p')_\mu,\quad
 H^\pm HW^\mp_\mu:\,\pm i\frac{g_W^{}}{2^{}}\sin(\beta-\alpha)(p+p')_\mu,\quad
\nonumber\\
&H^\pm AW^\mp_\mu:\,\frac{g_W^{}}{2}(p+p')_\mu,
\label{hhV}
\end{align}
where   $p_\mu$ and $p'_\mu$ are
outgoing four-momenta of the first and the second scalars, respectively,
and $g_Z^{}=g_W^{}/\cos\theta_W^{}$. 

\subsection{Type of Yukawa interaction}

The Yukawa interactions of the 2HDM Higgs bosons to the SM fermions are 
written as
\begin{align}
 {\mathcal L}^{\rm 2HDM}_{\rm Yukawa} = -\bar{Q}_{L}Y_u\tilde\Phi_u u_R
-\bar{Q}_{L}Y_d\Phi_d d_R - \bar{L}_{L}Y_{\ell}\Phi_{\ell}\ell_R
+{\rm h.c.}, \label{eq:Yukawa}
\end{align}
where $R$ and $L$ represent the right-handed and left-handed chirality
of fermions, respectively. 
$\Phi_{f=u,d,\ell}$ is chosen from $\Phi_1$ or $\Phi_2$ to make the
interaction term $Z_2$ invariant, according to the Table~\ref{tab:Z2}. 
The Type-I 2HDM is the case that all the quarks and charged leptons
obtain the masses from $v_2$, and the Type-II 2HDM is that up-type quark
masses are generated by $v_2$ but the masses of down-type quarks and
charged leptons are generated by $v_1$. 
In the Type-X 2HDM, both up- and down- type quarks couple to $\Phi_2$
while charged leptons couple to $\Phi_1$. 
The last case is the Type-Y 2HDM where up-type quarks and charged
leptons couple to $\Phi_2$ while up-type quarks couple to $\Phi_1$. 
We note that the Type-II 2HDM is predicted in the context of the
MSSM~\cite{Haber:1984rc,Gunion:1989we}
and that the Type-X 2HDM is used in some of radiative seesaw
models~\cite{Aoki:2008av,Aoki:2011zg}.

In terms of the mass eigenstates, Eq.~(\ref{eq:Yukawa}) is rewritten as 
\begin{align}
 {\mathcal L}^{\rm 2HDM}_{\rm Yukawa} = &-\sum_{f=u,d,\ell}
\left[\frac{m_f}{v}\xi_h^f\bar{f}fh+\frac{m_f}{v}\xi_H^f\bar{f}fH
-i\frac{m_f}{v}\xi_A^f\gamma_5\bar{f}fA\right]\nonumber\\
&-\left\{
\frac{\sqrt{2}V_{ud}}{v}\bar{u}\left[m_u\xi_A^uP_L+m_d\xi_A^dP_R
\right]dH^+ + \frac{\sqrt{2}m_\ell}{v}\xi_{A}^\ell\bar{v}_{L}\ell_RH^+
+{\rm h.c.} \right\},
\end{align}
where $P_{R,L}$ are the chiral projection operators.
The coefficients $\xi_\phi^f$ are summarized in
Table~\ref{tab:yukawa}.

\begin{table}[t]
 \begin{tabular}{c|ccccccccc}
  \hline
  & $\xi_h^u$ & $\xi_h^d$ & $\xi_h^\ell$ & $\xi_H^u$ & $\xi_H^d$ &
  $\xi_H^\ell$ & $\xi_A^u$ & $\xi_A^d$ & $\xi_A^\ell$ \\
  \hline
  \hline
  Type-I & $c_\alpha/s_\beta$ & $c_\alpha/s_\beta$ & $c_\alpha/s_\beta$
              & $s_\alpha/s_\beta$ & $s_\alpha/s_\beta$ & $s_\alpha/s_\beta$
                          & $\cot\beta$ & $-\cot\beta$& $-\cot\beta$ \\
  Type-II & $c_\alpha/s_\beta$ & $-s_\alpha/c_\beta$ & $-s_\alpha/c_\beta$
  & $s_\alpha/s_\beta$ & $c_\alpha/c_\beta$ & $c_\alpha/c_\beta$
  & $\cot\beta$ & $\tan\beta$& $\tan\beta$ \\
  Type-X & $c_\alpha/s_\beta$ & $c_\alpha/s_\beta$ & $-s_\alpha/c_\beta$
  & $s_\alpha/s_\beta$ & $s_\alpha/s_\beta$ & $c_\alpha/c_\beta$
  & $\cot\beta$ & $-\cot\beta$& $\tan\beta$ \\
  Type-Y & $c_\alpha/s_\beta$ & $-s_\alpha/c_\beta$ & $c_\alpha/s_\beta$
  & $s_\alpha/s_\beta$ & $c_\alpha/c_\beta$ & $s_\alpha/s_\beta$
  & $\cot\beta$ & $\tan\beta$& $-\cot\beta$ \\
  \hline
\end{tabular}
\caption{The coefficients for different type of Yukawa
 interactions~\cite{Aoki:2009ha}. 
$c_\theta=\cos\theta,~{\rm and }~s_\theta=\sin\theta$ for $\theta = 
\alpha,~\beta$.}
\label{tab:yukawa}
\end{table}

In the SM-like limit, all the $\phi VV$ vertices in Eq.~(\ref{hVV}) and 
$\phi h V$ in Eq.~(\ref{hhV}) in
which one additional Higgs boson is involved disappear, where $\phi$ 
represents $H$, $A$ or $H^{\pm}$. 
On the other hand, the Yukawa interactions of additional Higgs boson
remain even in this limit.
Therefore, Yukawa interactions of the additional Higgs bosons are
very important for the decay and production processes of additional
Higgs bosons in this limit.

\subsection{Decay widths and decay branching ratios}

For each type of Yukawa interaction, the decay widths and branching
ratios of additional Higgs bosons can be calculated for given values 
of $\tan\beta$, $\sin(\beta-\alpha)$ and the masses. 
The total decay widths of additional Higgs bosons are necessary for the
consistent treatment of the production and decays of additional Higgs
bosons.  
We refer to Ref.~\cite{Aoki:2009ha} where the total decay 
widths are discussed in details for $\sin(\beta-\alpha)\simeq1$. 
Explicit formulae for all the partial decay widths can be found, e.g.,
in Ref.~\cite{Aoki:2009ha}. 
Here, we review the characteristic behaviors of the decays of additional
Higgs bosons in each type of Yukawa interaction by presenting numerical
results of the branching ratios. 
For simplicity, we set $\sin(\beta-\alpha)=1$, the SM-like limit. 
In this limit, the decay modes of $H\to W^+W^-$, $ZZ$, $hh$ as well as
$A\to Zh$ are absent. 
Decay branching ratios of the SM-like Higgs boson become completely
the same as those in the SM at the leading order, so that we cannot
distinguish models by the precision measurement of the couplings of the
SM-like Higgs boson\footnote{%
The decay branching ratios can be different from the SM prediction at
the next-to-leading order~\cite{Guasch:2001wv,Hollik:2001px,%
Dobado:2002jz,Kanemura:2004mg,Kanemura:2014dja}.
}. 
As we discuss later, the branching ratios can drastically change if 
$\sin(\beta-\alpha)$ is slightly deviated from unity. 

For numerical evaluation, $\overline{\rm MS}$ masses of fermions at
their own mass scales are taken to be $m_b=4.2$~GeV, $m_c=1.3$~GeV, 
$m_s=0.12$~GeV, and the leading order QCD running of them to the mass of
the Higgs boson is taken into account. 
In addition, we include the off-diagonal CKM matrix elements in our
analysis, $|V_{cb}|=|V_{ts}|=0.04$~\cite{Beringer:1900zz}.

\begin{figure}[t]
\includegraphics[width=\textwidth]{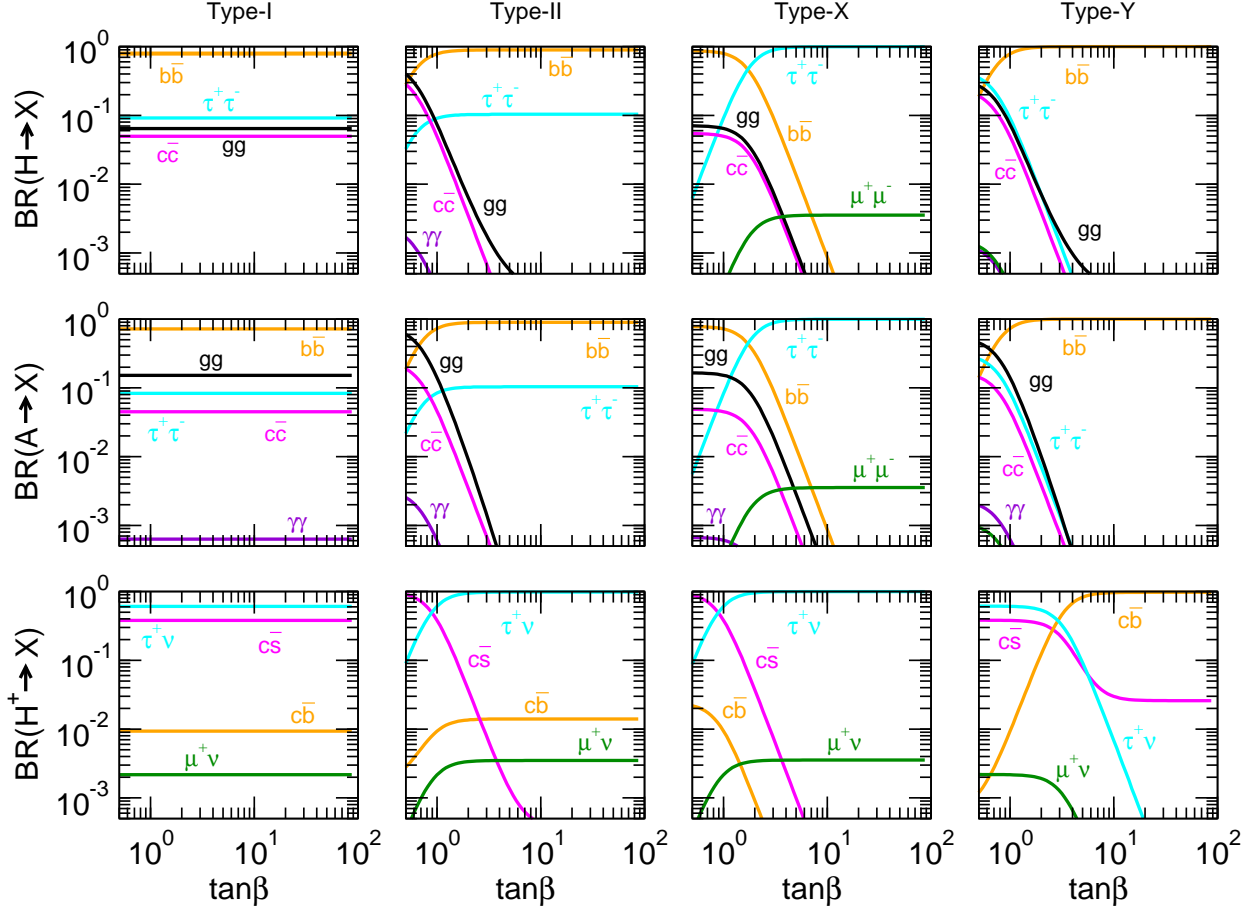} 
\caption{Branching ratios of $H$, $A$, $H^{\pm}$ as a function of
 $\tan\beta$ for $m_H=m_A=m_{H^\pm}=M=125$~GeV in the Type-I, II, X and
 Y 2HDM with $\sin(\beta-\alpha)=1$.}
\label{fig:Br_125}
\end{figure}

In the following, we show the branching ratios of additional Higgs
bosons in each type of Yukawa interaction, for the masses of 
125~GeV, 250~GeV and 500~GeV. 
In Fig.~\ref{fig:Br_125}, decay branching ratios of additional Higgs
bosons, $H$, $A$, and $H^\pm$ for $m_H=m_A=m_{H^\pm}=M=125$~GeV are
plotted as a function of $\tan\beta$ in each type of Yukawa
interaction.
Here, for the purpose of completeness, we do not take seriously the
direct and indirect exclusion limits, which are discussed later. 
Decay modes of $H,A\to t\bar{t}$ and $H^\pm\to tb$ are yet to
open. 
For Type-I, since all the Yukawa couplings are modified by the same
factor, the $\tan\beta$ dependence on the branching ratios is small. 
For large $\tan\beta$, all the Yukawa couplings are suppressed, 
which leads to very narrow widths of additional Higgs bosons.
The dominant decay modes are $b\bar{b}$ for the decays of $H$ and $A$,
and $\tau\nu$ and $cs$ for that of $H^\pm$.
For Type-II, the Yukawa interaction of down-type quarks and
charged leptons are scaled by $\tan\beta$, while up-type quarks are
by $\cot\beta$.
The decays of $H$ and $A$ are dominated by the $b\bar{b}$ mode
($\sim90\%$) and the $\tau^+\tau^-$ mode ($\sim10\%$) for wide
regions of $\tan\beta$, except in the small $\tan\beta$ regions where
$gg$ and $c\bar{c}$ decays become major modes.
The decay of $H^\pm$ is dominated by the $\tau\nu$ mode for
$\tan\beta\gtrsim1$. 
For $\tan\beta\lesssim 1$, the dominant decay mode becomes $cs$.
For Type-X, since leptonic decay modes are enhanced by $\tan\beta$,
$\tau^+\tau^-$ would be the dominant decay mode of $H$ and $A$ for
$\tan\beta\gtrsim 2$, while $\tau\nu$ is dominant in the $H^\pm$
decay for $\tan\beta\gtrsim 1$.
For the smaller $\tan\beta$ values, the dominant decay modes are
$b\bar{b}$ for $H$ and $A$, and $cs$ for $H^\pm$. 
For Type-Y, only the Yukawa couplings of down-type quarks are enhanced
by $\tan\beta$, $b\bar{b}$ would be the dominant decay mode of $H$ and
$A$ for $\tan\beta\gtrsim 1$. 
The dominant decay mode of $H^\pm$ is $cb$ for large
$\tan\beta$ values, and $\tau\nu$ and $cs$ for smaller
$\tan\beta$ ones.

\begin{figure}[t]
\includegraphics[width=\textwidth]{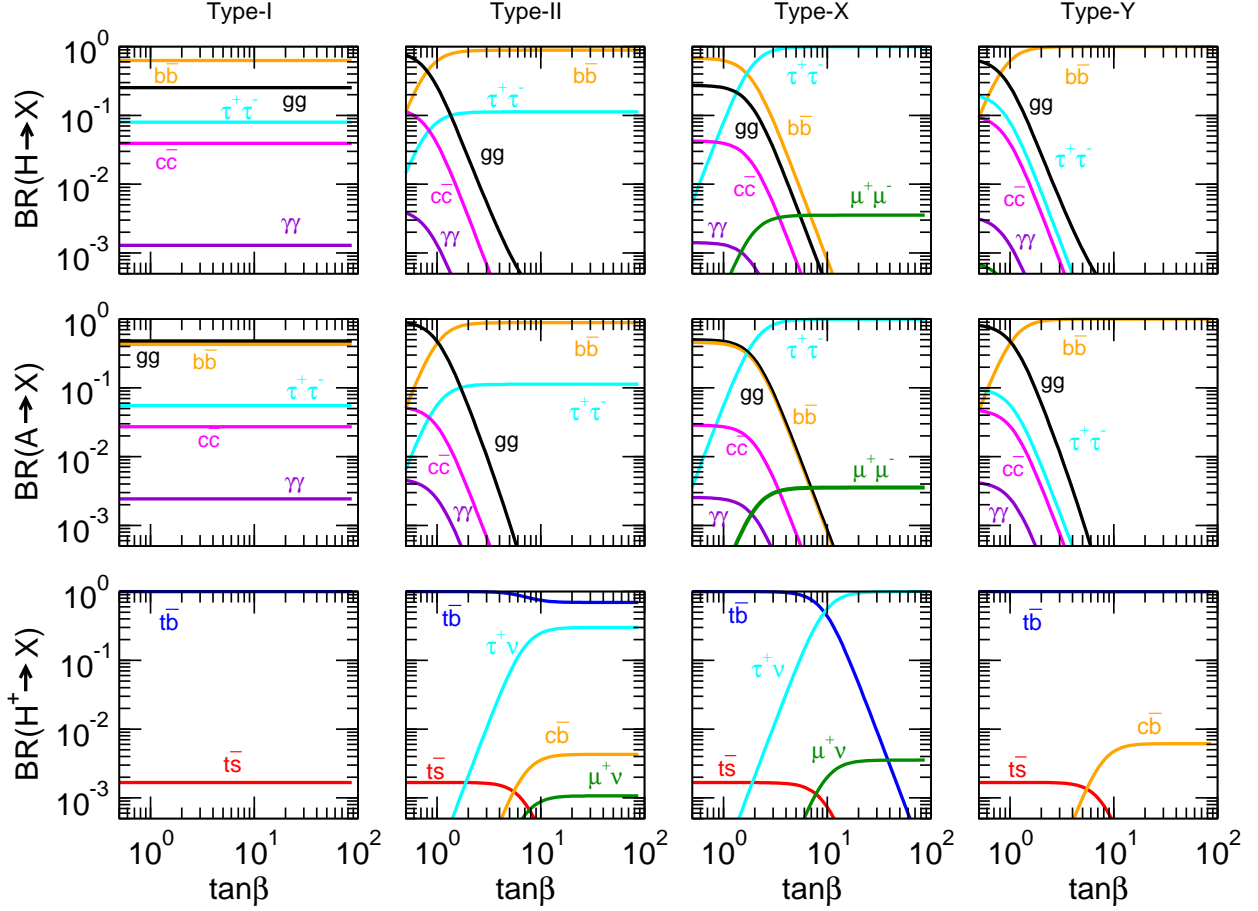} 
\caption{The same as Fig.~\ref{fig:Br_125}, but for
 $m_H=m_A=m_{H^\pm}=M=250$~GeV.}
\label{fig:Br_250}
\end{figure}

In Fig.~\ref{fig:Br_250}, the same branching ratios are evaluated for
$m_H=m_A=m_{H^\pm}=M=250$~GeV.
The decay branching ratios of $H$ and $A$ are almost unchanged from
the results for 125~GeV, but those of $H^\pm$ are changed due to the
new decay mode $tb$.
This decay mode dominates for all the $\tan\beta$ regions for the
Type-I, Type-II and Type-Y, and for $\tan\beta\lesssim 10$ for Type-X.
The $\tau\nu$ mode can be dominant and sub-dominant ($\sim0.3$) for
$\tan\beta\gtrsim 10$ for Type-X and Type-II, respectively.

\begin{figure}[t]
\includegraphics[width=\textwidth]{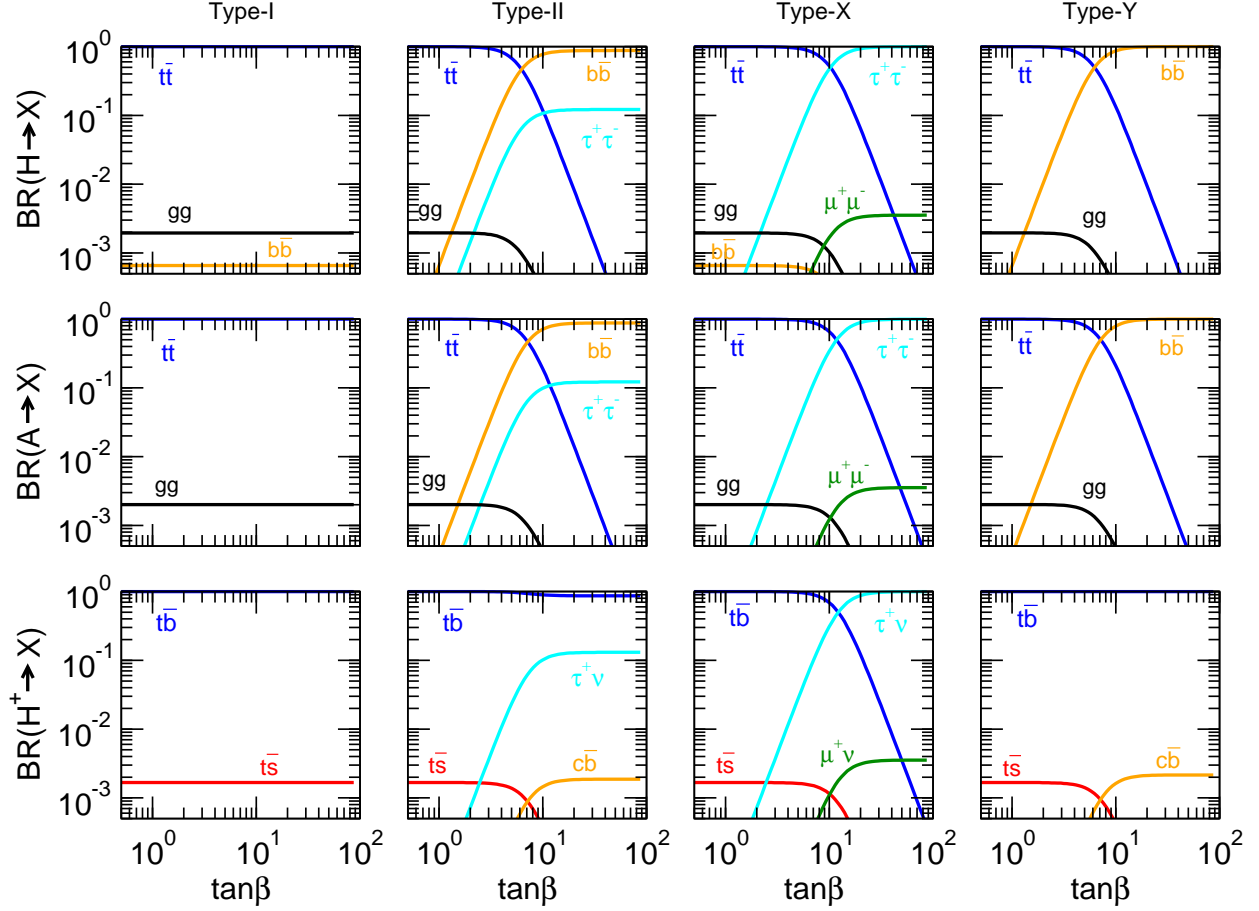} 
\caption{The same as Fig.~\ref{fig:Br_125}, but for
 $m_H=m_A=m_{H^\pm}=M=500$~GeV.}
 \label{fig:Br_500}
\end{figure}

In Fig.~\ref{fig:Br_500}, the same branching ratios are evaluated for
$m_H=m_A=m_{H^\pm}=M=500$~GeV.
In this case, the $t\bar{t}$ mode opens in the decays of $H$ and $A$.
The $t\bar{t}$ decay dominates in all the $\tan\beta$ regions for
Type-I, $\tan\beta\lesssim 5$ for Type-II, Type-X and Type-Y, while the
other modes are suppressed accordingly. 
The decays of $H^\pm$ are similar to those in the 250~GeV cases.

\section{Constraints on 2HDM parameters}\label{sec:constraints}

In this section, we briefly review the theoretical and experimental
constraints on the parameters in the 2HDMs.

\subsection{Constraints on the Higgs potential from perturbative
  unitarity and vacuum stability}

First, we introduce the constraints on the parameters by theoretical
arguments, namely perturbative unitarity and vacuum stability. 
The tree-level unitarity requires the scattering amplitudes
to be perturbative~\cite{Lee:1977yc},
{\it i.e.} $|a_i|<1/2$~\cite{Gunion:1989we},
where $a_i$ are the
eigenvalues of the $S$-wave amplitudes of two-to-two elastic
scatterings of the longitudinal component of weak gauge bosons and the
Higgs boson.
In the 2HDM with the softly-broken $Z_2$ symmetry, this condition gives
constraints on the quartic couplings in the Higgs 
potential~\cite{Kanemura:1993hm,Akeroyd:2000wc,Ginzburg:2005dt}. 
The eigenvalues for $14\times14$ scattering matrix for neutral states
are given as~\cite{Kanemura:1993hm}, 
\begin{subequations}
\begin{align}
&a_1^{\pm} = \frac{1}{16\pi}\left[\frac{3}{2}(\lambda_1+\lambda_2)
\pm\sqrt{\frac{9}{4}(\lambda_1-\lambda_2)^2+(2\lambda_3+\lambda_4)^2}
 \right],\\
&a_2^{\pm} = \frac{1}{16\pi}\left[\frac{1}{2}(\lambda_1+\lambda_2)
\pm\sqrt{\frac{1}{4}(\lambda_1-\lambda_2)^2+\lambda_4^2}
\right],\\
&a_3^{\pm} = \frac{1}{16\pi}\left[\frac{1}{2}(\lambda_1+\lambda_2)
\pm\sqrt{\frac{1}{4}(\lambda_1-\lambda_2)^2+\lambda_5^2}
\right],\\
&a_4 = \frac{1}{16\pi}(\lambda_3+2\lambda_4-3\lambda_5),\\
&a_5 = \frac{1}{16\pi}(\lambda_3-\lambda_5),\\
&a_6 = \frac{1}{16\pi}(\lambda_3+2\lambda_4+3\lambda_5),\\
&a_7 = \frac{1}{16\pi}(\lambda_3+\lambda_5),\\
&a_8 = \frac{1}{16\pi}(\lambda_3+\lambda_4),
\end{align}
\end{subequations}
and for singly charged states, one additional eigenvalue is
added~\cite{Akeroyd:2000wc},
\begin{align}
a_9 = \frac{1}{16\pi}(\lambda_3-\lambda_4).
\end{align}
Second, the requirement of vacuum stability that the Higgs potential
must be 
bounded from below
gives~\cite{Deshpande:1977rw,Nie:1998yn,Kanemura:1999xf} 
\begin{align}
 &\lambda_1>0,\quad \lambda_2>0,\quad
 \sqrt{\lambda_1\lambda_2}+\lambda_3+{\rm Min}(0,\lambda_4-|\lambda_5|)>0. 
\end{align}
The parameter space of the model is constrained by these conditions on
the coupling constants in the Higgs potential.

\subsection{Constraints on the Higgs potential from electroweak
  precision observables}

Further constraints on the Higgs potential of the 2HDM are from the
electroweak precision measurements. 
The $S$, $T$ and $U$ parameters are defined to disentangle new physics
effects in the radiative corrections to the gauge bosons two-point 
functions~\cite{Peskin:1991sw}. 
Those are sensitive to the effects of Higgs bosons through the loop
corrections~\cite{Toussaint:1978zm,Bertolini:1985ia}. 
The $T$ parameter corresponds to the $\rho$ parameter,
which is severely constrained by experimental observations as 
$\rho=1.0005^{+0.0007}_{-0.0006}$ where $U=0$ is
assumed~\cite{Beringer:1900zz}. 
Because of this constraint, the mass splitting among the additional
Higgs bosons are constrained in the 2HDM with the light SM-like Higgs
boson~\cite{Haber:2010bw,Kanemura:2011sj}.

\subsection{Flavour constraints on $m_{H^\pm}$ and $\tan\beta$ }

Flavour experiments constrain the 2HDM through the $H^\pm$
contribution to the flavour mixing observables by tree-level or loop
diagrams~\cite{Aoki:2009ha,Logan:2009uf,Su:2009fz}. 
Since the amplitudes of these processes contain the Yukawa interaction,
constraints from the flavour physics strongly depends on the type of
Yukawa interaction. 
In Ref.~\cite{Mahmoudi:2009zx}, the limits on the general couplings by
flavour physics are translated to the limits in the
($m_{H^\pm},\tan\beta$) plane in each type of Yukawa interaction in the
2HDM.
See also recent studies in
Refs.~\cite{Botella:2014ska,Cheng:2014ova,Bhattacharyya:2014nja}.

The strong exclusion limit is provided from the measurements of the
branching ratio of $B\to X_s\gamma$ processes~\cite{Amhis:2012bh}. 
For Type-II and Type-Y, a $\tan\beta$-independent lower limit of
$m_{H^\pm}\gtrsim 380$~GeV is obtained~\cite{oai:arXiv.org:1208.2788} by
combining with the NNLO calculation~\cite{Misiak:2006ab}. 
On the other hand, for Type-I and Type-X, $\tan\beta\lesssim 1$ is
excluded for $m_{H^\pm}\lesssim 800$~GeV, but no lower bound on
$m_{H^\pm}$ can be obtained. 

For all types of Yukawa interaction, lower $\tan\beta$ regions
($\tan\beta\le1$) are also excluded for $m_{H^\pm}\lesssim500$~GeV by
the measurement of $B_{d}^0$-$\bar{B}^0_{d}$ mixing~\cite{Amhis:2012bh},
because of the universal couplings of $H^\pm$ to the up-type quarks. 

Constraints for larger $\tan\beta$ regions are obtained only in the
Type-II 2HDM by using the leptonic meson decay
processes~\cite{Amhis:2012bh}, $B\to\tau\nu$~\cite{Hou:1992sy} and
$D_s\to\tau\nu$~\cite{Akeroyd:2009tn}. 
This is because the relevant couplings behave
$\xi_A^d\xi_A^\ell=\tan^2\beta$ in Type-II, but $\xi_A^d\xi_A^\ell=-1$
($\cot^2\beta$) for Type-X and Type-Y (Type-I). 
For Type-II, upper bounds of $\tan\beta$ are given at around 30 for 
$m_{H^\pm}\simeq350$~GeV and around 60 for
$m_{H^\pm}\simeq700$~GeV~\cite{Mahmoudi:2009zx}.

\subsection{Collider constraints on Higgs boson masses and $\tan\beta$}

Here, we briefly summarize constraints on the additional neutral and
charged Higgs bosons in the 2HDM from previous collider data at LEP,
Tevatron and LHC experiments. 
Most of the searches before have been performed in the context of the
MSSM, namely, the Type-II 2HDM. 
Some of the results can be used to analyze the constraints on the other
types of 2HDMs. 
There have also been other studies which directly investigate some types
of Yukawa interaction such as Type-I, Type-X and Type-Y. 

From the LEP experiment, lower mass bounds on $H$ and $A$ have been
obtained as $m_H>92.8$~GeV and $m_A>93.4$~GeV in a CP-conservation
scenario~\cite{Abdallah:2004wy,Schael:2006cr}. 
Combined searches for $H^\pm$ give the mass bound of $m_{H^\pm}>80$~GeV
assuming ${\mathcal B}(H^+\to\tau^+\nu)+{\mathcal B}(H^+\to
c\bar{s})=1$~\cite{Achard:2003gt,Abdallah:2003wd,Abbiendi:2013hk}.

CDF and D0 Collaborations at the Fermilab Tevatron have searched for
the processes of $p\bar{p}\to b\bar{b}H/A$, followed by $H/A\to
b\bar{b}$ or
$H/A\to\tau^+\tau^-$~\cite{Aaltonen:2011nh,Abazov:2011up,Aaltonen:2012zh}. 
By utilizing the $\tau^+\tau^-$ ($b\bar{b}$) decay mode, which can be
sensitive to the cases of Type-II (Type-II and Type-Y), upper bounds of
$\tan\beta$ have been obtained from around 25 to 80 (40 to 90) for $m_A$
from 100~GeV to 300~GeV, respectively. 
For the $H^\pm$ search at the Tevatron, the decay modes of
$H^\pm\to\tau\nu$ and $H^\pm\to cs$ have been investigated using the
production from the top quark decay of $t\to
bH^\pm$~\cite{Abazov:2008rn,Abazov:2009aa,Aaltonen:2009ke}. 
Upper bounds on the decay branching ratio ${\mathcal B}(t\to bH^\pm)$ have
been obtained, which can be translated into the bound on $\tan\beta$ in
various scenarios. 
In the Type-I 2HDM, for $H^\pm$ heavier than the top quark, upper bounds
on $\tan\beta$ have been obtained to be from around 20 to 70 for
$m_{H^\pm}$ from 180~GeV to 190~GeV, respectively~\cite{Abazov:2008rn}. 

At the LHC, direct searches for the additional Higgs bosons have been
 performed by using the recorded events at a center-of-mass
energy of 7~TeV with 4.9~fb$^{-1}$ and 8~TeV with 19.7~fb$^{-1}$ in 2011
and 2012, respectively. 
The CMS experiment has searched $H$ and $A$ decaying to the
$\tau^+\tau^-$ final state, and upper limits on $\tan\beta$ have been
obtained for the MSSM scenario or the Type-II 2HDM from 4 to 60 for
$m_A$ from 140~GeV to 900~GeV, respectively~\cite{CMS_neutral_new}. 
Similar searches have been also performed by ATLAS~\cite{Aad:2012cfr}.
In Type-II and Type-Y 2HDMs, the CMS experiment has also searched the
bottom-quark associated production of $H$ or $A$ which decays into the
$b\bar{b}$ final state~\cite{CMS_neutral_old}, and has obtained the
upper bounds on $\tan\beta$; i.e., $\tan\beta\gtrsim 16$ (28) is
excluded at $m_{A}=100$~GeV (350~GeV). 
ATLAS has reported the $H^\pm$ searches via the $\tau$+jets
final state~\cite{Aad:2012tj,ATLAS_charged_constraints}. 
In the Type-II 2HDM, for $m_{H^{\pm}}\lesssim m_{t}$, wide parameter
regions have been excluded for $100$~GeV $\lesssim
 m_{H^\pm}\lesssim140$~GeV with $\tan\beta\gtrsim 1$. 
In addition, for $m_{H^\pm}\gtrsim 180$~GeV, the parameter regions of
$\tan\beta\gtrsim 50$ at $m_{H^\pm}=200$~GeV and $\tan\beta\gtrsim 65$
 at $m_{H^\pm}=300$~GeV have been excluded, respectively. 
The searches for $H^\pm$ in the $cs$ final-state have been
performed by ATLAS~\cite{Aad:2013hla}, and the upper limit on the
branching ratio of $t\to bH^\pm$ decay is obtained assuming the 100\%
branching ratio of $H^\pm\to cs$.
For $\sin(\beta-\alpha)<1$, searches for $H\to W^+W^-$, $hh$ and $A\to
 Zh$ signals give constraints on the 2HDMs with Type-I and Type-II
 Yukawa interactions~\cite{ATLAS:2013zla,CMS:2013eua}. 

\subsection{Prospect for the searches at the LHC}\label{sec:LHC}

In the previous subsections, we have seen the current bounds on the
additional Higgs bosons via the flavour and collider experiments. 
However, until the time when ILC experiments start, the LHC will be
further operated with higher energies and luminosity. 
Therefore, it is important to summarize future prospects for 
additional Higgs boson searches in the 2HDMs at the LHC with the highest
energy of 14~TeV. 

According to Refs.~\cite{Asner:2013psa,KTYY}, we evaluate the expected
discovery potential of additional Higgs bosons at the LHC with the
integrated luminosity of $L=300$~fb$^{-1}$ and 3000~fb$^{-1}$ by using
the signal and background analysis for various
channels~\cite{ATLAS:1999vwa}, which are combined with the production 
cross sections and the decay branching ratios for each type of Yukawa
interaction. 
Processes available for the searches are
\begin{itemize}
 \item $H/A (+ b\bar{b})$ inclusive and associated production followed
       by the $H/A\to\tau^+\tau^-$ decay~\cite{Baglio:2010ae}.
 \item $H/A+b\bar{b}$ associated production followed by the $H/A\to
       b\bar{b}$ decay~\cite{Dai:1994vu,DiazCruz:1998qc,Baglio:2010ae}.
 \item $gb\to tH^\pm$ production followed by the $H^\pm\to
       tb$ decay~\cite{Borzumati:1999th,Plehn:2002vy}.
 \item $q\bar{q}\to HA\to 4\tau$
       process~\cite{Kanemura:2011kx,Liu:2013gba}.
\end{itemize}
For the production cross sections, we utilize the Born-level
cross sections convoluted with the CTEQ6L parton distribution
functions~\cite{Pumplin:2002vw}. 
The scales of the strong coupling constant and parton distribution
functions are chosen to the values used in
Ref.~\cite{Djouadi:2005gi,Djouadi:2005gj}. 
For the last process, we follow the analysis in
Ref.~\cite{Kanemura:2011kx} by re-evaluating the signal events for
the different mass, and combine the statistical significance of all
channels for the decay patterns of $4\tau$. 
The similar analysis on the $HH^\pm$ and $AH^\pm$ production processes
resulting the signature of $3\tau$ plus large missing transverse
momentum gives comparable exclusion curves to the $4\tau$
analysis~\cite{Kanemura:2011kx}. 

\begin{figure}[t]
 \begin{center}
  \includegraphics[width=0.45\textwidth]{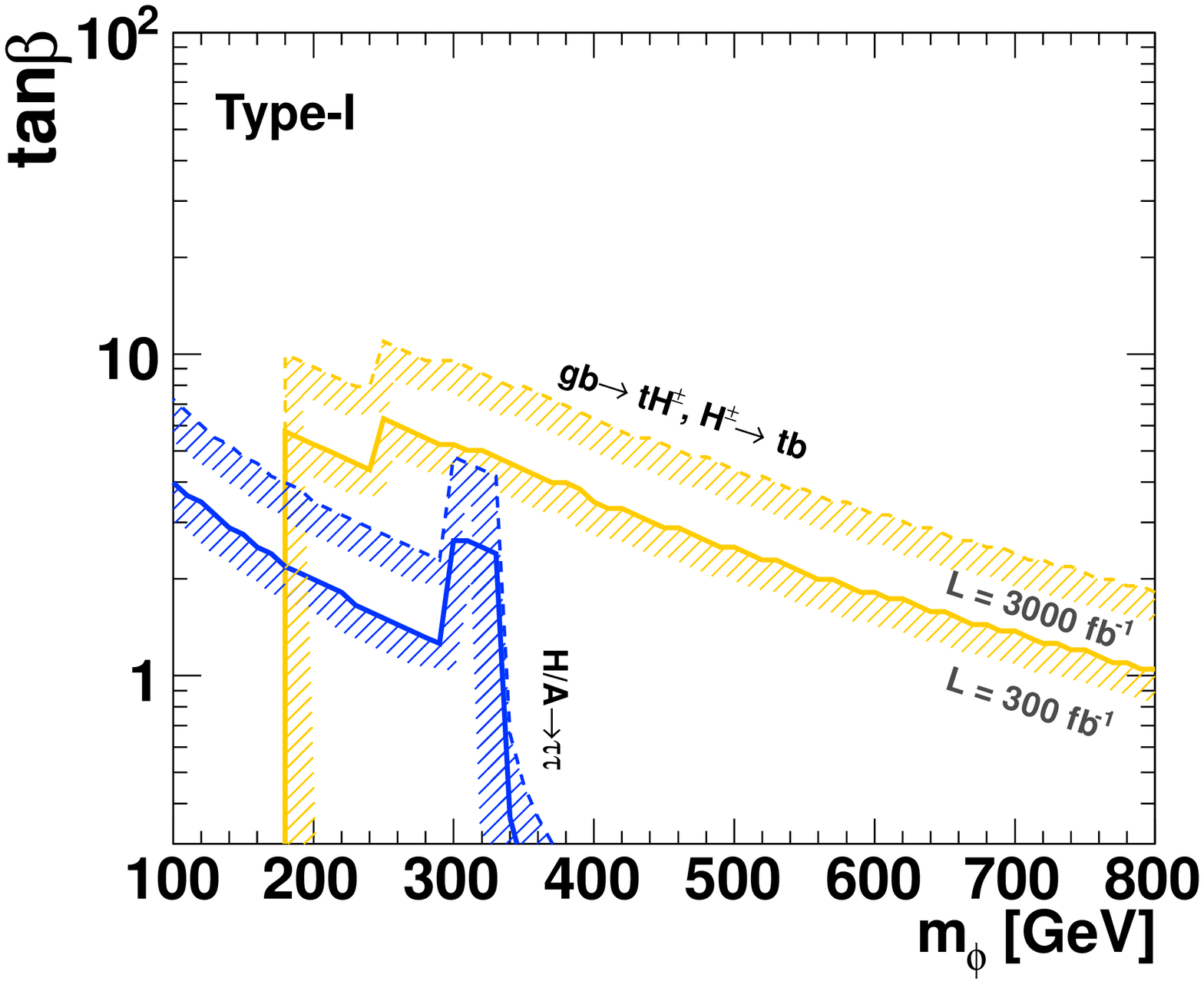}
  \quad
  \includegraphics[width=0.45\textwidth]{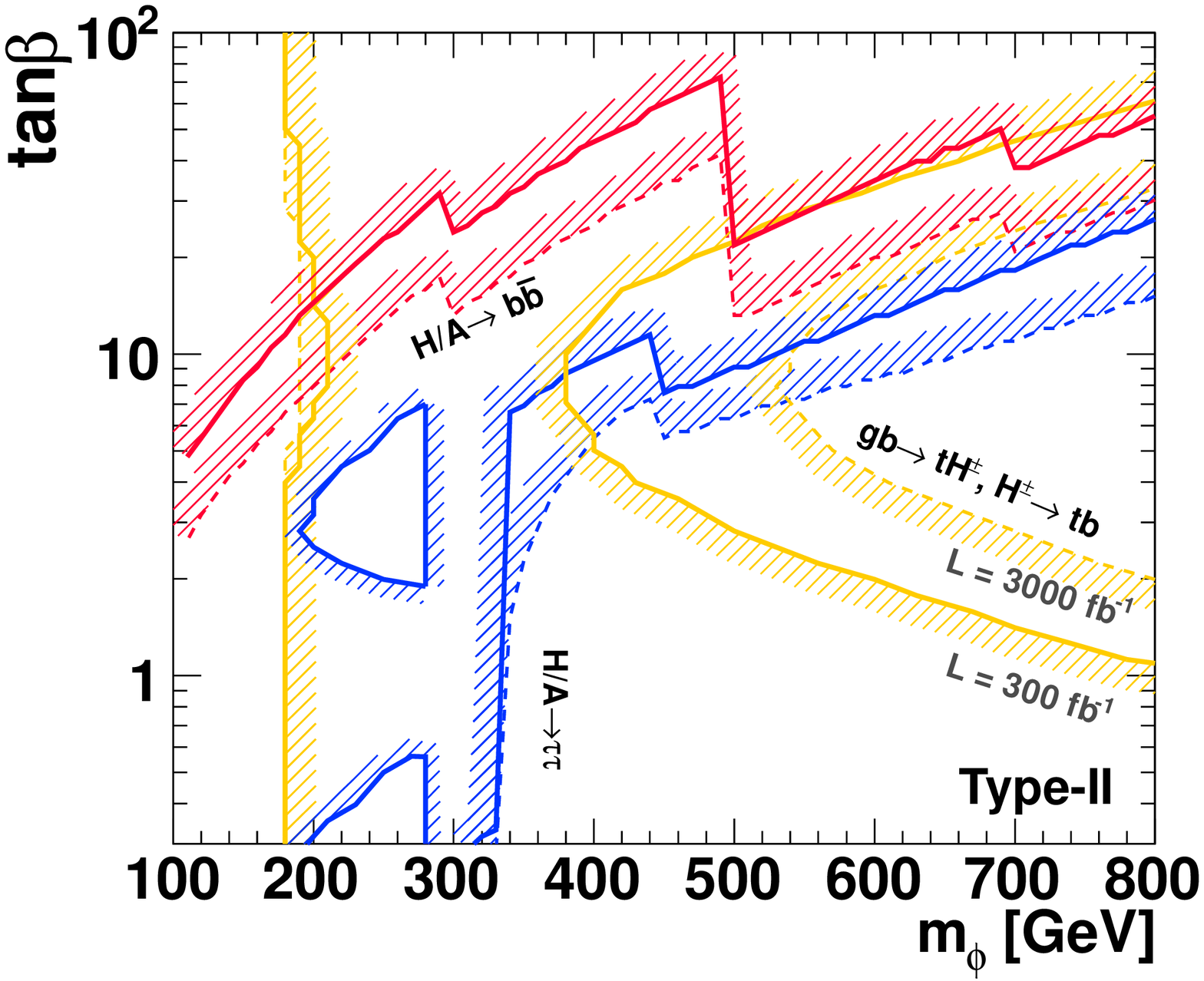}
  \\
  \includegraphics[width=0.45\textwidth]{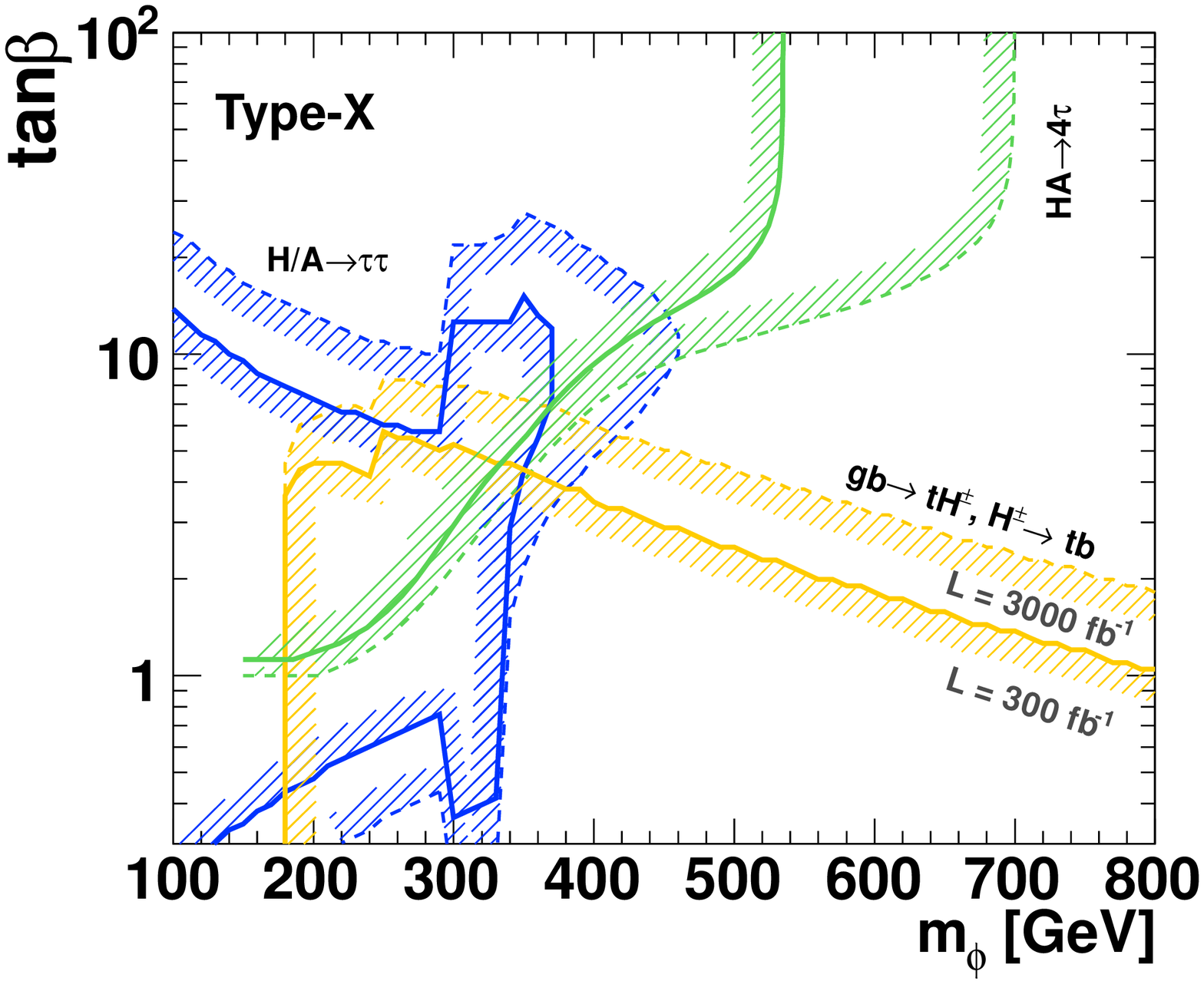}
  \quad
  \includegraphics[width=0.45\textwidth]{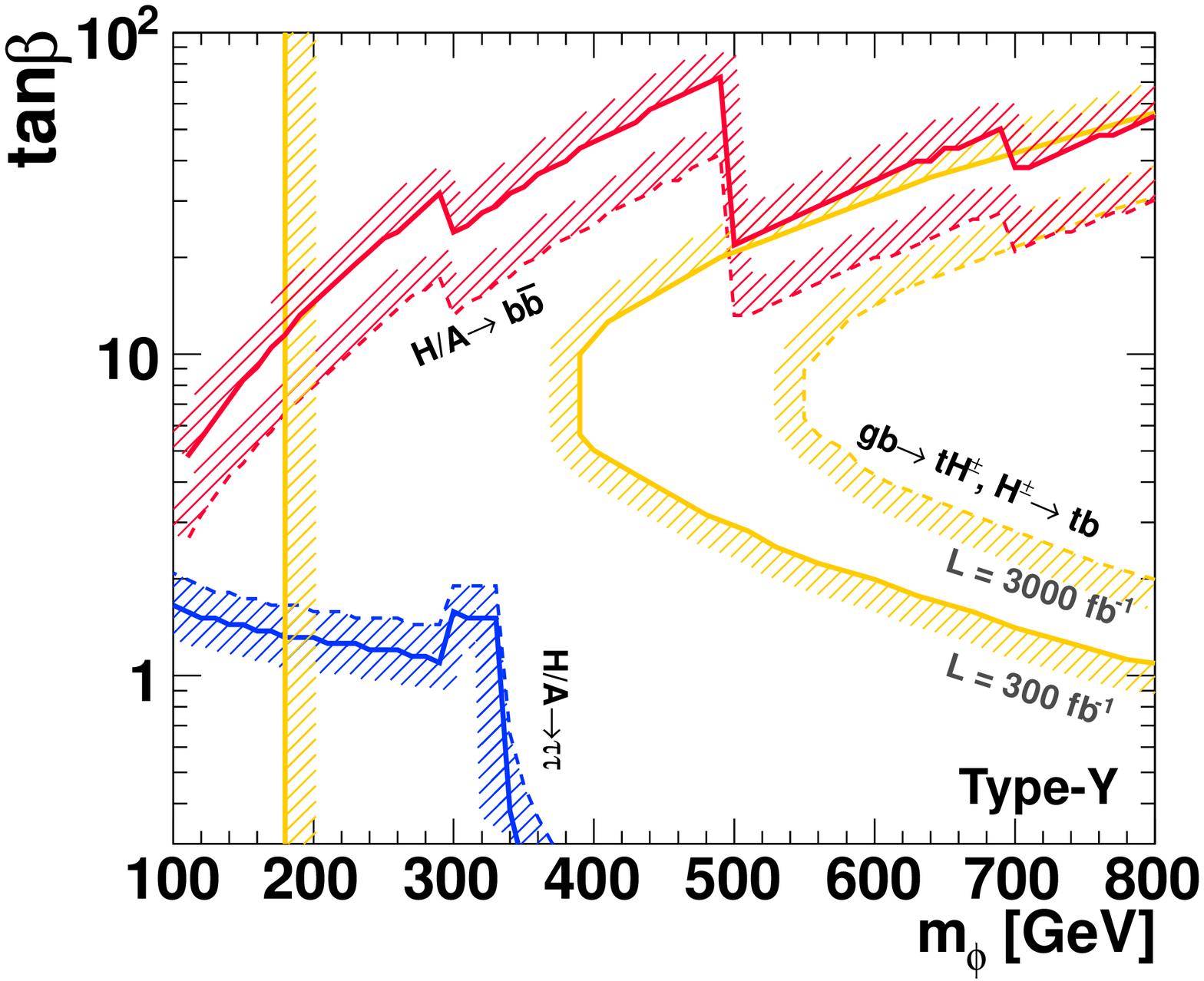}
  \caption{
  Expected exclusion regions ($2\sigma$ CL) in the
  plane of $\tan\beta$ and the mass scale $m_{\phi}$ of the additional
  Higgs bosons at the LHC.
  Curves are evaluated by using the signal and background analysis given
  in Ref.~\cite{ATLAS:1999vwa} for each process, where the signal events
  are rescaled to the prediction in each case~\cite{Asner:2013psa,KTYY},
  except the $4\tau$ process for which we follow the analysis in
  Ref.~\cite{Kanemura:2011kx}. 
  Thick solid lines are the expected exclusion contours by
  $L=300$~fb$^{-1}$ data, and thin dashed lines are for
  $L=3000$~fb$^{-1}$ data. 
  For Type-II, the regions indicated by circles may not be excluded by
  $H/A\to\tau^+\tau^-$ search by using the 300~fb$^{-1}$ data due to the
  large SM background. 
  }\label{fig:LHCreach}
 \end{center}
\end{figure}

In Fig.~\ref{fig:LHCreach}, we show the contour plots of the expected
exclusion regions [$2\sigma$ confidence level (CL)] in the
$(m_\phi,\tan\beta)$ plane, where $m_\phi$ represents common masses of
additional Higgs bosons, at the LHC $\sqrt{s}=14$~TeV with the
integrated luminosity of 300~fb$^{-1}$ (thick solid lines) and
3000~fb$^{-1}$ (thin dashed lines). 
The value of $M$ is also taken to the same as $m_\phi$.
From the top-left panel to the bottom-right panel, the results for
Type-I, Type-II, Type-X and Type-Y are shown separately. 
According to the analysis in Ref.~\cite{ATLAS:1999vwa}, we
change the reference values of the expected numbers of signal and
background events at certain values of the mass of additional Higgs
bosons~\cite{KTYY}.
This makes sharp artificial edges of the curves in
Fig.~\ref{fig:LHCreach}. 

For Type-I, $H/A$ production followed by their $\tau^+\tau^-$ decay can
be probed for the parameter regions of $\tan\beta\lesssim 3$ and
$m_{H,A}\le350$~GeV, where the inclusive production cross section is
enhanced by the relatively large top Yukawa coupling and 
also the $\tau^+\tau^-$ branching ratio is sizable. 
The $tH^\pm$ production followed by the $H^\pm\to tb$ decay can be
used to search $H^\pm$ in relatively smaller $\tan\beta$ regions. 
The mass reach for the discovery of $H^\pm$ can be up to 800~GeV for
$\tan\beta\lesssim1$ (2) for the integrated luminosity of 300~fb$^{-1}$
(3000~fb$^{-1}$). 

For Type-II, the inclusive and the bottom-quark-associated production
processes of $H/A$ followed by the $\tau^+\tau^-$ decay or the $b
\overline b$ decay can be used to search $H$ and $A$ in relatively large
$\tan\beta$ regions. 
They can also be used in relatively small $\tan\beta$ regions with
$m_{H,A}\lesssim350$~GeV. 
Because of the difficulty of separating the signal from the SM
background, the lighter mass regions (200 $\sim$ 300~GeV) may not be
excluded with the 300~fb$^{-1}$ data as loopholes are seen in the
figure. 
$H^\pm$ can be probed by the $tH^\pm$ production followed by
the $H^\pm\to tb$ decay for $m_{H^\pm}\gtrsim180$~GeV with relatively
small and large $\tan\beta$ regions. 
The regions of $m_{H^\pm}\gtrsim350$~GeV (500~GeV) can be excluded with the
300~fb$^{-1}$ (3000~fb$^{-1}$) data. 

For Type-X, $H$ and $A$ can be searched via the inclusive production and
$HA$ pair production processes by using their dominant decays into
$\tau^+\tau^-$. 
The inclusive production can exclude the regions of $\tan\beta\lesssim
10$ with $m_{H,A}\lesssim 350$~GeV, and the regions of up to
$m_{H,A}\simeq500$~GeV (700~GeV) with $\tan\beta\gtrsim10$
can be excluded by using the pair production with the
300~fb$^{-1}$ (3000~fb$^{-1}$) data. 
The search for $H^\pm$ is the similar to that for Type-I. 

For Type-Y, the inclusive production of $H$ and $A$ followed by their
$\tau^+\tau^-$ decays can be searched for the regions of
$\tan\beta\lesssim 2$ and $m_{H,A}\le350$~GeV, where the inclusive
production cross section is enhanced due to a large top Yukawa coupling
constant and the $\tau^+\tau^-$ branching ratio is sizable. 
The bottom-quark associated production of $H$ and $A$ followed by
$H/A\to b\bar{b}$ decays can be searched for the regions of
$\tan\beta\gtrsim 30$ up to $m_{H,A}\simeq800$~GeV. 
This process is also relevant for Type-II, but the constraint is weaker
than $H/A\to\tau^+\tau^-$ mode. 
The search of $H^\pm$ is similar to that for Type-II. 

If all the curves are combined by assuming that all the masses of
additional Higgs bosons are the same, the mass below 400~GeV~(350~GeV)
can be excluded by the 300~fb$^{-1}$ data, and the mass below
550~GeV~(400~GeV) can be excluded by the 3000~fb$^{-1}$ data for any
value of $\tan\beta$ for Type-II and Type-Y (Type-X). 
Only for Type-I, a universal mass bound cannot be given, namely the
regions with $\tan\beta\gtrsim 5$~(10) cannot be excluded by the
300~fb$^{-1}$~(3000~fb$^{-1}$) data. 
However, in the general 2HDM, the mass spectrum of additional Higgs
boson is less constrained, and has more degrees of freedom. 
Therefore, we can still find allowed parameter regions where we keep
$m_H$ to be relatively light but taking $m_A (\simeq m_{H^\pm})$ rather
heavy for the rho parameter constraint~\cite{Kanemura:2011sj}. 
Thus, the overlaying of these exclusion curves for different additional
Higgs bosons may be applied to only the case with $m_H=m_A=m_{H^\pm}$.

At the LHC, the discovery reach of $H^\pm$ is extensive in all types of
Yukawa interaction, because of the large cross section of the $gb\to
tH^\pm$ process followed by the $H^\pm\to tb$ decay. 
If $H^\pm$ is discovered at the LHC, the determination of its mass would
follow immediately~\cite{ATLAS:1999vwa,Rindani:2013mqa}. 
Hence, the next progress would be the determination of the type of Yukawa
interaction.
At the LHC, although some methods have been proposed by using the
observables related to the top-quark
spin~\cite{Gong:2012ru,Rindani:2013mqa}, we could not completely
distinguish the types of Yukawa interaction, because the Type-I and
Type-X, or Type-II and Type-Y posses the same coupling structure for the
$tbH^\pm$ interaction. 
Therefore, we have to look at the other process like the neutral Higgs
boson production processes. 
However, as we have seen in Fig.~\ref{fig:LHCreach}, there can be no
complementary process for the neutral Higgs boson searches in some
parameter regions; e.g., $m_{H,A}\gtrsim350$~GeV with relatively small
$\tan\beta$ depending on the type of the Yukawa interaction. 
On the other hand, at the ILC, as long as $m_{H,A}\lesssim 500$~GeV, 
the neutral Higgs bosons can be produced and investigated almost
independent of $\tan\beta$. 
Therefore, it would be an important task of the ILC to search for the
additional Higgs bosons with the mass of $350$-$500$~GeV, and to
determine the models and parameters, even after the LHC. 

We also note that the above results are obtained in the SM-like limit,
$\sin(\beta-\alpha)=1$. 
However, in the general 2HDM, $\sin(\beta-\alpha)$ is also a free
parameter. 
It is known that a deviation from the SM-like limit induces decay modes
of $H\to W^+W^-$, $ZZ$, $hh$ as well as $A\to
Zh$~\cite{Gunion:1989we,Gunion:1990kf,Craig:2013hca,%
Chen:2013emb,Baglio:2014nea}.
Especially, for Type-I with a large value of $\tan\beta$, branching
ratios of these decay modes can be dominant even with a small deviation
from the SM-like limit~\cite{Aoki:2009ha,Craig:2013hca}. 
For example, if $\sin^2(\beta-\alpha)=0.96$, the decay mode of 
$H\to W^+W^-$ is dominant in $\tan\beta\gtrsim2$ for Type-I, and the decay
branching ratio can be up to $\sim0.2$ depending on the value of
$\tan\beta$ for the other types~\cite{Aoki:2009ha}. 
Therefore, searches for additional Higgs bosons in these decay modes can
give significant constraints on the deviation of $\sin(\beta-\alpha)$
from the SM-like limit~\cite{ATLAS:2013zla,CMS:2013eua}, which is
independent of coupling constants of $hVV$. 

\section{Prospect for the searches for the additional Higgs bosons at the
 ILC}\label{sec:ilc} 

In this section, we perform the detailed studies on the production
cross section of additional Higgs bosons at the ILC and their collider
signatures via the subsequent decays of them. 
We compare the results among the four types of the Yukawa interaction
in the general 2HDM, and see how the type of Yukawa interaction can be
discriminated and how the parameters can be determined from the collider
signatures or kinematical distributions in the observed processes.

\subsection{Cross Sections}

The main production mechanisms of additional Higgs bosons 
are $e^+ e^- \to H A$ and $e^+ e^- \to H^+ H^-$, where a pair of
additional Higgs bosons is produced via gauge interactions. 
These processes open when the collision energy is above the sum of the
masses of the two scalars. 
For energies below the threshold, the single production processes, $e^+
e^- \to H(A) f \bar{f}$ and $e^+ e^- \to H^\pm f \bar{f}'$ are the
leading contributions~\cite{Kanemura:2000cw}. 
The single production processes are enhanced when the relevant Yukawa
coupling constants of $\phi f \bar{f}^{(')}$ are large. 
The cross sections of these processes have been studied
extensively~\cite{Kanemura:2000cw,Moretti:2002pa,Kiyoura:2003tg,%
Behnke:2013lya},
mainly for the MSSM or for the Type-II 2HDM. 

Here, we give numerical results in the general 2HDMs but with
softly-broken discrete symmetry with all types of Yukawa interaction. 
We consider the processes of 
\begin{subequations}
\begin{align}
 &e^+e^-\to \tau^+\tau^- H,\label{eq:tautauH}\\
 &e^+e^-\to b\bar{b} H,\label{eq:bbH}\\
 &e^+e^-\to t\bar{t} H,\label{eq:ttH}\\
 &e^+e^-\to \tau^-\nu H^+,\label{eq:taunuH}\\
 &e^+e^-\to \bar{t}b H^+.\label{eq:tbH}
\end{align}
\end{subequations}
The cross sections of the processes  where $H$ is replaced by 
$A$ in Eqs.~(\ref{eq:tautauH}-\ref{eq:ttH}),  
and those of the charge conjugated processes of the processes in
Eqs.~(\ref{eq:taunuH}, \ref{eq:tbH}) are not explicitly shown. 

For energies above the threshold of the pair production,
$\sqrt{s}>m_H+m_A$, the contribution from $e^+e^-\to HA$ can be
significant in the processes in Eqs.~(\ref{eq:tautauH}-\ref{eq:ttH}). 
Similarly for $\sqrt{s}>2m_{H^\pm}$, the contribution from $e^+e^-\to
H^+H^-$ can be significant in the processes in
Eqs.~(\ref{eq:taunuH}, \ref{eq:tbH}). 
Below the threshold, the processes including diagrams of 
$e^+e^-\to f \bar{f}^\ast$ and $e^+e^-\to f^\ast\bar{f}$ dominate.

\begin{figure}[t]
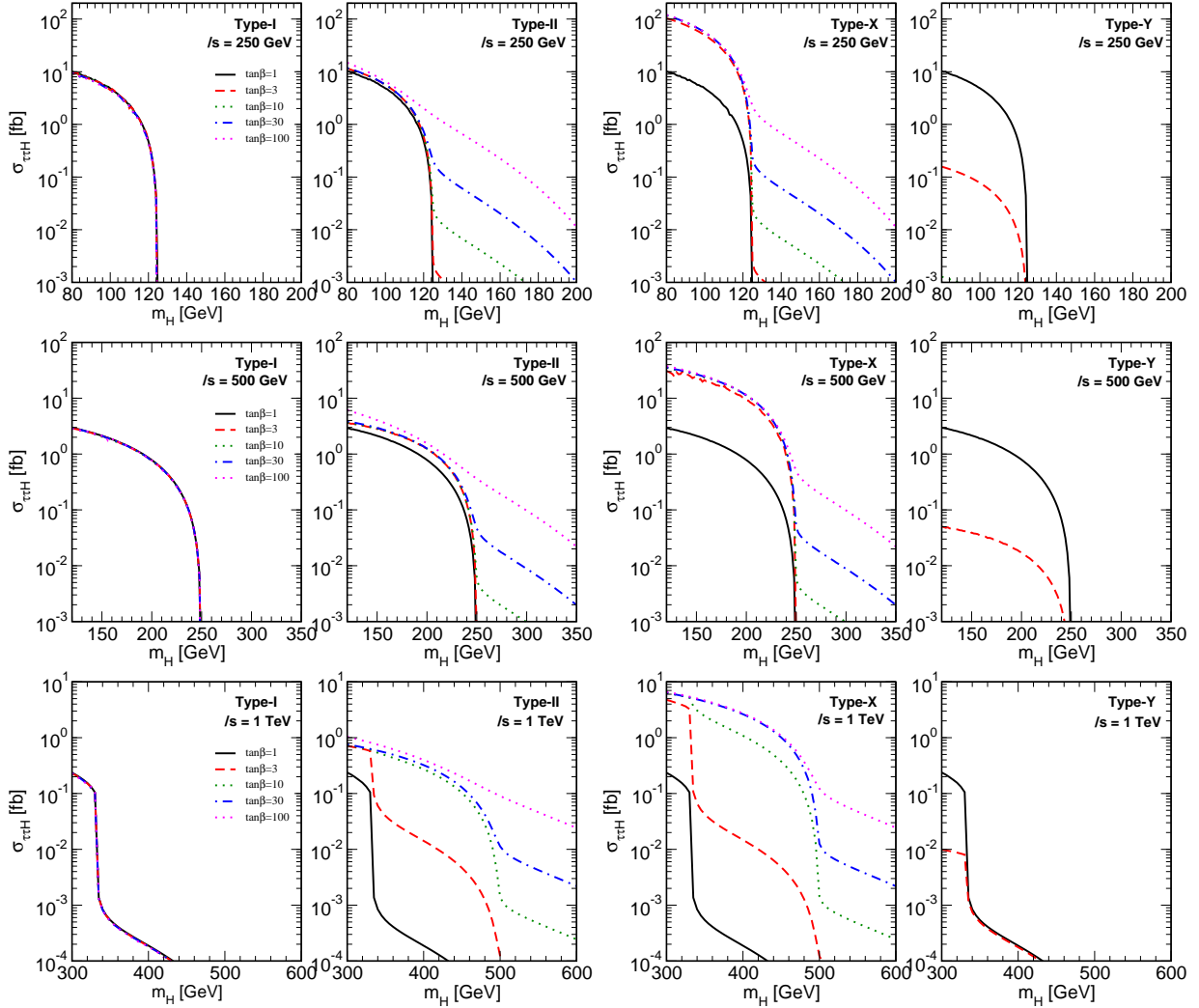

 \begin{center}
  \includegraphics[width=.49\textwidth,clip]{Sig_tataH_250_III.eps}
  \includegraphics[width=.49\textwidth,clip]{Sig_tataH_250_XY.eps}
  \includegraphics[width=.49\textwidth,clip]{Sig_tataH_500_III.eps}
  \includegraphics[width=.49\textwidth,clip]{Sig_tataH_500_XY.eps}
  \includegraphics[width=.49\textwidth,clip]{Sig_tataH_1TeV_III.eps}
  \includegraphics[width=.49\textwidth,clip]{Sig_tataH_1TeV_XY.eps}
  \caption{Cross sections of $e^+e^-\to \tau^+\tau^- H$ process as a
  function of $m_H=m_A$ at the ILC $\sqrt{s}=250$~GeV, 500~GeV and
  1~TeV. 
  Several values of $\tan\beta$ are examined with fixing
  $\sin(\beta-\alpha)=1$.}
  \label{fig:SigtataH}
 \end{center}
\end{figure}

In Fig.~\ref{fig:SigtataH}, the cross sections of
$e^+e^-\to\tau^+\tau^-H$ are shown as a function of $m_H$ for various 
situations. 
The cross sections for $\sqrt{s}=250$~GeV, 500~GeV and 1~TeV are shown
in the figures of the first, second and third rows, while figures in the
first to the fourth columns show the results in Type-I to Type-Y,
respectively. 
In the first row, curves are the cross sections of $e^+e^-\to
\tau^+\tau^-H$ for $\tan\beta=1$, 3, 10, 30 and 100 at the
ILC $\sqrt{s}=250$~GeV. 
The cross sections rapidly fall down at the mass threshold
$\sqrt{s}=m_H+m_A$. 
As stated above, for energies above the threshold of the $HA$ production,
$\sqrt{s}>m_H+m_A$, the cross sections come mainly from the pair
production $e^+e^-\to HA$ followed by the $A\to\tau^+\tau^-$ decay. 
Since the $HA$ production cross section does not depend on the type of
Yukawa interaction nor the value of $\tan\beta$, 
the $\tan\beta$ dependence in the process of $e^+e^-\to HA$ with
$H/A\to\tau^+\tau^-$ only comes from the decay branching ratios of $H$
and $A$, which are shown in Fig.~\ref{fig:Br_125}.
Below the threshold, $\sqrt{s}<m_H+m_A$, only the single production
processes contribute which are sensitive to $\tan\beta$, depending on
the type of Yukawa interaction. 
For Type-II and Type-X with large $\tan\beta$, the cross sections of
$e^+e^-\to\tau^+\tau^-H$ via the single production mechanism are
enhanced by the Yukawa couplings of $H\tau\tau/A\tau\tau$, while for
Type-I and Type-Y the cross sections are negligible. 
Figures in the second and third rows show the similar results but for
$\sqrt{s}=500$~GeV and 1~TeV, respectively. 
For the latter case, the decay of $H/A\to t\bar{t}$ opens for
$m_H\gtrsim 350$~GeV, and then the decay into $\tau^+\tau^-$ is
suppressed to a large extent. 

\begin{figure}[t]
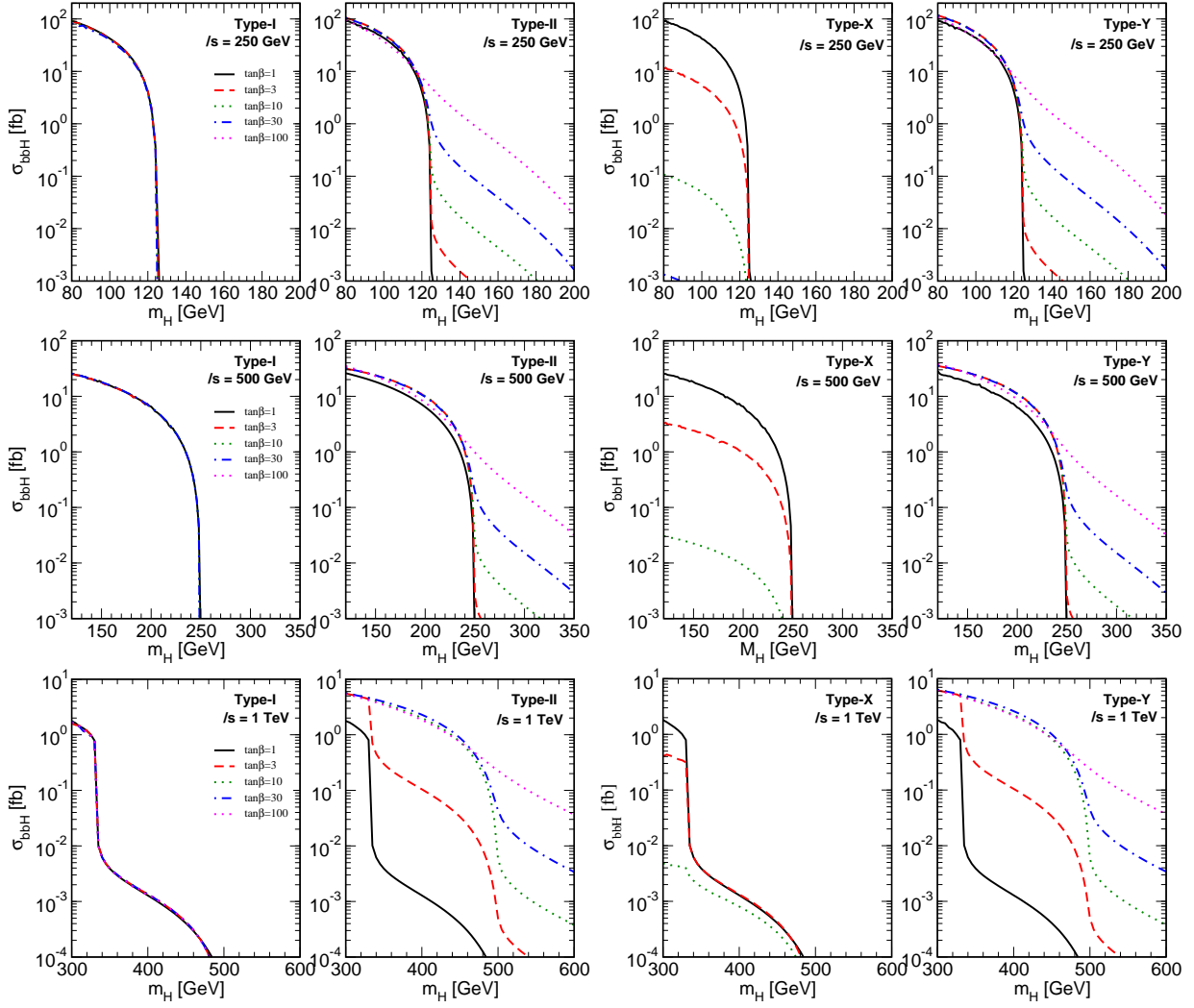

 \begin{center}
  \includegraphics[width=.49\textwidth,clip]{Sig_bbH_250_III.eps}
  \includegraphics[width=.49\textwidth,clip]{Sig_bbH_250_XY.eps}
  \includegraphics[width=.49\textwidth,clip]{Sig_bbH_500_III.eps}
  \includegraphics[width=.49\textwidth,clip]{Sig_bbH_500_XY.eps}
  \includegraphics[width=.49\textwidth,clip]{Sig_bbH_1TeV_III.eps}
  \includegraphics[width=.49\textwidth,clip]{Sig_bbH_1TeV_XY.eps}
  \caption{Cross sections of $e^+e^-\to b\bar{b}H$ process at the
  ILC $\sqrt{s}=250$~GeV, 500~GeV and 1~TeV, evaluated as the same
  manner as Fig.~\ref{fig:SigtataH}.}
  \label{fig:SigbbH}
 \end{center}
\end{figure}

In Fig.~\ref{fig:SigbbH}, the cross sections of
$e^+e^-\to b\bar{b}H$ are shown as a function of $m_H$ for various
situations in the same manner as Fig.~\ref{fig:SigtataH}. 
In the first row, cross sections of $e^+e^-\to
b\bar{b}H$ are plotted for $\tan\beta=1$, 3, 10, 30 and 100 at the ILC
$\sqrt{s}=250$~GeV. 
For this process, Type-II and Type-Y have enhanced single production
cross section for large $\tan\beta$, due to the enhanced Yukawa couplings
of $H$ and $A$ to $b$ quarks.
Figures in the second and third rows show the similar results but for
$\sqrt{s}=500$~GeV and 1~TeV, respectively. 
For $m_{H,A}\gtrsim 350$~GeV, the cross sections decrease because the
decay of $H/A\to t\bar{t}$ becomes dominant. 

\begin{figure}[t]
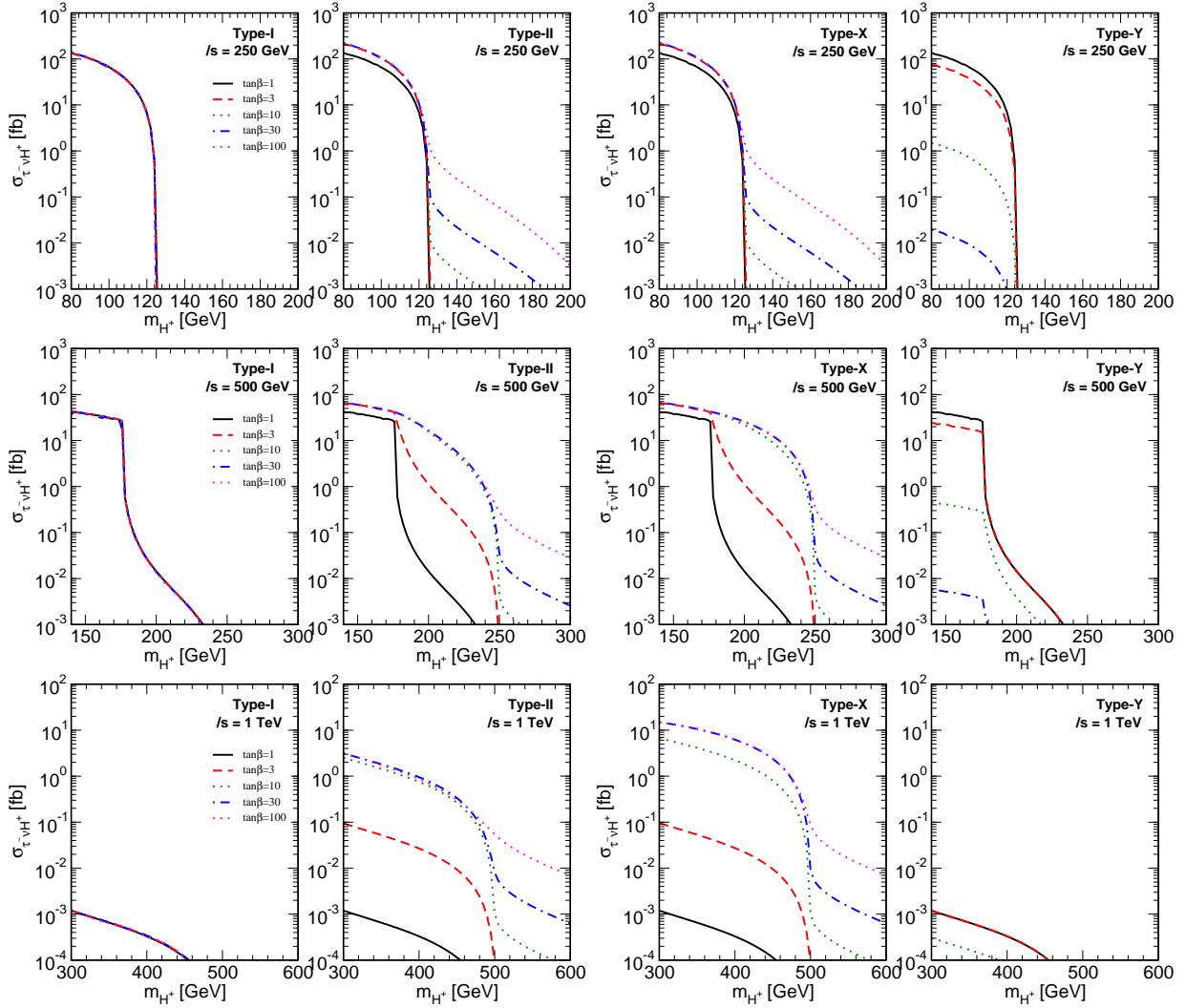

 \begin{center}
  \includegraphics[width=.49\textwidth,clip]{Sig_tanuH_250_III.eps}
  \includegraphics[width=.49\textwidth,clip]{Sig_tanuH_250_XY.eps}
  \includegraphics[width=.49\textwidth,clip]{Sig_tanuH_500_III.eps}
  \includegraphics[width=.49\textwidth,clip]{Sig_tanuH_500_XY.eps}
  \includegraphics[width=.49\textwidth,clip]{Sig_tanuH_1TeV_III.eps}
  \includegraphics[width=.49\textwidth,clip]{Sig_tanuH_1TeV_XY.eps}
  \caption{Cross sections of $e^+e^-\to \tau^-\nu H^+$ process as a
  function of $m_{H^\pm}$ at the ILC $\sqrt{s}=250$~GeV, 500~GeV and
  1~TeV.
  Several values of $\tan\beta$ are examined with fixing
  $\sin(\beta-\alpha)=1$.}
  \label{fig:SigtanuH}
 \end{center}
\end{figure}

In Fig.~\ref{fig:SigtanuH}, cross sections of $e^+e^-\to
\tau^-\nu H^+$ are shown as a function of $m_{H^\pm}$ for various
situations in the same manner as Fig.~\ref{fig:SigtataH}. 
In the first row, cross sections of $e^+e^-\to
\tau^-\nu H^+$ are plotted for $\tan\beta=1$, 3, 10, 30 and 100 at the
ILC $\sqrt{s}=250$~GeV. 
For energies below the threshold, $\sqrt{s}<2m_{H^\pm}$, the single
production process can be sizable for Type-II and Type-X, due to the
enhanced $\tau\nu H^\pm$ couplings by $\tan\beta$. 
In the second row, for $\sqrt{s}=500$~GeV, there is a sharp edge at
around $m_{H^\pm}=180$~GeV for Type-I, Type-Y and also for Type-II and
Type-X with small $\tan\beta$, because the decay of $H^\pm\to tb$
opens. 
In the third row, for $\sqrt{s}=1$~TeV, only for Type-II and Type-X the
cross sections increase with $\tan\beta$.

\begin{figure}[t]
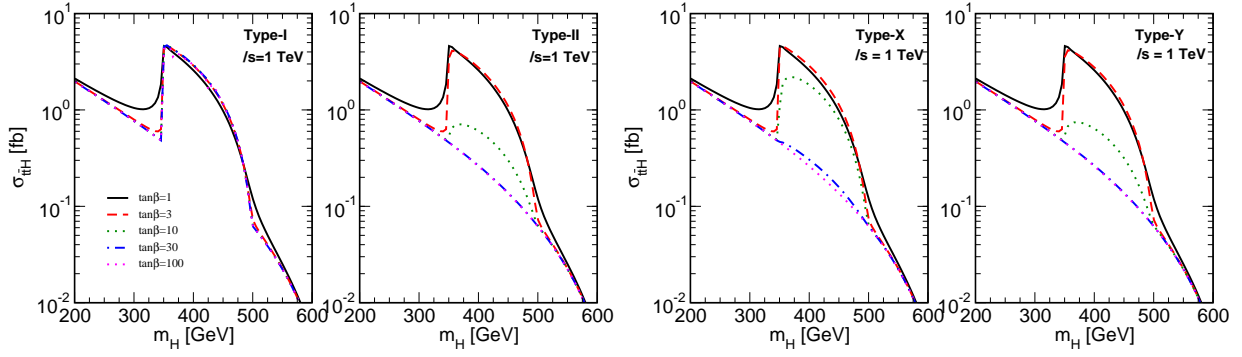

 \begin{center}
  \includegraphics[width=.49\textwidth,clip]{Sig_ttH_1TeV_III.eps}
  \includegraphics[width=.49\textwidth,clip]{Sig_ttH_1TeV_XY.eps}
  \caption{Cross sections of $e^+e^-\to t\bar{t}H$ process at the
  ILC $\sqrt{s}=1$~TeV.
  }
  \label{fig:SigttH}
 \end{center}
\end{figure}

In Fig.~\ref{fig:SigttH}, cross sections of $e^+e^-\to
t\bar{t}H$ are shown as a function of $m_{H}$ for various
situations for $\sqrt{s}=1$~TeV.
Figures from left to right show the results in Type-I to
Type-Y, respectively. 
The cross sections rise sharply at the top quark pair threshold,
$m_{H}\simeq 350$~GeV. 
Below the top pair threshold, $m_A<2m_t$, $e^+e^-\to HA\to H t\bar{t}$
process is kinematically suppressed, but only the single production
mechanism through the Yukawa interaction to the top quark can
contribute. 
For $350$~GeV $\le m_H\le500$~GeV, as long as the decay branching ratio
of $A\to t\bar{t}$ is sizable, the cross section is enhanced via the
$HA$ production process. 
For $m_H\ge500$~GeV, $HA$ pair production is kinematically forbidden,
and the single production becomes the leading mechanism. 
In all types, the Yukawa couplings of $H$ and $A$ to the top quark are
suppressed for large $\tan\beta$. 

\begin{figure}[t]
 \begin{center}
  \includegraphics[width=.49\textwidth,clip]{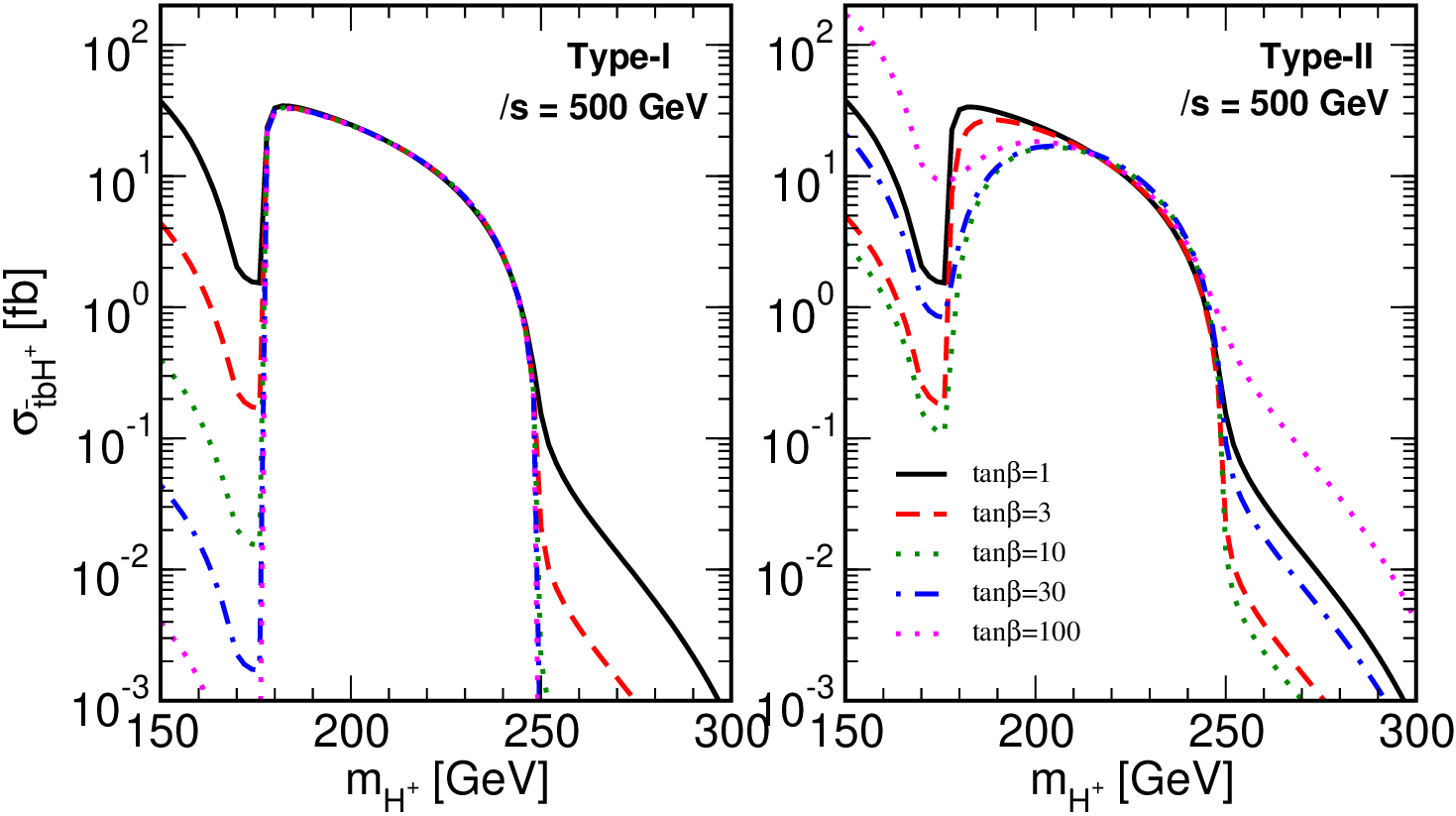}
  \includegraphics[width=.49\textwidth,clip]{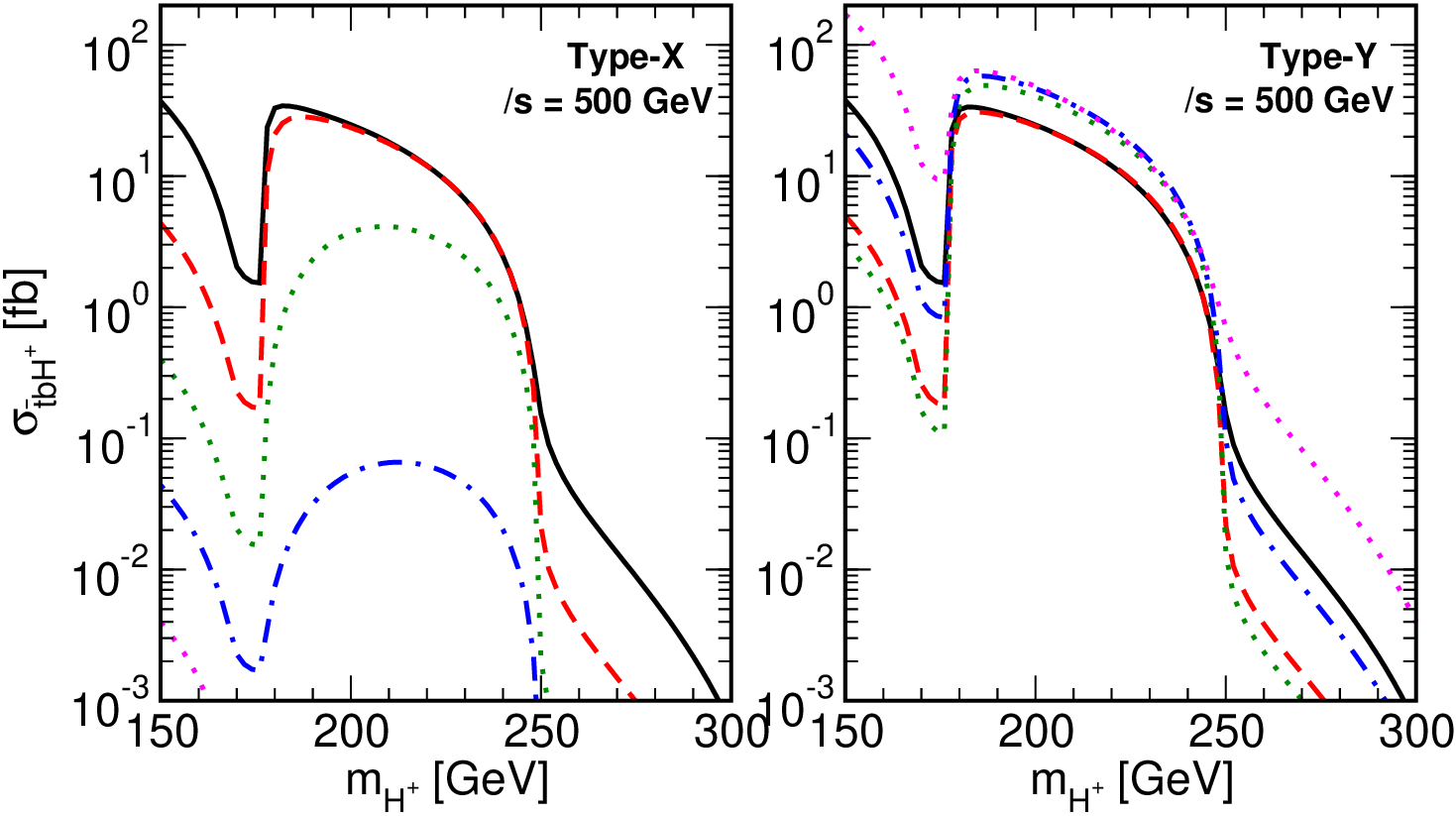}
  \includegraphics[width=.49\textwidth,clip]{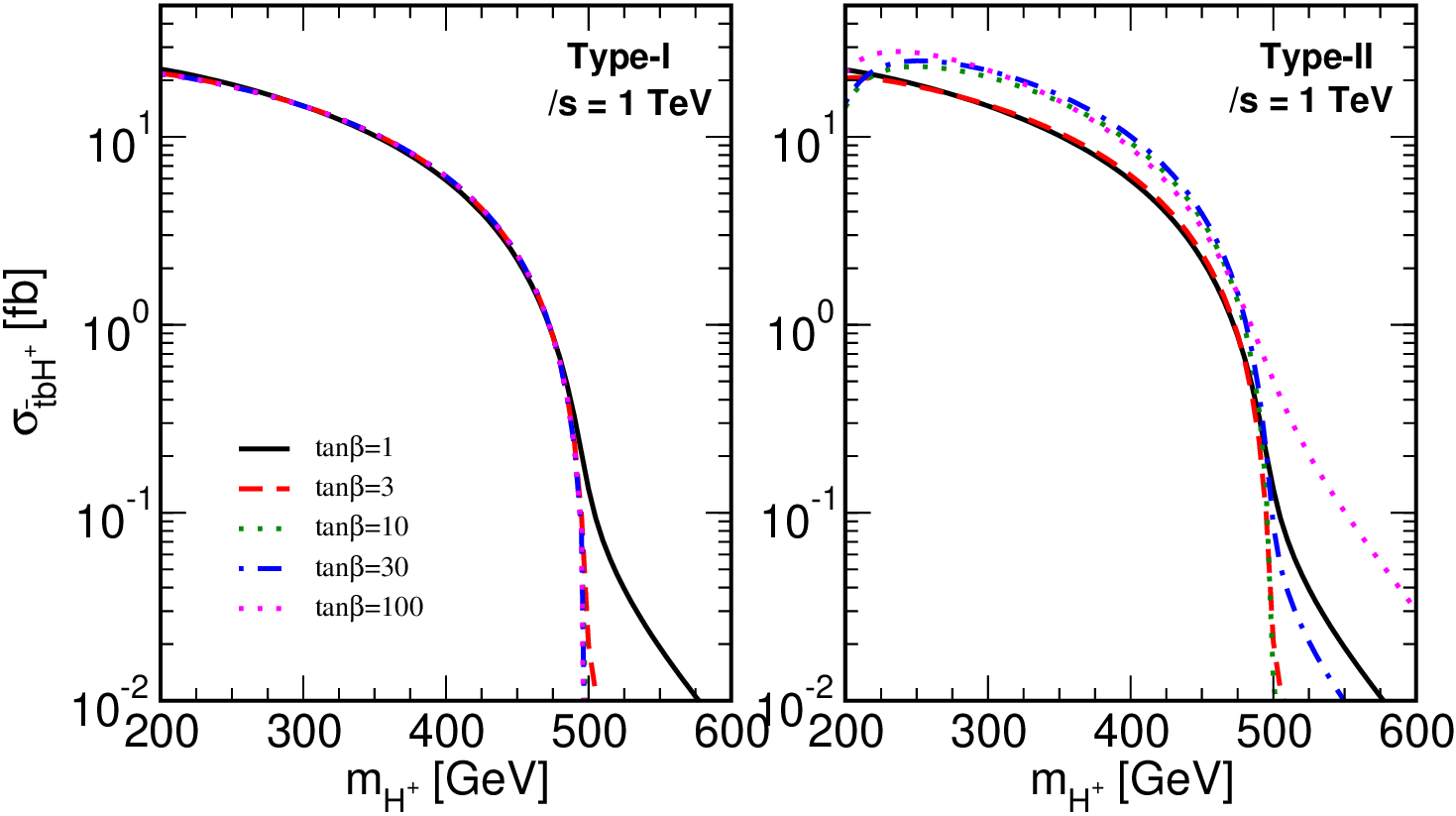}
  \includegraphics[width=.49\textwidth,clip]{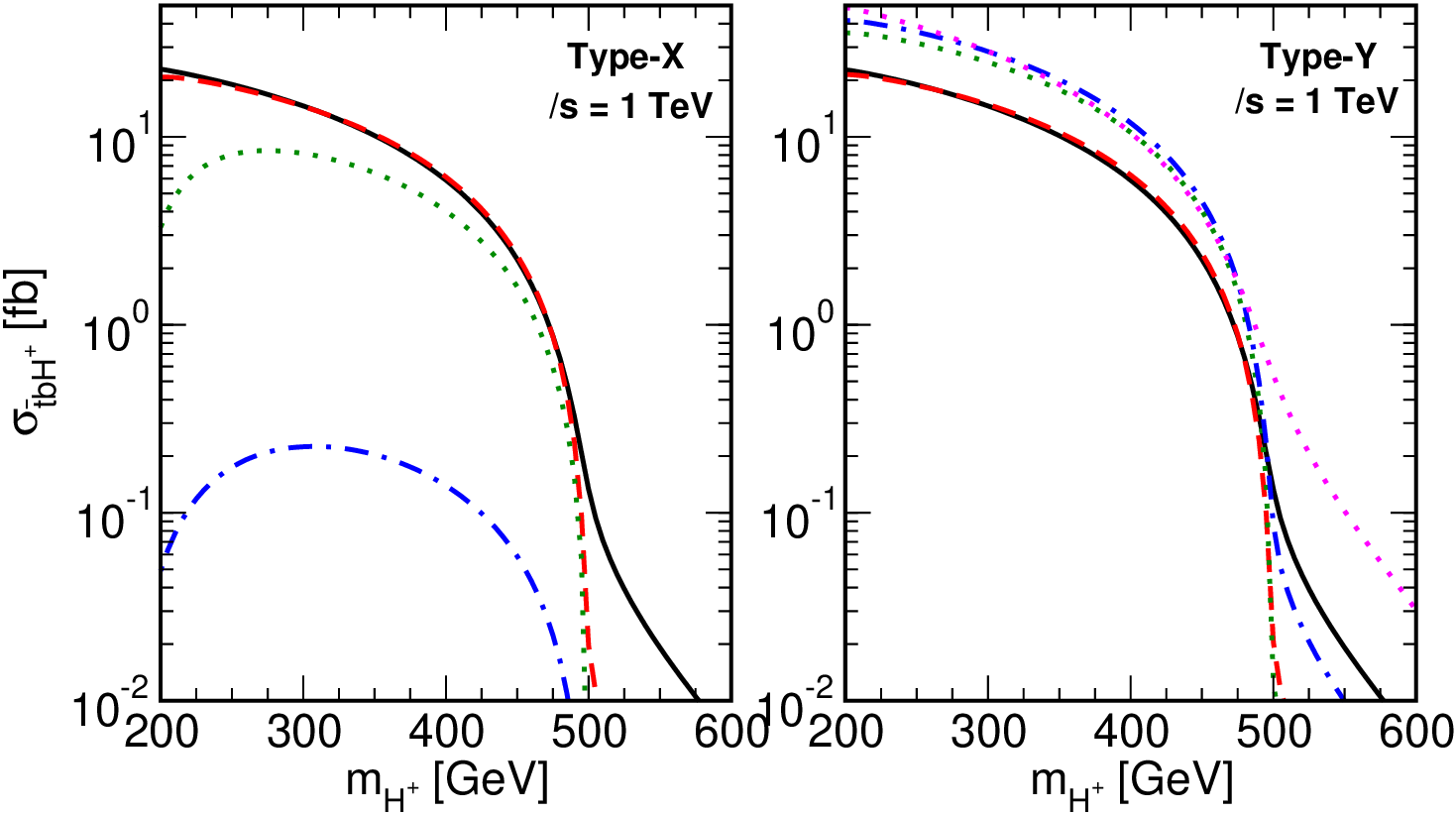}
  \caption{Cross sections of $e^+e^-\to t\bar{b}H^-$ process at the
  ILC $\sqrt{s}=500$~GeV and 1~TeV.
  }
  \label{fig:SigtbH}
 \end{center}
\end{figure}

In Fig.~\ref{fig:SigtbH}, cross sections of $e^+e^-\to \bar{t}bH^+$
 are plotted as a function of $m_{H^\pm}$.
In the first row, the results for $\sqrt{s}=500$~GeV are shown.
For $m_t+m_b\le m_{H^\pm}\le250$~GeV, the pair production $e^+e^-\to
 H^+H^-$ followed by the decay of $H^-\to \bar{t}b$ gives the largest 
contribution. 
The cross section of $e^+e^-\to H^+H^-$ does not depend on $\tan\beta$,
 but only the branching ratio of the decay $H^\pm\to tb$ does. 
For $m_{H^{\pm}}\le m_t-m_b$ and $\sqrt{s}\ge2m_t$, there is a
 production mechanism of $\bar{t}bH^+$ from $e^+e^-\to t\bar{t}$
followed by the decay of $t\to bH^+$. 
The partial decay width of $t\to bH^\pm$ can be found e.g.\ in 
Ref.~\cite{Aoki:2009ha}. 
For $m_{H^{\pm}}\ge250$~GeV, only the single production mechanism
contributes for Type-II and Type-Y, which is enhanced by $\cot\beta$ via
the top quark Yukawa coupling or by $\tan\beta$ via the bottom quark Yukawa
coupling.
In the second row, the same results but for $\sqrt{s}=1$~TeV are shown.


\subsection{Contour Plot}

Now we discuss the collider signatures of additional Higgs
boson production at the ILC.
Both the pair and single production processes of additional Higgs bosons
tend to result in four-particle final-states (including neutrinos) when
the decays of the additional Higgs bosons are taken into account. 
To evaluate the net production rates of them, the production
cross sections and the decay branching ratios of additional Higgs bosons
have to be taken into account consistently.
We calculate the cross sections of various four-particle final-states
for given masses of additional Higgs bosons and $\tan\beta$ with setting
$\sin(\beta-\alpha)=1$, and draw contour curves where the cross sections
are $0.1$~fb~\cite{Kiyoura:2003tg}. 
This value is chosen commonly for all processes as it could be regarded
as a typical order of magnitude of the cross section of the additional
Higgs boson production. 
In addition, this value can also be considered as a criterion for
observation with the expected integrated luminosity at the
ILC~\cite{Djouadi:2007ik,Behnke:2013lya}. 
Certainly, the detecting efficiencies are different for different
four-particle final-states.
Moreover, the decay of unstable particles such as tau leptons and
top quarks have to be considered if they are involved. 
Expected background processes and a brief strategy of observing the
signatures are discussed later. 
We here restrict ourselves to simply compare the various four-particle 
production processes in four types of Yukawa interaction in the 2HDMs
with taking the criterion of 0.1~fb as a magnitude of the cross
sections. 
Our calculation is performed at the tree level by {\tt
Madgraph}~\cite{Alwall:2011uj}, by taking into account both the pair and
single production of additional Higgs bosons followed by their
subsequent decays. 
We note that in Ref.~\cite{Kiyoura:2003tg}, the cross sections without
including the decay of additional Higgs bosons have been studied in the
MSSM, while in our paper we study the cross sections of the
four-particle final-states by including the decays of additional Higgs
bosons in the 2HDMs with four types of Yukawa interaction. 

\begin{figure}[t]
 \includegraphics[width=0.328\textwidth]{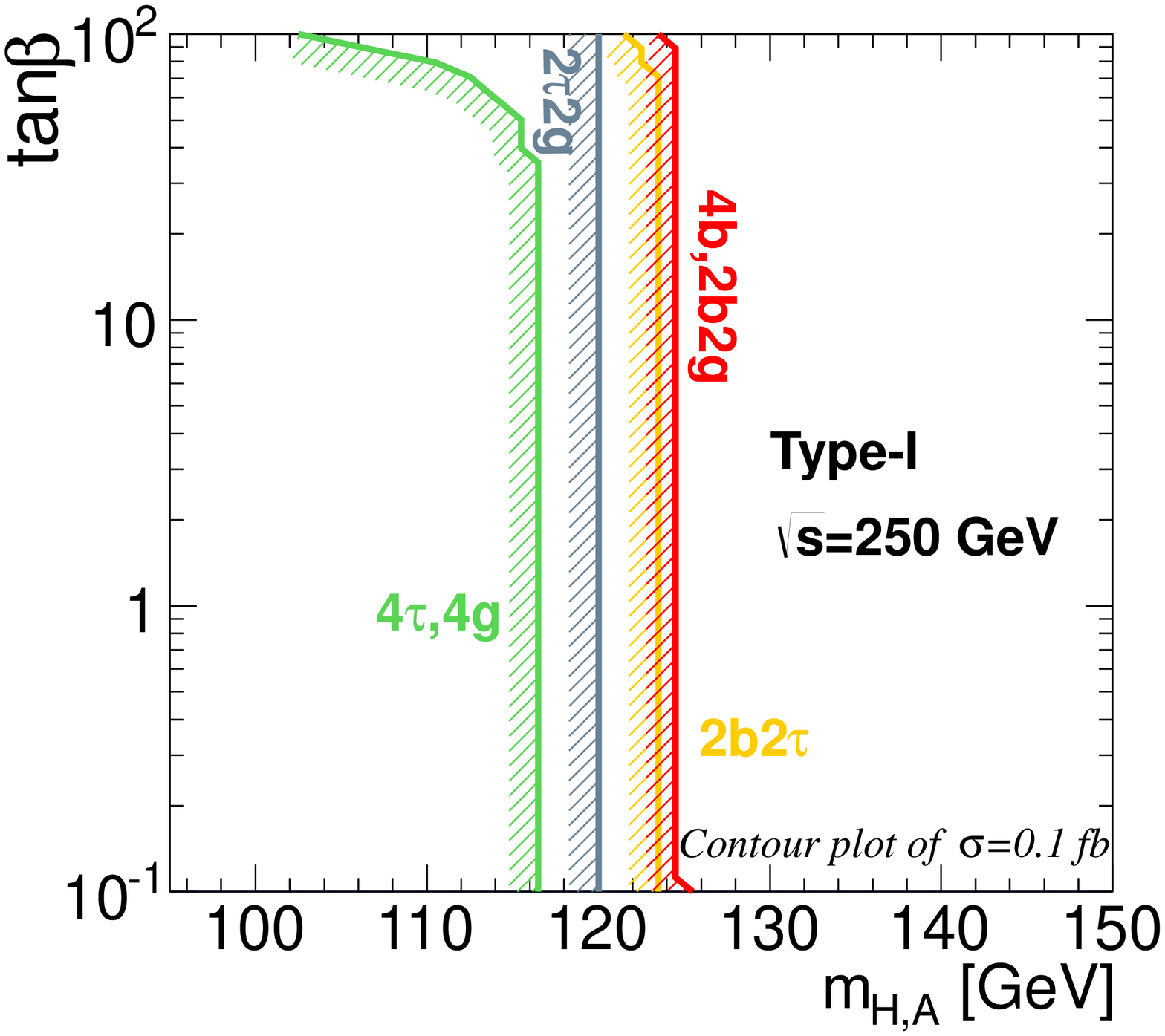} 
 \includegraphics[width=0.328\textwidth]{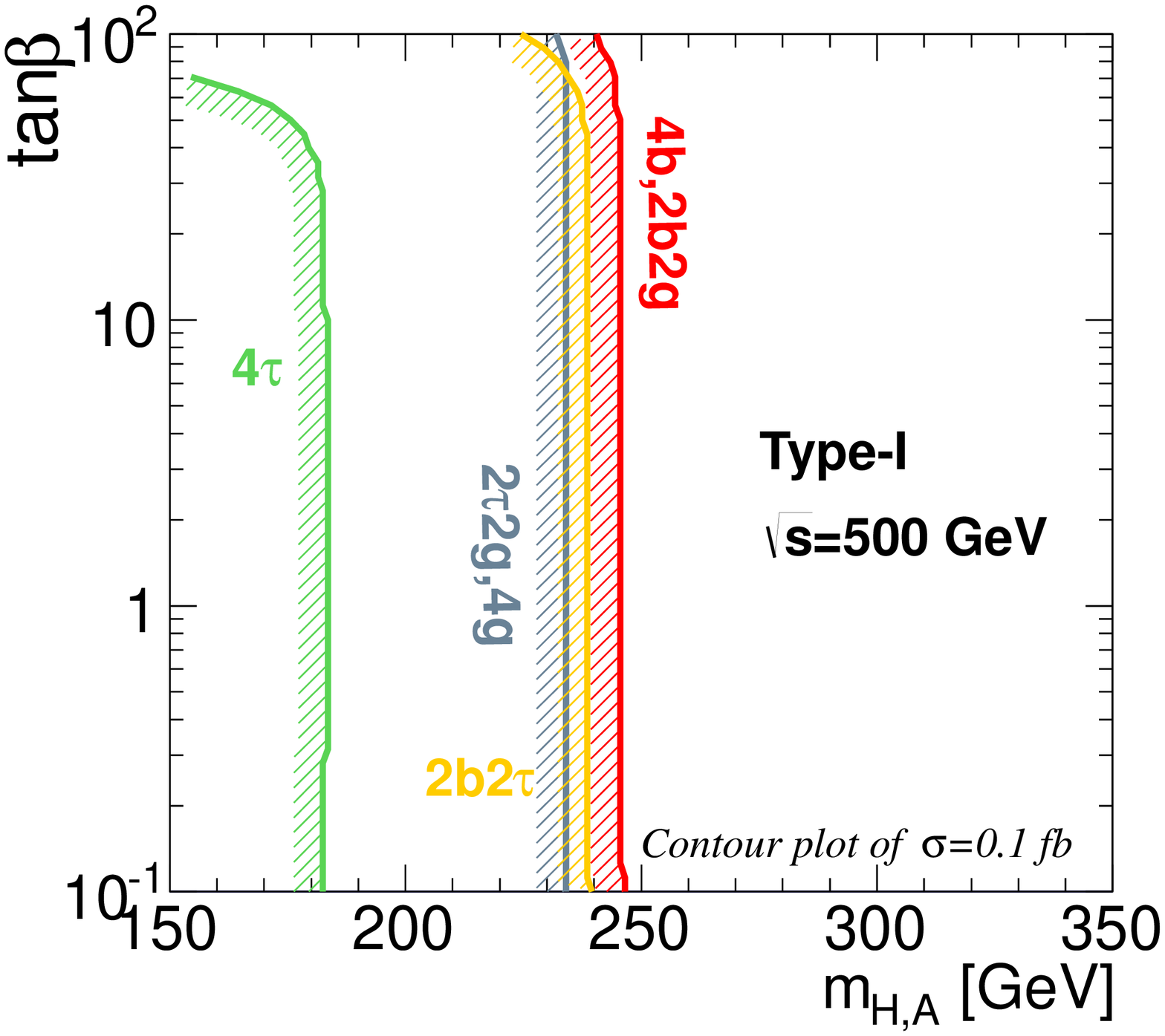} 
 \includegraphics[width=0.328\textwidth]{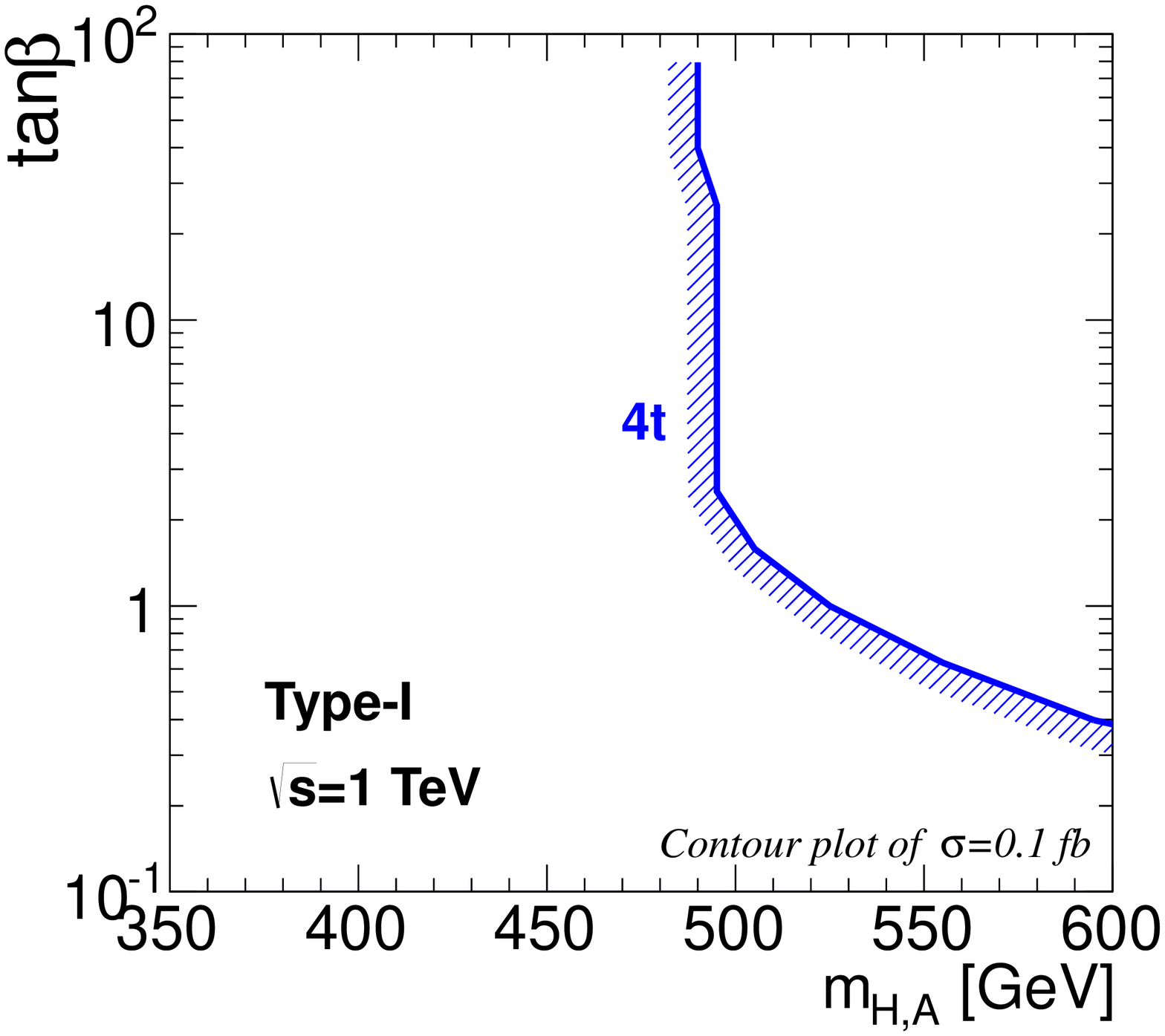} 
 \includegraphics[width=0.328\textwidth]{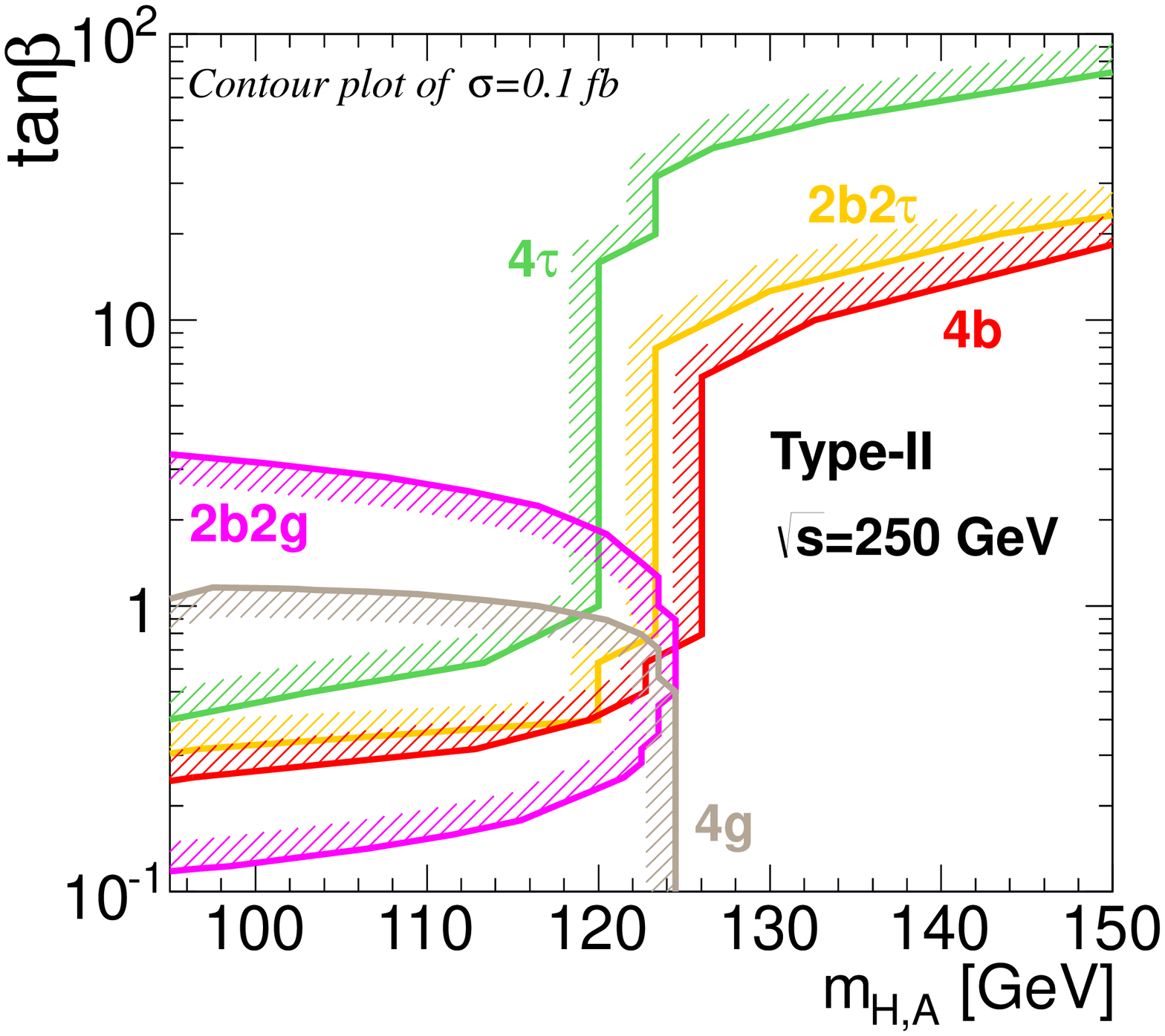} 
 \includegraphics[width=0.328\textwidth]{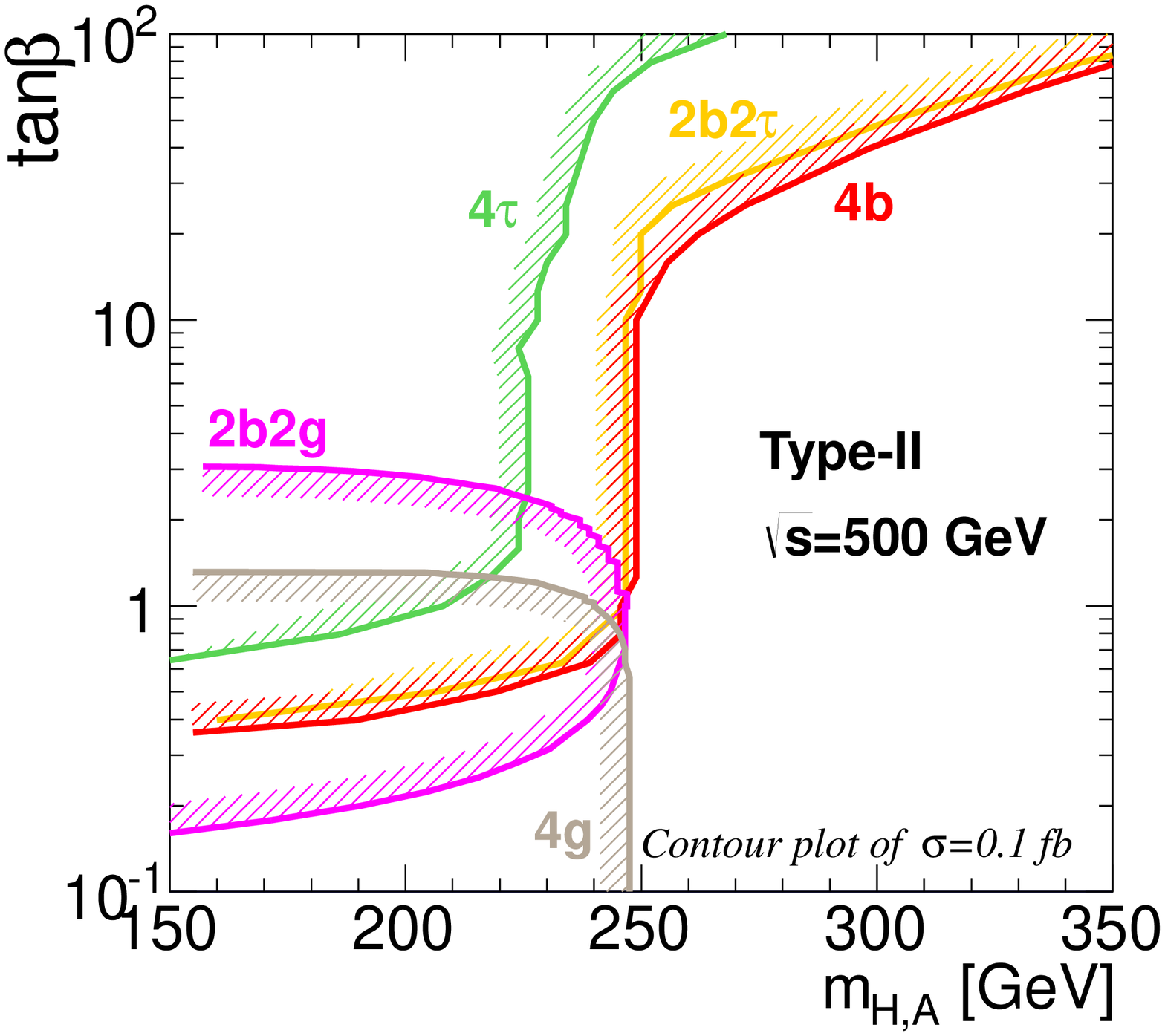} 
 \includegraphics[width=0.328\textwidth]{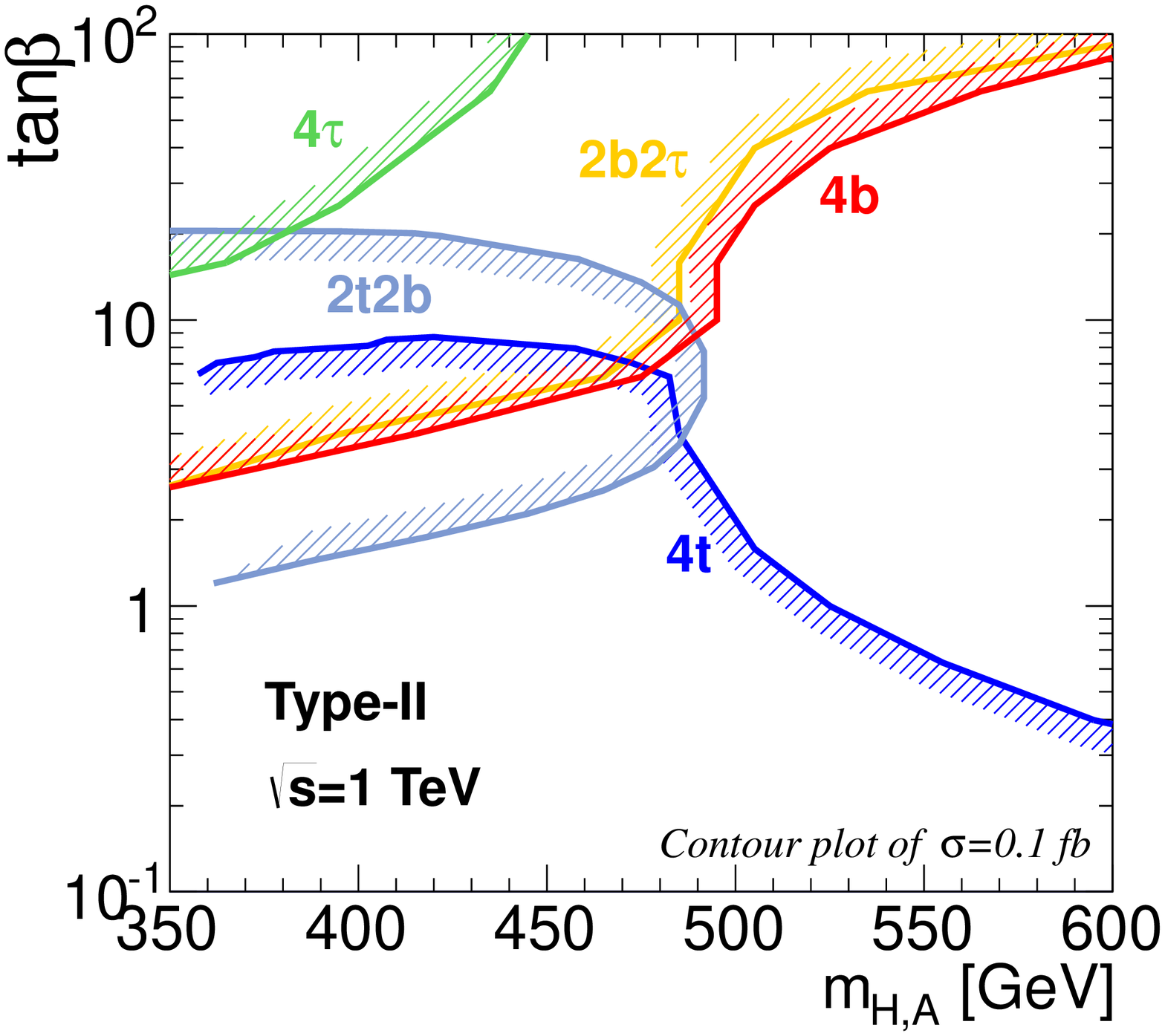} 
 \includegraphics[width=0.328\textwidth]{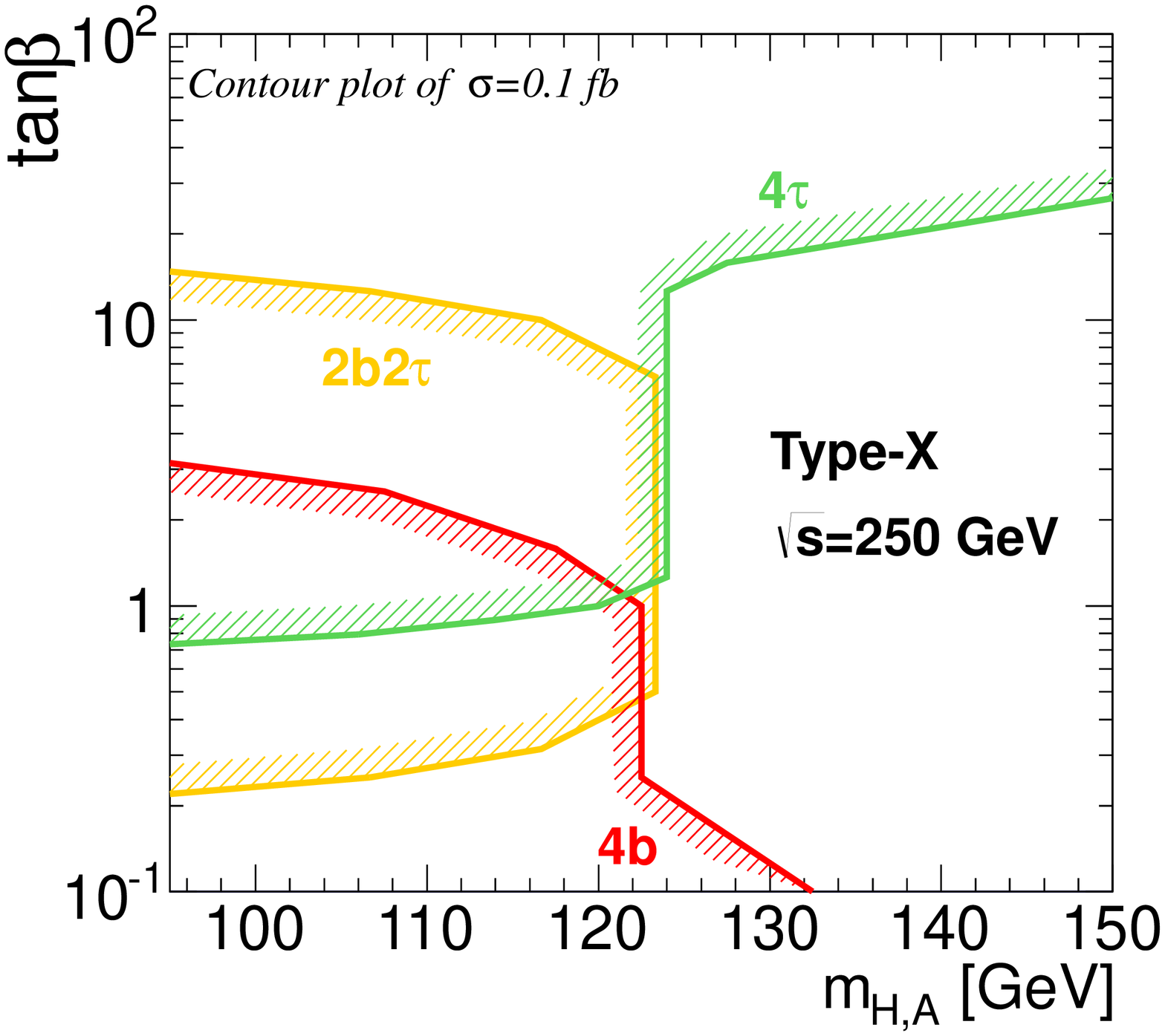} 
 \includegraphics[width=0.328\textwidth]{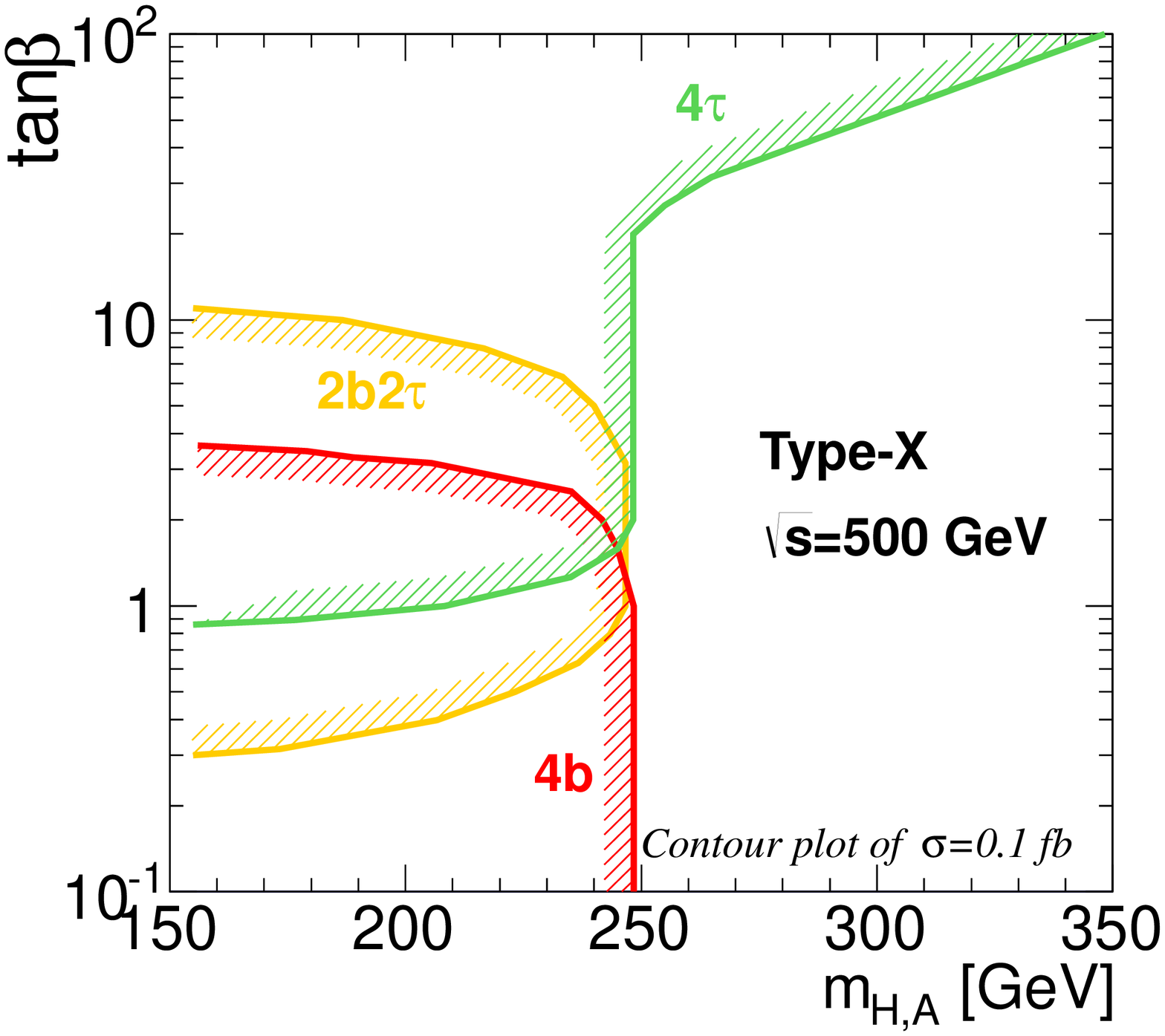} 
 \includegraphics[width=0.328\textwidth]{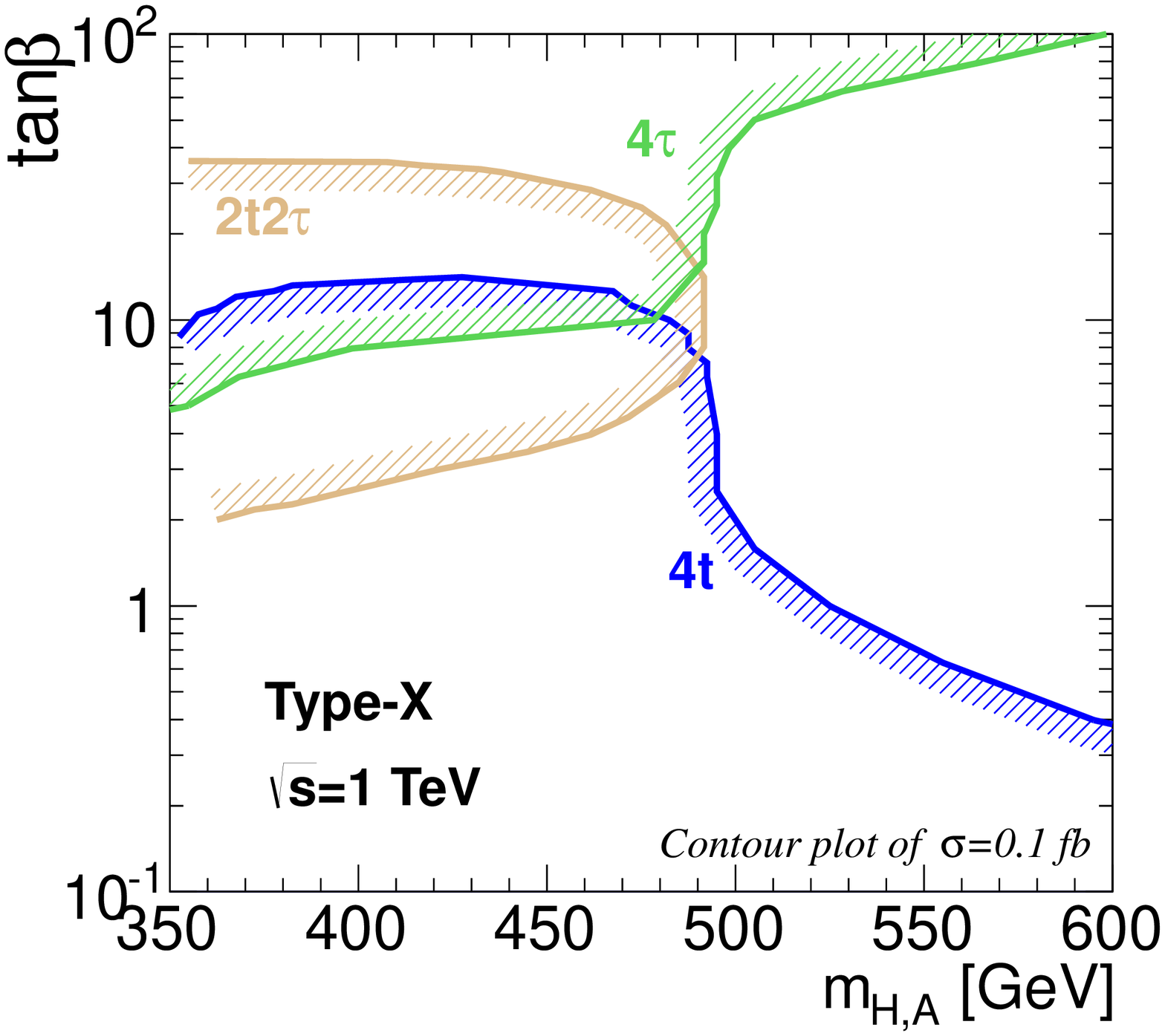} 
 \includegraphics[width=0.328\textwidth]{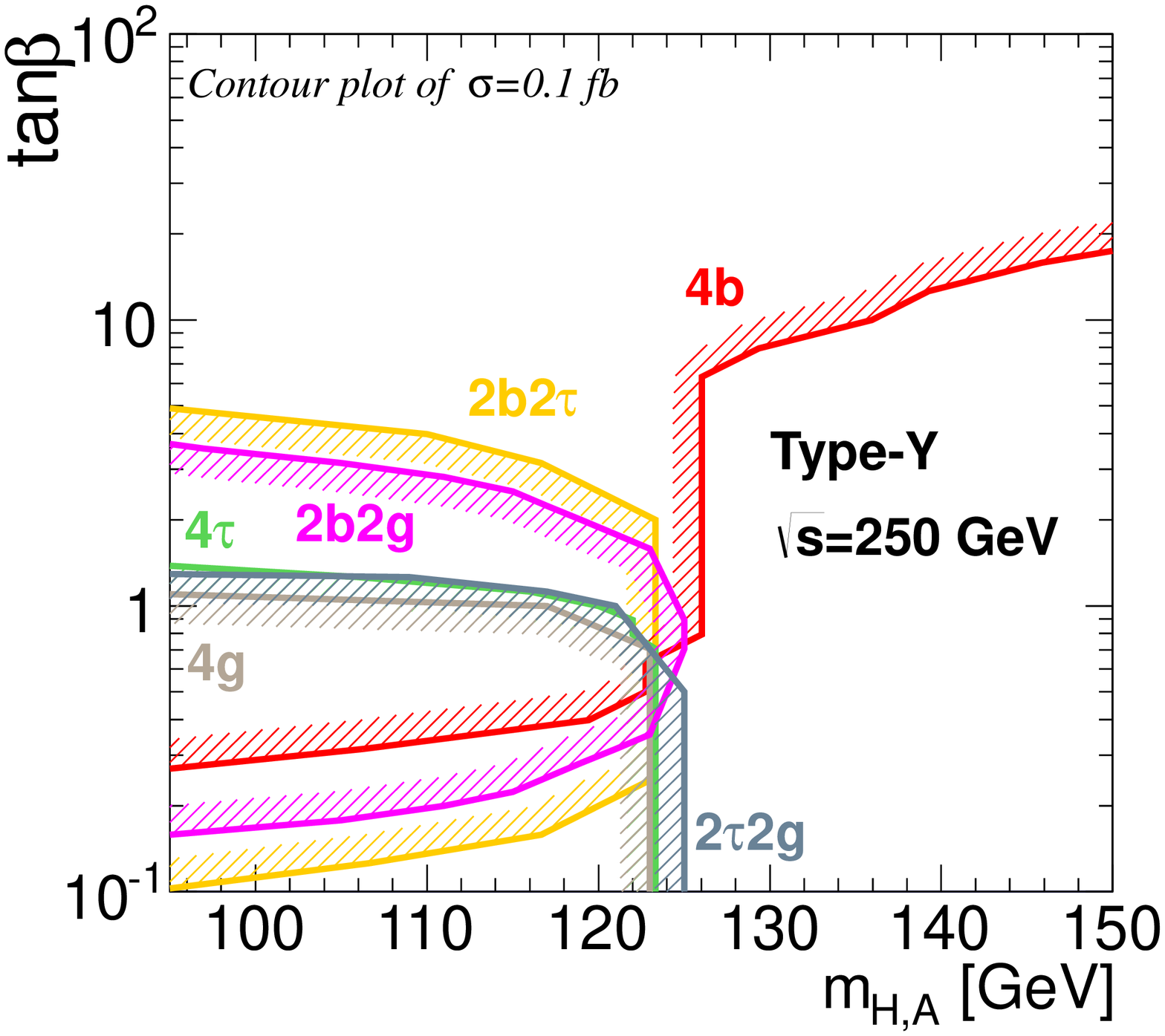} 
 \includegraphics[width=0.328\textwidth]{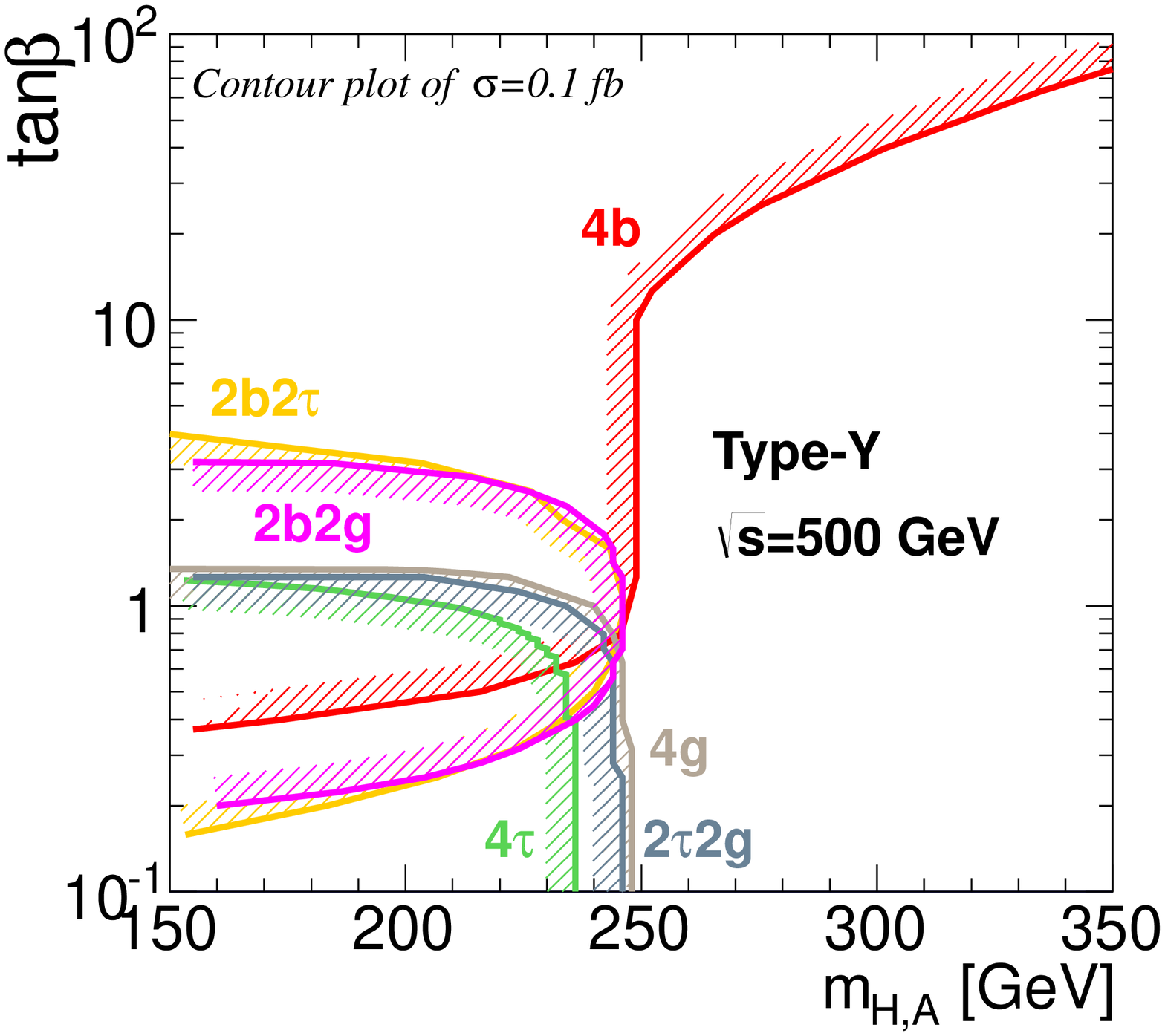} 
 \includegraphics[width=0.328\textwidth]{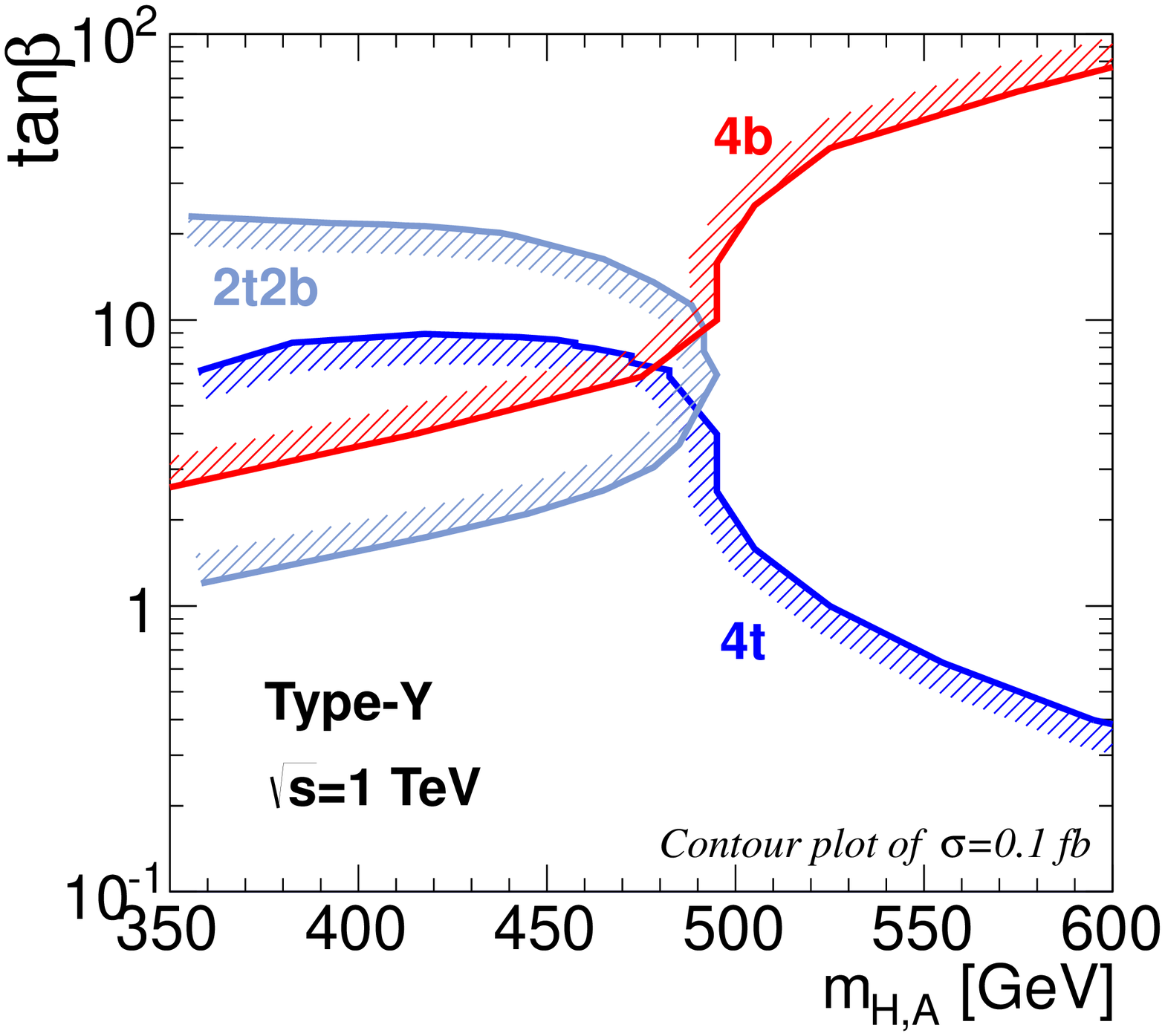} 
 \caption{Contour plots of the four-particle production cross sections
 through the $H$ and/or $A$ production processes at the ILC with
 $\sqrt{s}=250$~GeV, 500~GeV and 1~TeV in the $(m_{H,A},\tan\beta)$
 plane.
 Contour of $\sigma=0.1$~fb is drawn for each signature. } 
 \label{fig:CP1TeV}
\end{figure}

\afterpage{\clearpage}
In Fig.~\ref{fig:CP1TeV}, contour plots of the cross sections of
four-particle production processes through $H$ and/or $A$
are shown in the $(m_{H/A},\tan\beta)$ plane.
The results for $\sqrt{s}=250$~GeV, 500~GeV and 1~TeV are shown
in the figures in the first, second and third columns, while figures in
the first to the fourth rows show the results in Type-I to Type-Y,
respectively. 
We restrict ourselves to consider the degenerated mass case, $m_H=m_A$.
Discussions on the non-degenerated mass cases as well as the case where
$\sin(\beta-\alpha)$ is slightly less than unity are given later. 

The figures in the first row are for Type-I.
The signatures come dominantly from $HA$ pair production
followed by their subsequent decays. 
For $m_{H/A}\lesssim 350$~GeV, the $t\bar{t}$ decay mode does not open,
and then the decays are mostly into $b\bar{b}$, $\tau^+\tau^-$ and $gg$
as shown in Fig.~\ref{fig:Br_125} and Fig.~\ref{fig:Br_250}.
Thus, $4b$, $2b2\tau$ and $4\tau$ signatures as well as the signatures
with gluons $2b2g$, $2\tau2g$ and $4g$ are expected to be observed. 
For $m_{H/A}\gtrsim 350$~GeV where the $t\bar{t}$ decay mode opens,
only the $4t$ signature is expected to be significant. 
Because the $HA$ pair production cross section sharply fall down at the
threshold, the signatures are not expected above the mass threshold for
each collider energy.
Only in the small $\tan\beta$ regions ($\tan\beta<1$), the contour of
the $4t$ signature is extended to above the mass threshold, because
of the large top Yukawa coupling enhancing the single production
cross section associated with top-quark pair, $t\bar{t}H$ and
$t\bar{t}A$. 

The figures in the second row are for Type-II. 
Since the bottom and tau Yukawa interaction are enhanced
by $\tan\beta$, $4b$, $2b2\tau$ and $4\tau$ signatures are expected to
be seen even below the mass threshold through the single production
processes. 
For $m_{H/A}\lesssim 350$~GeV, in small $\tan\beta$ regions, $gg$ decay
mode can be dominant, therefore $4g$ and $2b2g$ signatures which tend to
be four-jet events would be significant. 
Although the SM backgrounds obscure such signatures, the
invariant-mass distributions of dijets may help to distinguish them.
For $m_{H/A}\gtrsim 350$~GeV, $4t$ and $2t2b$ signatures are expected
for $\tan\beta\lesssim 10$ because of the large top Yukawa coupling
constants. 

The figures in the third row are for Type-X.
The $4\tau$ signature can be expected for large $\tan\beta$ regions
even below the pair production mass threshold.
The detailed studies for the $4\tau$ signature can be found in
Ref.~\cite{Kanemura:2012az}. 
For relatively small $\tan\beta$ regions, $4b$ or $4t$ signature is
expected depending on the masses of $H$ and $A$. 
In between, $2b2\tau$ or $2t2\tau$ signature can have sizable rates. 

Finally, the figures in the fourth row are for Type-Y.
The $4b$ signature is dominant for large $\tan\beta$ regions, while
for the small $\tan\beta$ regions with $m_{H/A}\lesssim 350$~GeV,
various signatures including $\tau^+\tau^-$, $gg$ and $c\bar{c}$ can be
expected because all these decay branching ratios are comparably
sizable. 
To avoid too much overlapping, we ignore the curves for the signatures
including $c\bar{c}$, which are however comparable with those of the
$4g$, $2g2\tau$ and $4\tau$ signatures. 
For $m_{H/A}\gtrsim 350$~GeV, the $4t$ and $2t2b$ signatures are
expected to appear for $\tan\beta\lesssim 10$. 

\begin{figure}[t]
 \includegraphics[width=0.328\textwidth]{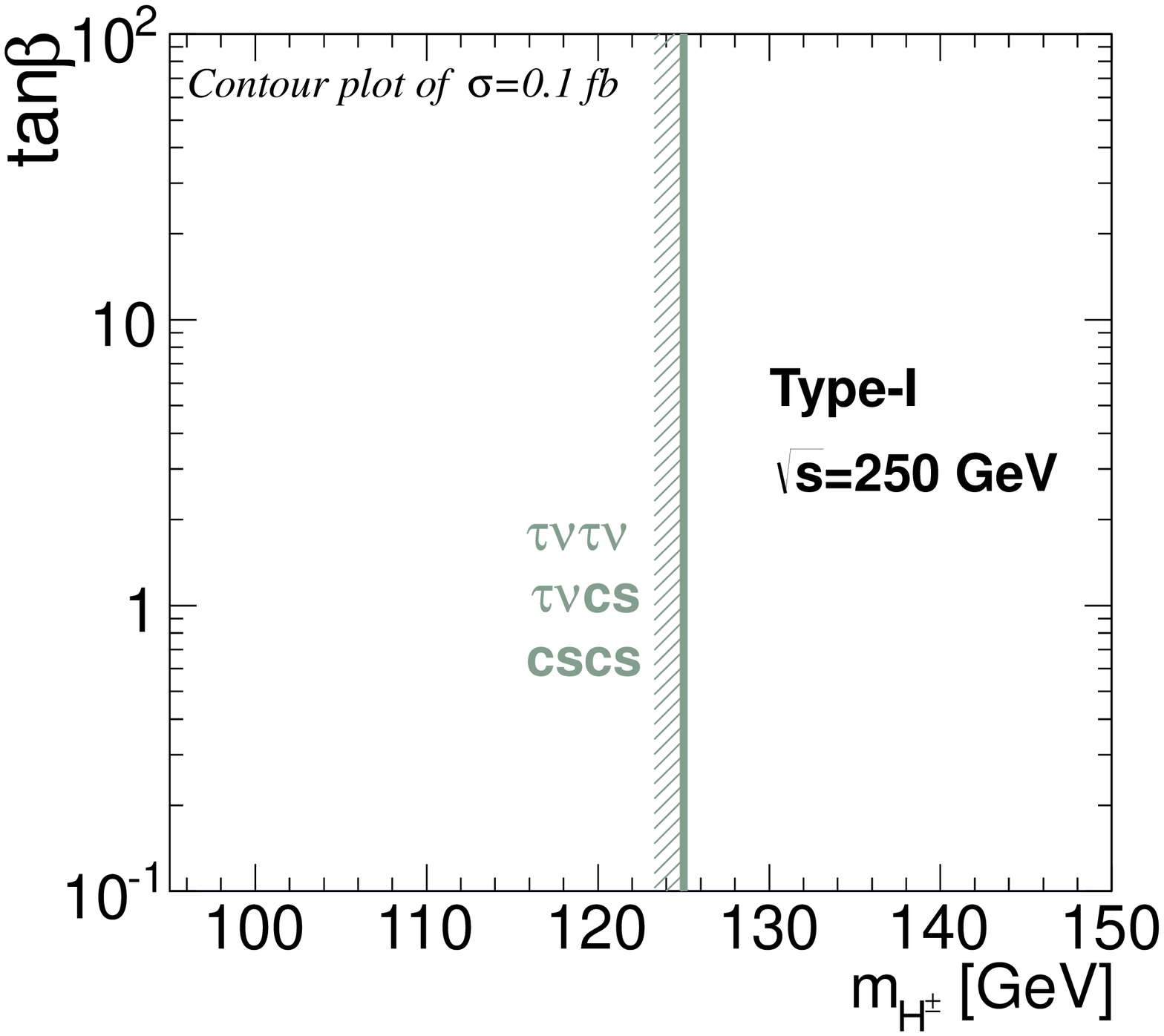} 
 \includegraphics[width=0.328\textwidth]{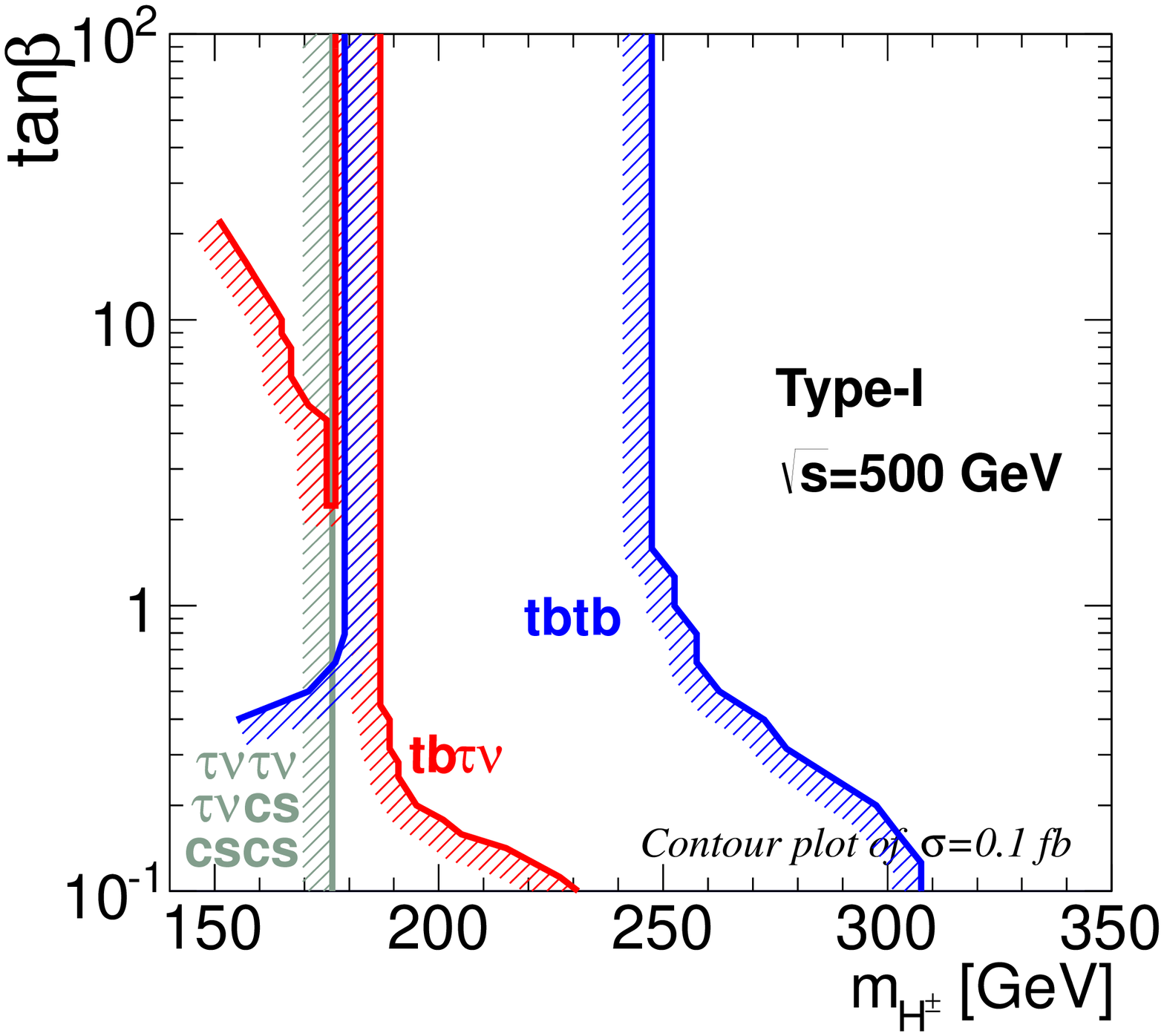} 
 \includegraphics[width=0.328\textwidth]{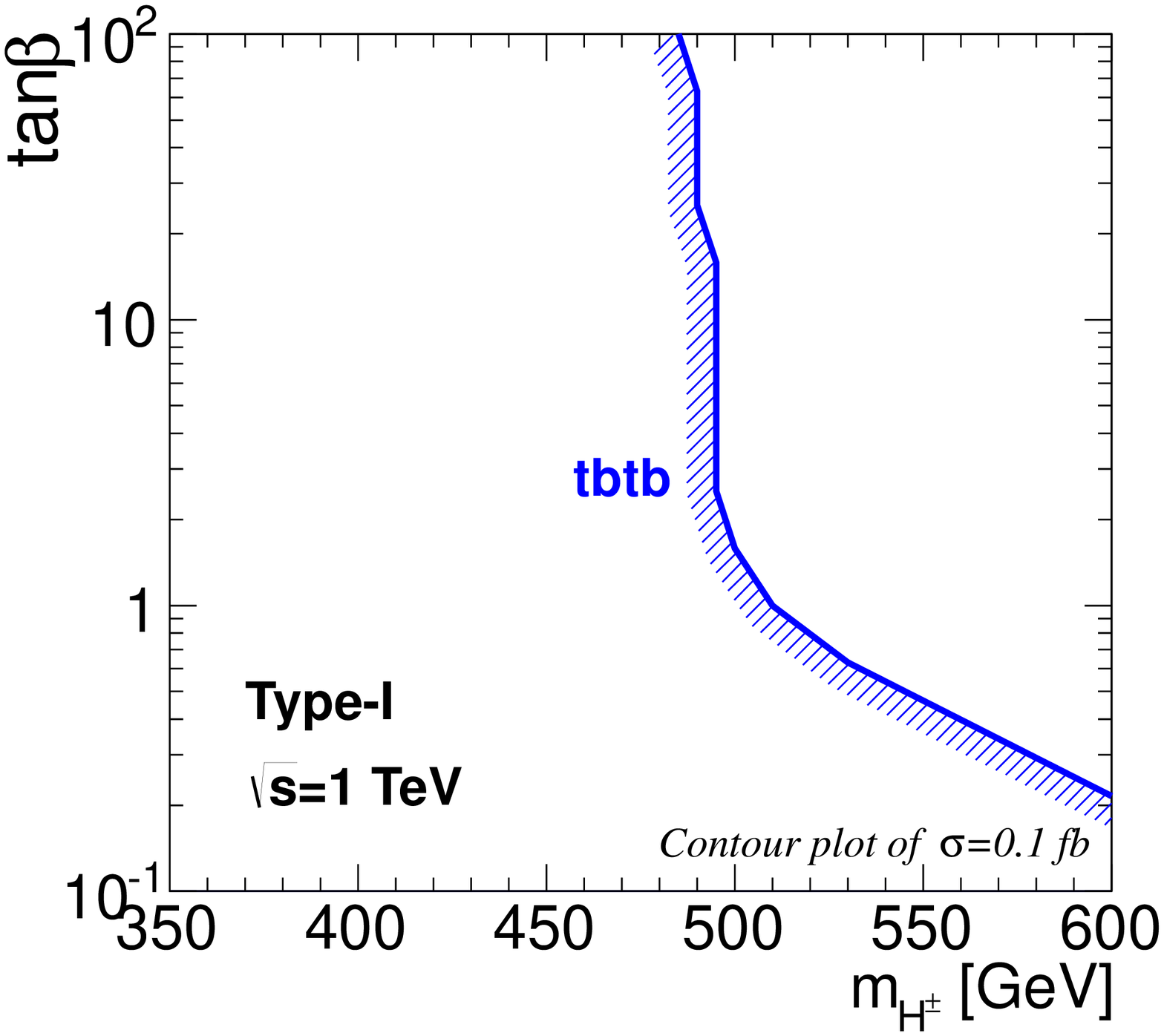} 
 \includegraphics[width=0.328\textwidth]{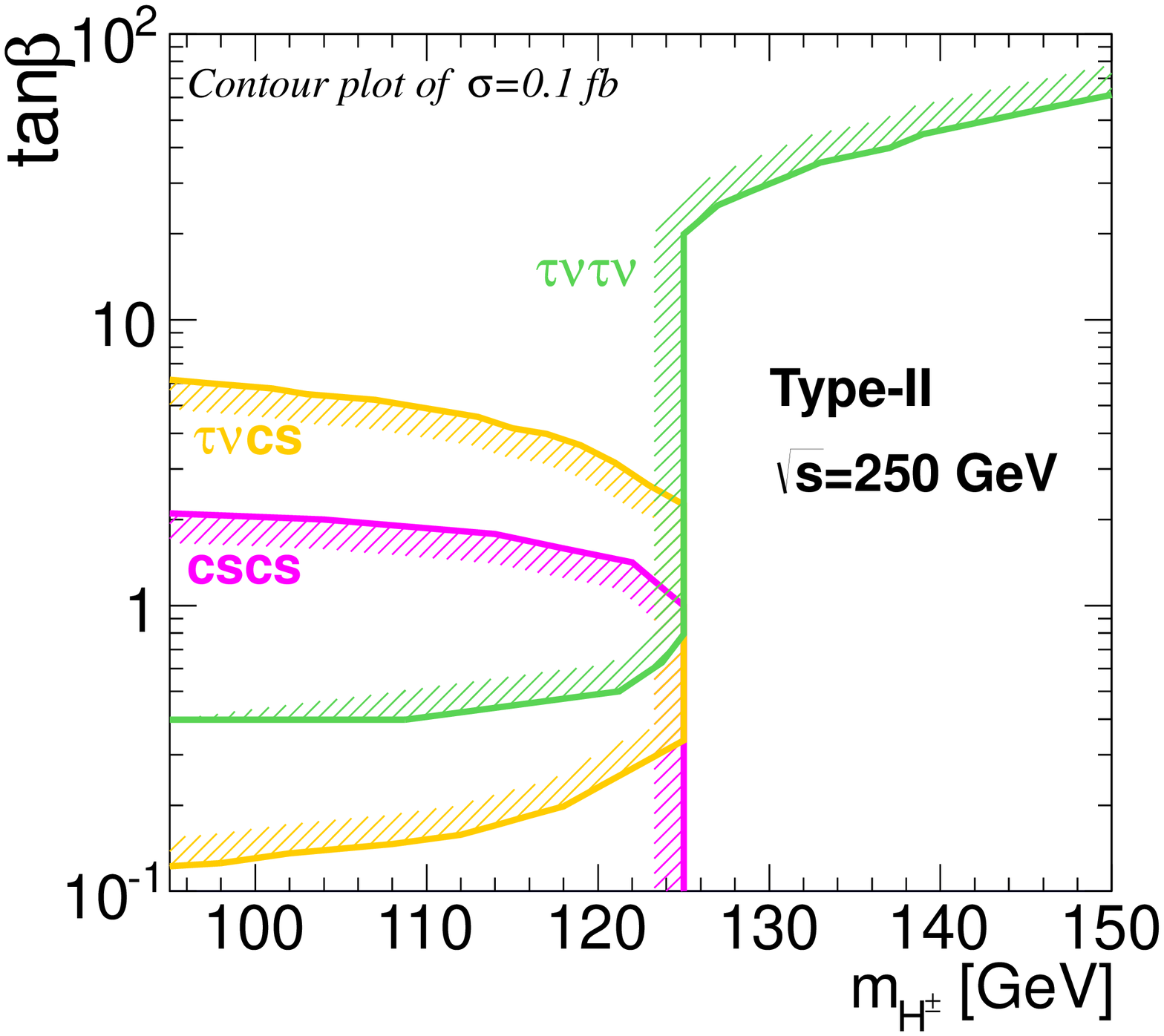} 
 \includegraphics[width=0.328\textwidth]{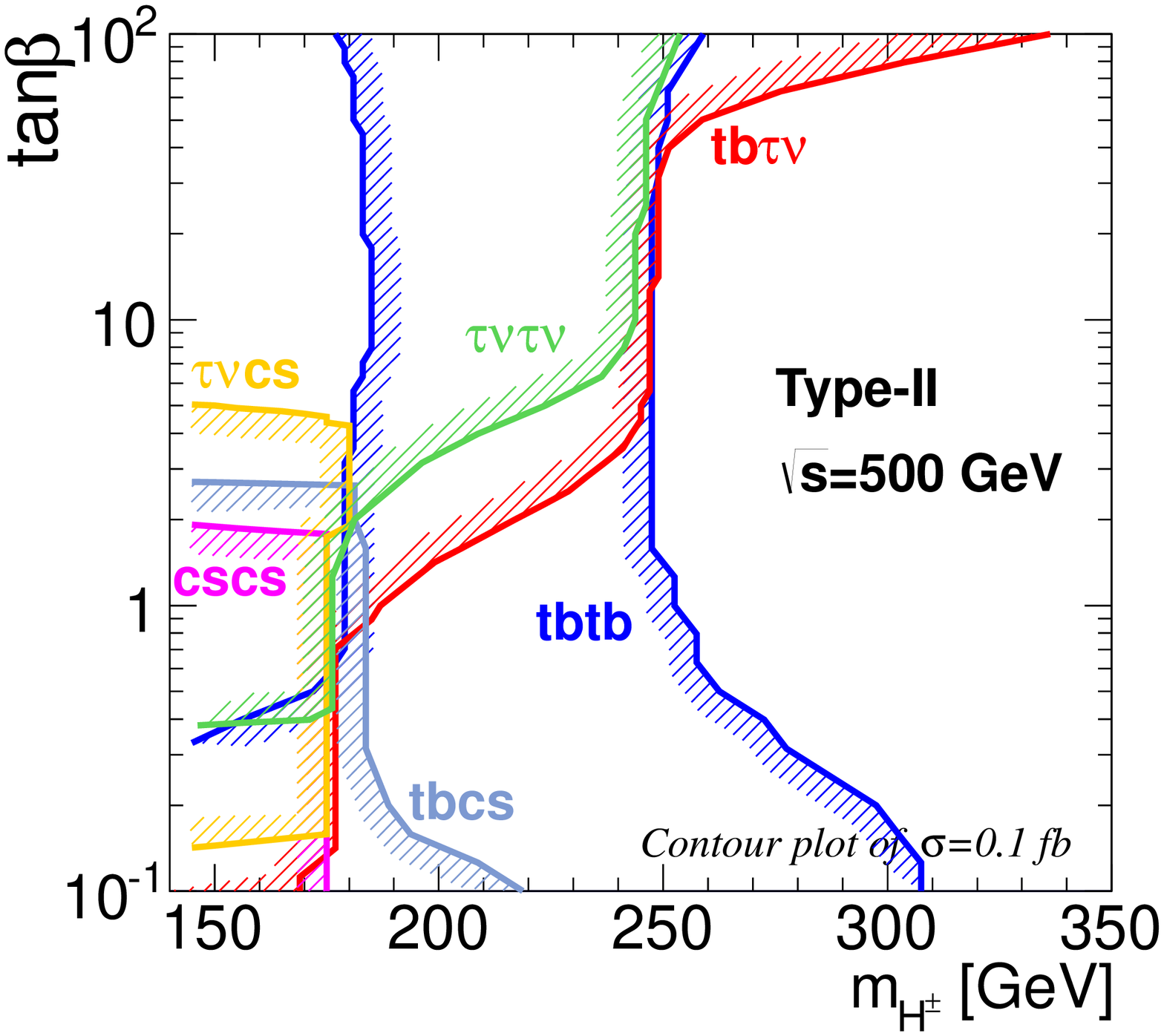} 
 \includegraphics[width=0.328\textwidth]{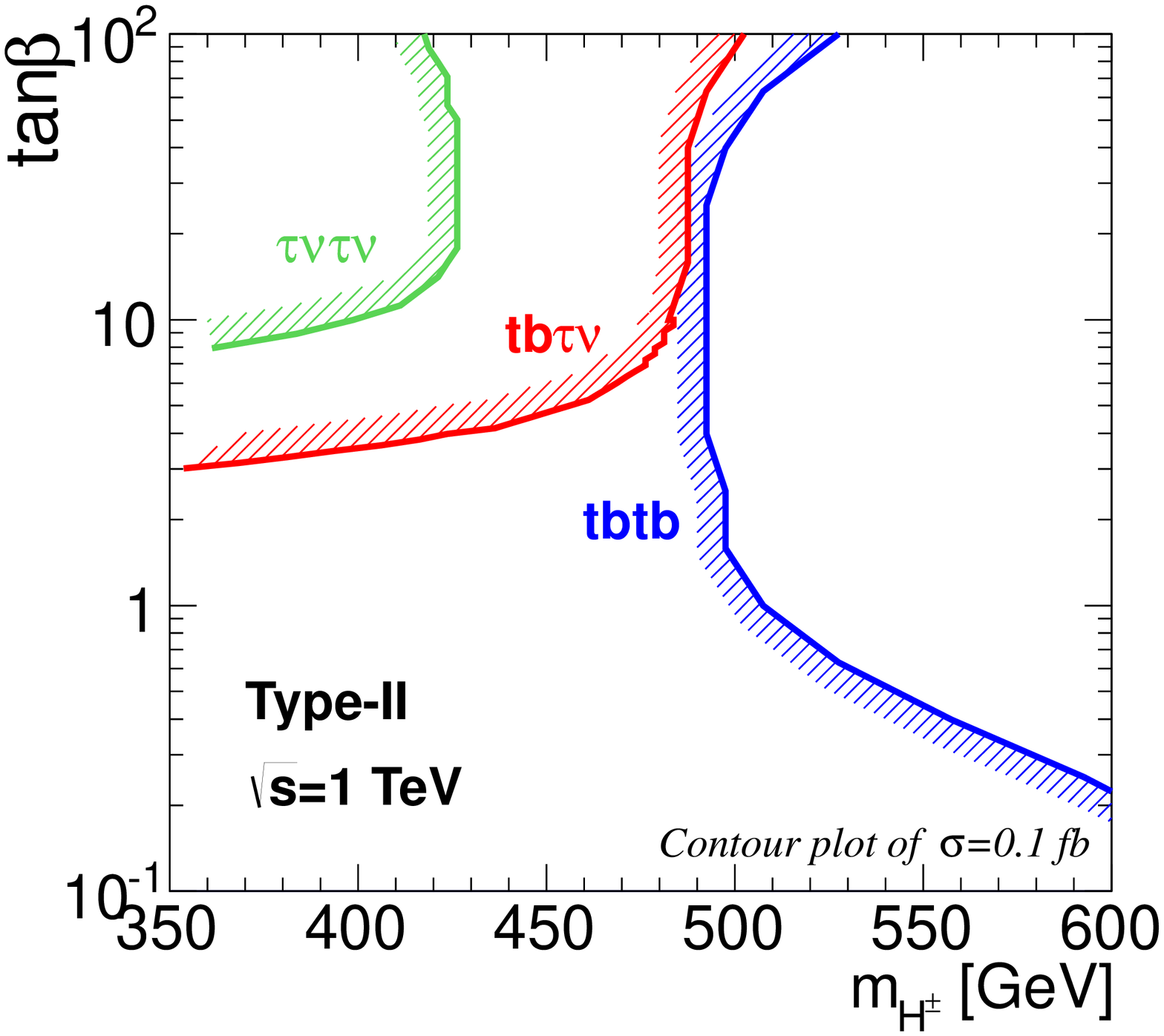} 
 \includegraphics[width=0.328\textwidth]{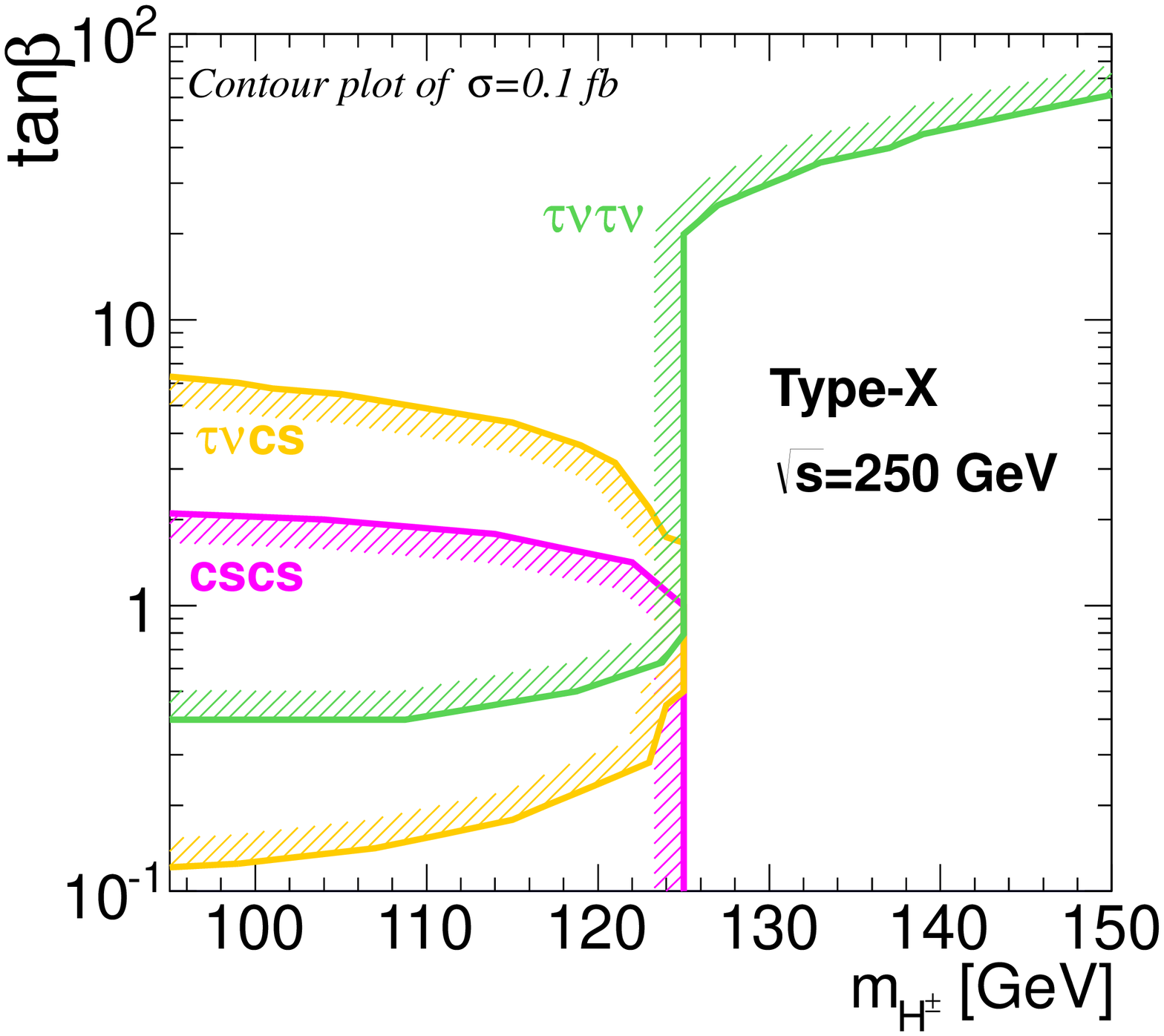} 
 \includegraphics[width=0.328\textwidth]{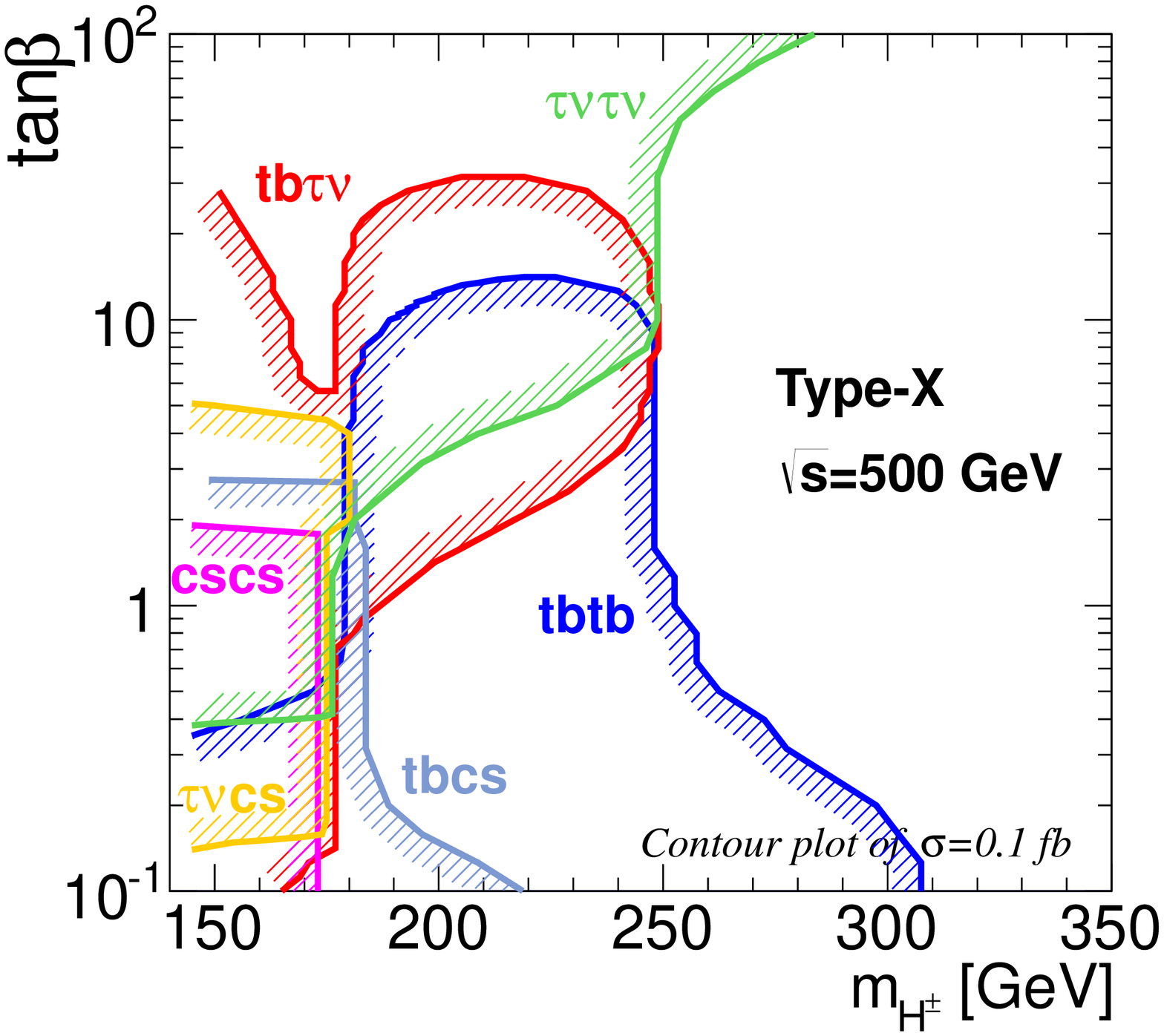} 
 \includegraphics[width=0.328\textwidth]{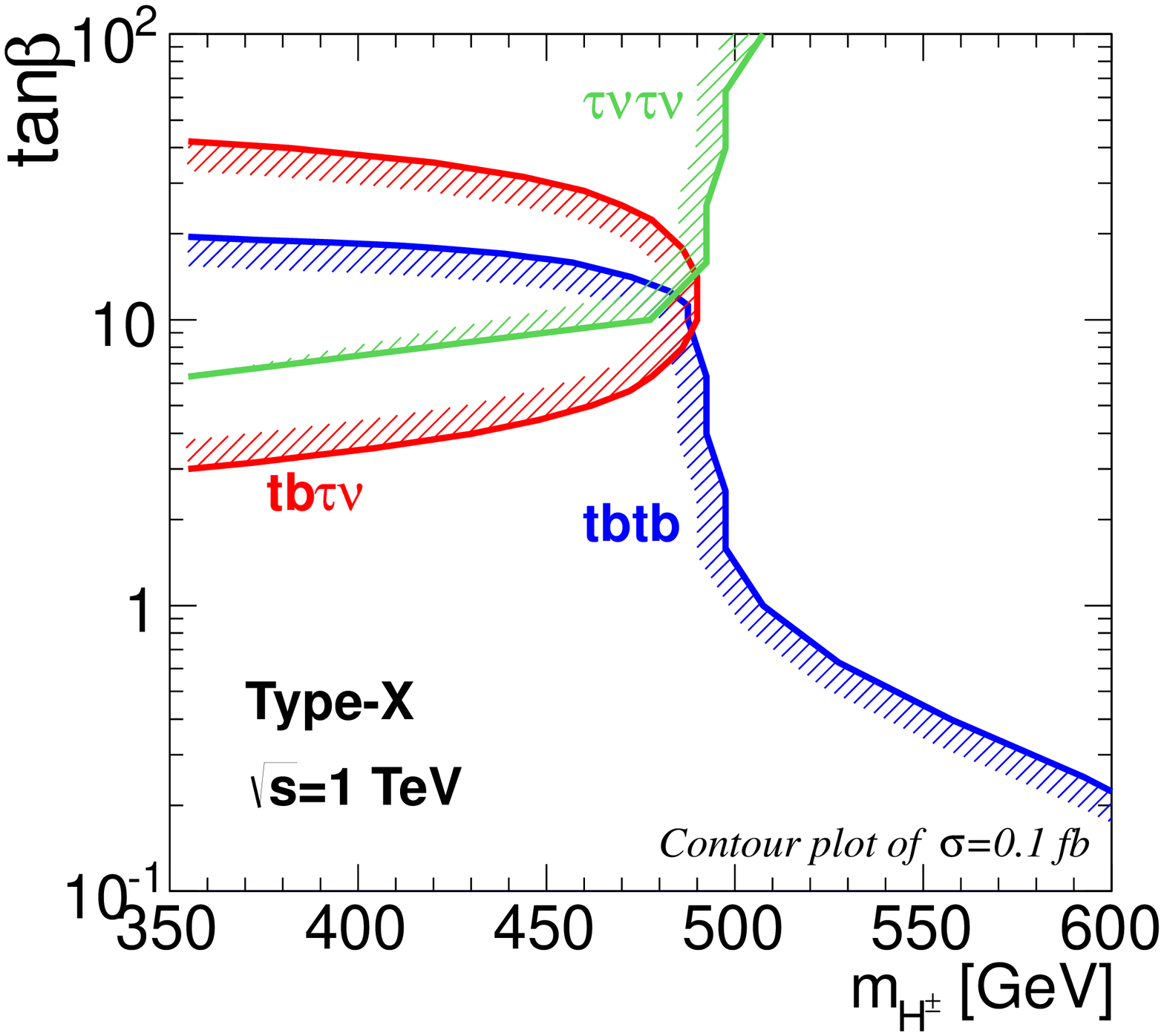} 
 \includegraphics[width=0.328\textwidth]{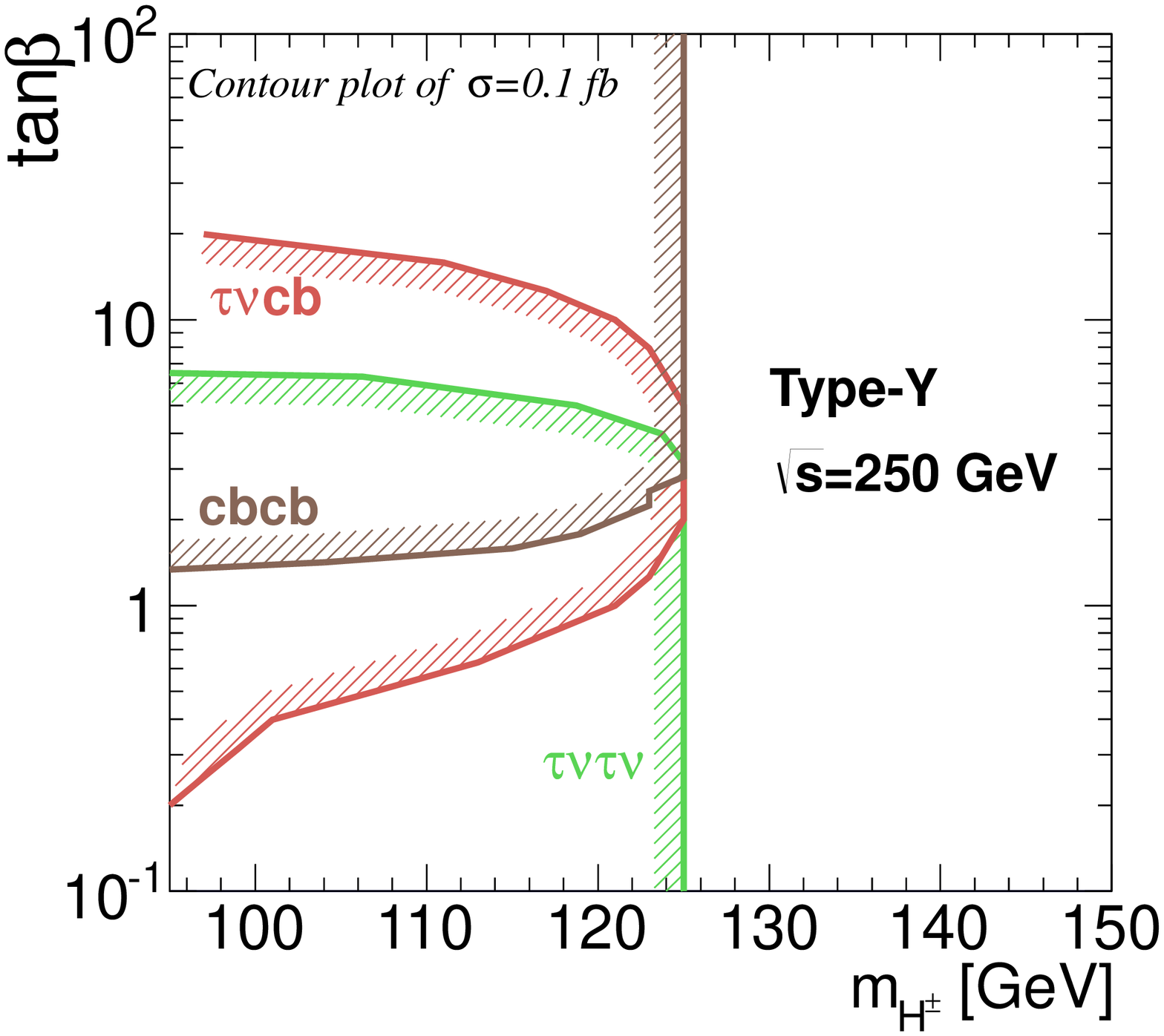} 
 \includegraphics[width=0.328\textwidth]{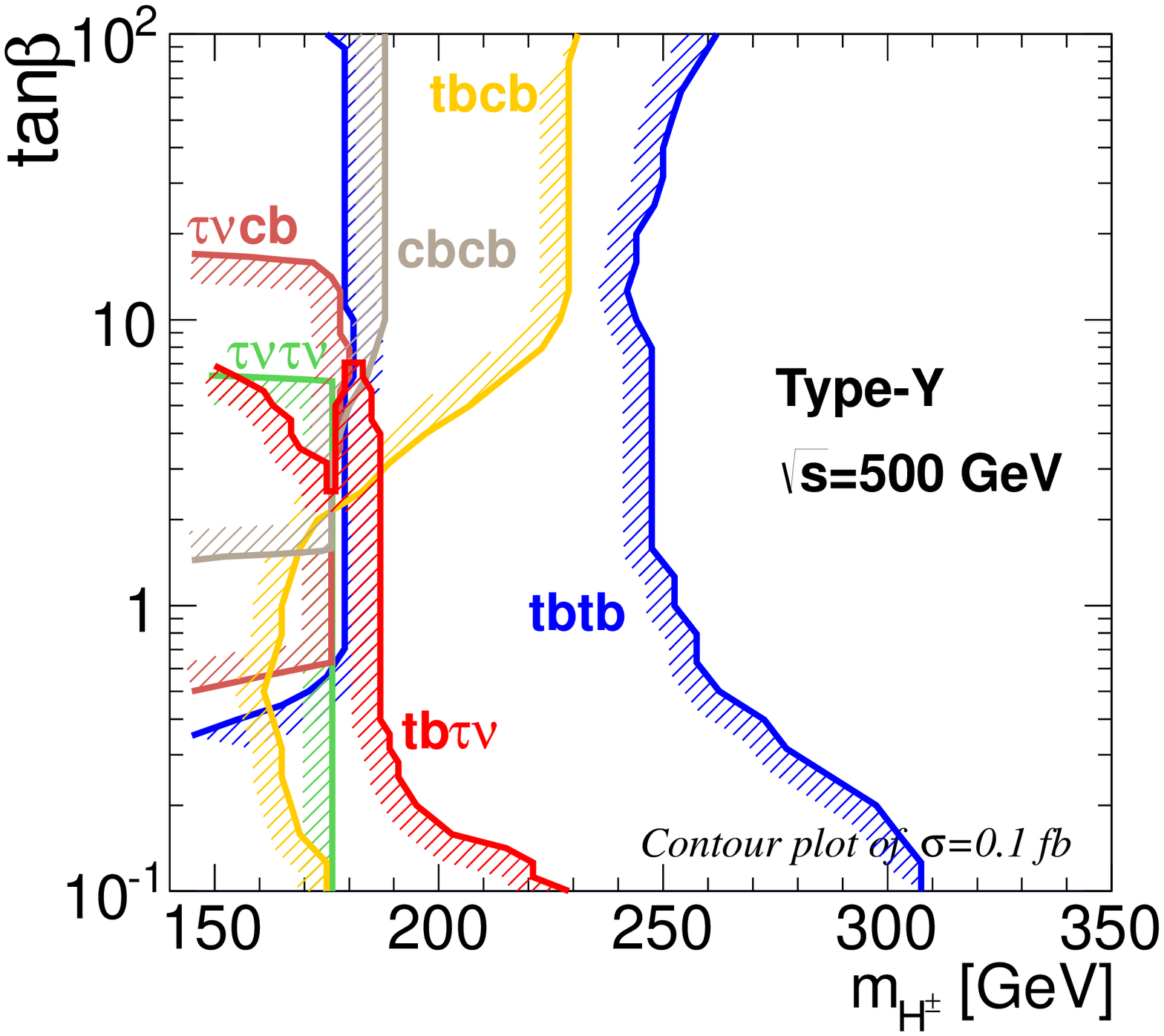} 
 \includegraphics[width=0.328\textwidth]{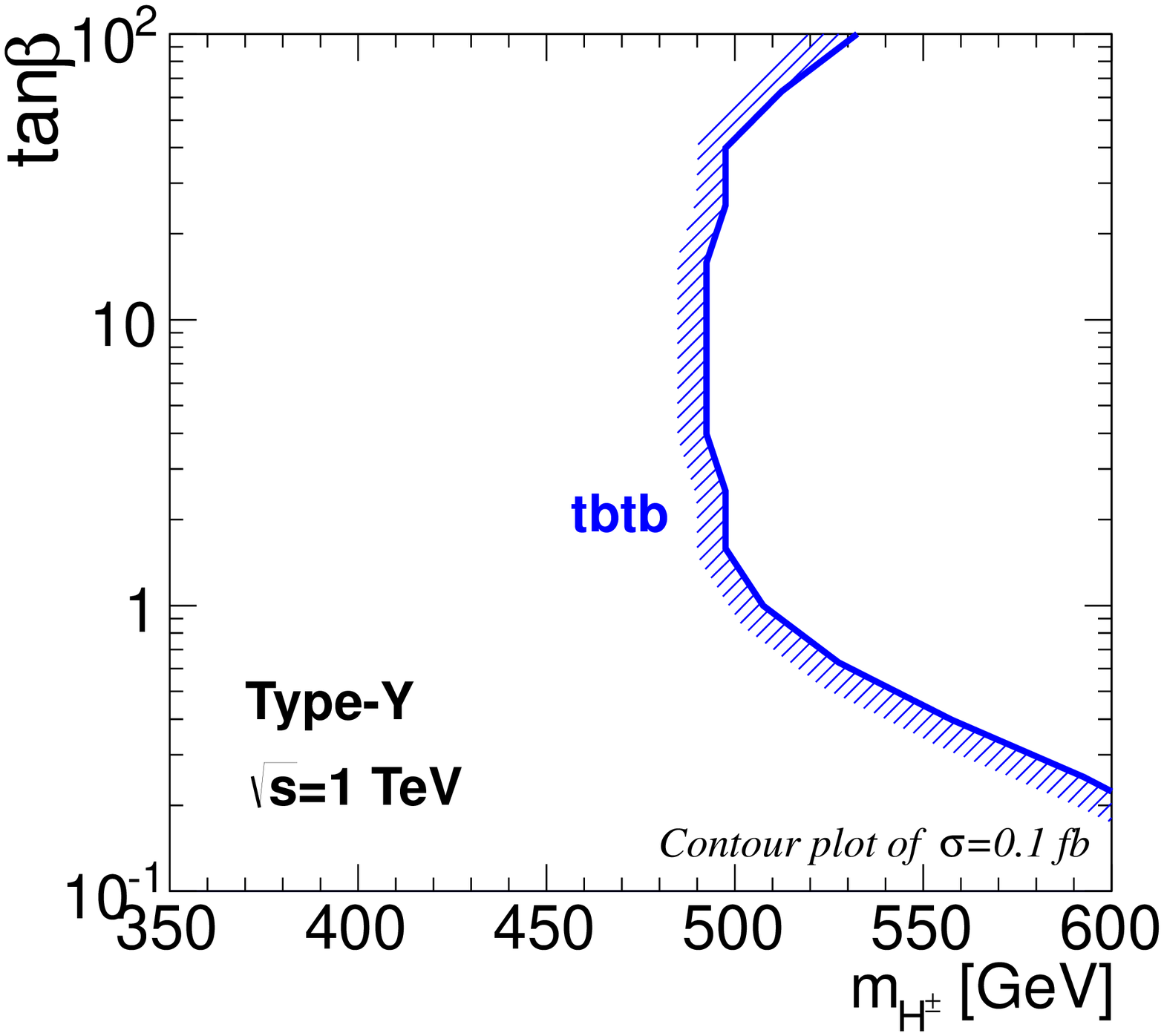} 
 \caption{Contour plots of the four-particle production cross sections
 through the $H^\pm$ production process at the ILC
 $\sqrt{s}=250$~GeV, 500~GeV and 1~TeV in the $(m_{H^\pm},\tan\beta)$
 plane.
 Contour of $\sigma=0.1$~fb is drawn for each signature. }
 \label{fig:CPCH1TeV}
\end{figure}

\afterpage{\clearpage}
In Fig.~\ref{fig:CPCH1TeV}, contour plots of the four-particle
production cross sections through $H^\pm$ are shown in the
$(m_{H^\pm},\tan\beta)$ plane in the same manner as
Fig.~\ref{fig:CP1TeV}. 

The figures in the first row are for Type-I. 
For $m_{H^\pm}\lesssim 180$~GeV below the $H^\pm\to tb$ threshold,
$H^{\pm}\to\tau\nu$ and $cs$ are the dominant decay modes, as
illustrated in Fig.~\ref{fig:Br_125}.
Therefore, the $\tau\nu\tau\nu$, $\tau\nu cs$ and $cscs$ signatures are
expected to appear as long as $\sqrt{s}\ge2m_{H^\pm}$. 
For $m_{H^\pm}\lesssim 180$~GeV and $\sqrt{s}\ge 350$~GeV, $H^\pm$ can
be produced through the decay of top quarks in the top quark pair
production process. 
In the middle column at $\sqrt{s}=500$~GeV, the signature of $tb\tau\nu$
comes from this contribution followed by the decay of
$H^\pm\to\tau\nu$.
For $m_{H^\pm}\gtrsim 180$~GeV, the dominant decay mode quickly switches
into $tb$. 
Therefore the $tbtb$ signature becomes the largest. 

The figures in the second row are for Type-II.
For the mass below the $tb$ threshold, $H^+H^-$ pair production
tends to be the $\tau\nu\tau\nu$ signature in the large $\tan\beta$
regions, and the $\tau\nu cs$, $cscs$ signatures in the medium to small
$\tan\beta$ regions. 
In addition, because of the large Yukawa coupling of top quarks,
single $tbH^\pm$ production followed by $H^\pm\to\tau\nu$ and $cs$
decays gives sizable $tb\tau\nu$ and $tbcs$ signatures, respectively.
On the other hand, for the mass above the $tb$ threshold, the 
$tbtb$ signature is the dominant signature for any values of $\tan\beta$
because of the enhanced $tbH^\pm$ Yukawa interaction. 
The $tb\tau\nu$ and $\tau\nu\tau\nu$ signatures are still visible in
large $\tan\beta$ regions, because of the large $H^\pm\to\tau\nu$
branching ratio. 

The figures in the third row are for Type-X.
As is the case for Type-II, for the mass below the $tb$ threshold,
the $\tau\nu\tau\nu$ signature in the large $\tan\beta$ regions, and the
$\tau\nu cs$, $cscs$ signatures in the medium to small $\tan\beta$
regions are expected. 
Through the $tbH^\pm$ production which is sizable only in the small
and medium $\tan\beta$ regions, the $tb\tau\nu$ and $tbcs$ signatures
are expected to be seen. 
Above the $tb$ threshold, the signatures are $tbtb$ for small and
medium $\tan\beta$ and $\tau\nu\tau\nu$ for large $\tan\beta$.
In between, $tb\tau\nu$ can also be large.

The figures in the fourth row are for Type-Y.
In this case, for the mass below the $tb$ threshold the dominant
decay mode of $H^\pm$ is $cb$ for large $\tan\beta$.
Therefore, $cbcb$ signature is expected for large $\tan\beta$ regions.
In small $\tan\beta$ regions, $\tau\nu$ and $cs$ would be the
dominant. 
Therefore, $\tau\nu\tau\nu$, $\tau\nu cs$ and $cscs$ signatures are
expected to be significant. 
To avoid overlapped plotting, we ignore to plot the contours which
include the $cs$ mode. 
Above the $tb$ threshold, since the $tb$ decay mode is 
dominant for any values of $\tan\beta$, the $tbtb$ signature would be
the only visible mode. 

\subsection{SM background processes}

Here, we discuss the SM background processes and their cross
sections. 
In Table~\ref{tab:BG}, total cross sections without kinematical cuts are 
calculated by {\tt Madgraph}~\cite{Alwall:2011uj}. 
The cross-section for the signatures including gluons is neglected,
because the partonic calculation is meaningless unless an infrared safe
observable is defined, such as the cross-section for jets production.
In general, for the four-particle production processes, the SM
background cross sections are larger for $\sqrt{s}=250$~GeV, but
decrease with the collision energy. 
The typical orders of cross sections are of the order of 1~fb to 10~fb 
for the $Z/\gamma$ mediated processes,
and of the order of 10 to 100~fb 
for the processes which are also mediated by $W^\pm$.
For the four-quark production processes, gluon exchange diagrams also
contribute.
Some of the background cross sections are larger than the expected
signal cross sections. 
In order to reduce the background events, efficient kinematical cuts are
required. 
Since the additional Higgs bosons are expected to have narrow decay
widths and since there are many background contributions from the decays
of $Z$ bosons, a cut on the invariant mass of the decay particles is
useful. 

The cross section of the $4t$ production is very small in the SM, see
Table~\ref{tab:BG}. 
Therefore, a clean signature can be expected to be detected in this
mode. 
However, because of the decays of top quarks, more complicated
background processes can be involved, and the event reconstruction
is not straightforward. 
Detailed studies on the signal and background processes for
$tbtb$ production can be found in
Ref.~\cite{Moretti:2002pa}, and the signal-to-background analysis for
the $4\tau$ production can be found in Ref.~\cite{Kanemura:2012az} with
the reconstruction method of the masses of additional Higgs bosons. 

\begin{table}[t]
 \begin{center}
  \begin{tabular}{c||c|c|c}
   Signature & $\sqrt{s}=250$~GeV & $\sqrt{s}=500$~GeV &
   $\sqrt{s}=1$~TeV \\
   \hline
   $4b$       & 18 & 7.2 & 2.9  \\
   $4\tau$    & 4.4 & 1.6 & 0.63  \\
   $2\tau2b$  & 28 & 10 & 3.5  \\
   $2\tau 2\nu$ &210 & 94.4& $35.8$ \\
   $tb\tau\nu$ &$5.7\times10^{-4}$ & 122.7& 40 \\
   $2t2b$     & $-$ & 1.7 & 5.1 \\
   $2t2\tau$  & $-$ & 0.14 & 0.34 \\
   $4t$ & $-$ & $-$ & $3.8\times10^{-3}$ \\
  \end{tabular}
  \caption{Background cross sections in unit of fb for the four-particle
  processes at the ILC.
  Total cross sections without kinematical cuts are calculated by {\tt
  Madgraph}~\cite{Alwall:2011uj}.} \label{tab:BG}
 \end{center}
\end{table}
%

\section{Discussions}\label{sec:dis}

In this section, we further discuss future prospects for the additional
Higgs boson searches and the parameter determinations at the LHC and
the ILC, and their complementarity in the general framework of the 2HDM
with the softly-broken discrete symmetry.
As we have seen in Sec.~\ref{sec:LHC}, ability of the LHC for 
discovery or exclusion of additional Higgs bosons is high. 
However, there are still wide regions in the parameter space where the
LHC cannot discover all the additional Higgs bosons, or where the
type of Yukawa interaction cannot be determined even if they are
discovered. 
In the previous section, we have seen that at the ILC, as long as the
masses of these bosons are within a kinematical reach, various
signatures are expected to be used for the discrimination of the type of
Yukawa interaction.
Here, as an example, we give some concrete scenarios to show the
complementarity of direct searches for the
additional Higgs bosons in the 2HDMs at the LHC and the ILC. 

We take six sets of $(m_\phi, \tan\beta)$ as benchmark scenarios, 
where $m_\phi$ represents the common mass of $H$, $A$ and
$H^\pm$, namely $m_\phi=220$~GeV and 400~GeV, and $\tan\beta=2$, 7 and
20, for all types of Yukawa interaction.
We fix the value of $\sin(\beta-\alpha)$ to be unity. 
In Table~\ref{tab:Benchmark1}, we summarize the expected signatures
of $H/A$ and $H^\pm$ to be observed at the LHC with 300~fb$^{-1}$, 
3000~fb$^{-1}$ and at the ILC with $\sqrt{s}=500$~GeV, according to our
estimation in the last sections for the benchmark scenarios with
$m_\phi=220$~GeV. 
In Table~\ref{tab:Benchmark2}, the expected signatures of $H/A$ and
$H^\pm$ are summarized at the LHC with 300~fb$^{-1}$, 3000~fb$^{-1}$ and
at the ILC with $\sqrt{s}=1$~TeV for the benchmark scenarios with
$m_\phi=400$~GeV. 
We note again that at the ILC 
signatures are assumed to be detected by a
criterion whether the cross section is greater than 0.1~fb. 
We present the results for each type of Yukawa interaction, Type-I to
Type-Y from the left column to right column, respectively. 

\begin{table}[t]
 \begin{tabular}{c||l|cc|cc|cc|cc}
  $(m_\phi, \tan\beta)$ & & \multicolumn{2}{c|}{Type-I} &
  \multicolumn{2}{c|}{Type-II} & \multicolumn{2}{c|}{Type-X} &
  \multicolumn{2}{c}{Type-Y} \\ \hline\hline
  & & $H, A$ & $H^\pm$ & $H, A$ & $H^\pm$ & $H, A$ & $H^\pm$ & $H, A$ &
  $H^\pm$ \\ \hline  \hline
  & LHC300 & $-$ & $-$ & $\tau\tau$, $bb$ & $tb$ & $4\tau$ & $-$ &
  $bb$ & $tb$ \\ 
  (220~GeV, 20) & LHC3000 & $-$ & $-$ & $\tau\tau$, $bb$  & $tb$ &
  $4\tau$ & $-$ & $bb$ & $tb$ \\[1mm] \cline{2-10}
  & ILC500 & \shortstack{{}\\$4b,2b2\tau,4g$,\\$2b2g,2\tau2g$} & $tbtb$ &
  \shortstack{$4b,2b2\tau$,\\$4\tau$} &
  \shortstack{$tbtb,tb\tau\nu$,\\$\tau\nu\tau\nu^{}$} & $4\tau$ &
  \shortstack{$tb\tau\nu$,\\$\tau\nu\tau\nu$} & $4b$ & $tbtb,tbcb$ \\
  \hline \hline
  & LHC300 & $-$ & $-$ & $\tau\tau$ & $tb$ & $4\tau$ & $-$ &
  $-$ & $tb$ \\
  (220~GeV, 7) & LHC3000 & $-$ & $tb$ & $\tau\tau$ & $tb$ &
  $\tau\tau,4\tau$ & $-$ & $-$ & $tb$ \\[1mm] \cline{2-10}
  & ILC500 & \shortstack{{}\\$4b,2b2\tau,4g$,\\$2b2g,2\tau2g$} & $tbtb$ &
  \shortstack{$4b,2b2\tau$,\\$4\tau$} & 
  \shortstack{$tbtb,tb\tau\nu$,\\$\tau\nu\tau\nu$} &
  $2b2\tau,4\tau$ & 
  \shortstack{$tbtb,tb\tau\nu$,\\$\tau\nu\tau\nu$} & $4b$ &
  $tbtb,tbcb$ \\ \hline \hline
  & LHC300 & $-$ & $tb$ & $\tau\tau$ & $tb$ &
  $\tau\tau,4\tau$ & $tb$ & $-$ & $tb$ \\
  (220~GeV, 2) & LHC3000 & $\tau\tau$ & $tb$ & $\tau\tau$ &
  $tb$ & $\tau\tau,4\tau$ & $tb$ & $-$ & $tb$\\[1mm] \cline{2-10}
  & ILC500 & \shortstack{{}\\$4b,2b2\tau,4g$,\\$2b2g,2\tau2g$} & $tbtb$ &
	  \shortstack{$4b,2b2\tau$,\\
  $4\tau,2b2g$} & \shortstack{$tbtb$,\\$tb\tau\nu$} &
  \shortstack{$4b,2b2\tau$,\\$4\tau$} &
  \shortstack{$tbtb$,\\$tb\tau\nu$} &
  \shortstack{$4b,2b2\tau$,\\$2b2g$} & $tbtb$ \\
 \hline \hline
 \end{tabular}
\caption{Expected signatures to be observed at the LHC and ILC
 for the benchmark scenarios with $m_\phi=220$~GeV. 
 Observable final-states are listed as the signatures of additional
 Higgs bosons, $H$, $A$ and $H^{\pm}$. 
 LHC300, LHC3000, ILC500 represent the LHC run of 300~fb$^{-1}$,
 3000~fb$^{-1}$ luminosity, ILC run of 500~GeV, respectively.}
 \label{tab:Benchmark1}
\end{table}

In Table~\ref{tab:Benchmark1}, the expected signals are summarized for 
each benchmark scenario with a relatively light mass, $m_\phi=220$~GeV. 
Let us look at the scenario of $(m_\phi, \tan\beta) = (220~{\rm
GeV}, 20)$. 
At the LHC with 300~fb$^{-1}$ and 3000~fb$^{-1}$, 
no signature is predicted for Type-I, while different signatures
are predicted for Type-II, Type-X and Type-Y. 
Therefore those three types can be discriminated at the LHC. 
On the other hand, at the ILC with $\sqrt{s}=500$~GeV, all the four
types of the Yukawa interaction including Type-I predict signatures 
which are different from each other.
Therefore, at the ILC, complete discrimination of the type of Yukawa
interaction can be performed. 
This benchmark scenario demonstrates necessity of the ILC (500~GeV)
to completely separate the all four types of Yukawa interaction. 

Next, we turn to the second scenario, $(m_\phi, \tan\beta) = (220~{\rm
GeV}, 7)$. 
At the LHC with 300~fb$^{-1}$, Type-I cannot be observed, while Type-II,
Type-X and Type-Y are expected to be observed with different signatures. 
At the LHC with 3000~fb$^{-1}$, the signature of Type-I can also be
observed with the same final state as Type-Y.
Type-I and Type-Y can be basically separated, because for Type-Y the
signals can be observed already with 300~fb$^{-1}$ while for Type-I that
can be observed only with 3000~fb$^{-1}$. 
Therefore, at the LHC with 3000~fb$^{-1}$, the complete discrimination
can be achieved. 
At the ILC, the four types of Yukawa interaction can also be separated
by a more variety of the signatures for both channels with the neutral
and charged Higgs bosons. 

Finally, we discuss the scenario of $(m_\phi, \tan\beta) = (220~{\rm
GeV}, 2)$. 
At the LHC with 300~fb$^{-1}$, signals for all the four types of Yukawa
interaction can be observed. 
However, the signatures of Type-I and Type-Y are identical, so that the
two types cannot be discriminated. 
With the 3000~fb$^{-1}$ data at the LHC, the difference between the
Type-I and Type-Y emerges in the $H/A$ signature.
Therefore the two types can be discriminated at this stage. 
Again, at the ILC, the four types can also be separated with a more
variety of the signatures for both channels with the neutral and charged
Higgs bosons. 

\begin{table}[t]
 \begin{tabular}{c||l|cc|cc|cc|cc}
  $(m_\phi, \tan\beta)$ & & \multicolumn{2}{c|}{Type-I} &
  \multicolumn{2}{c|}{Type-II} & \multicolumn{2}{c|}{Type-X} &
  \multicolumn{2}{c}{Type-Y} \\ \hline\hline
  & & $H, A$ & $H^\pm$ & $H, A$ & $H^\pm$ & $H, A$ & $H^\pm$ & $H, A$ &
  $H^\pm$ \\ \hline \hline
  & LHC300 & $-$ & $-$ & $\tau\tau$ & $tb$ & $4\tau$ & $-$ & $-$ & $tb$ \\ 
  (400~GeV, 20) & LHC3000 & $-$ & $-$ & $\tau\tau$ & $tb$ &
  $\tau\tau,4\tau$ & $-$ & $-$ & $tb$ \\[1mm] \cline{2-10}
    & ILC1TeV & $4t$ & $tbtb$ & \shortstack{{}\\$4b,2b2\tau$,\\$2t2b$} &
  \shortstack{$tbtb,tb\tau\nu$,\\$\tau\nu\tau\nu$} & $4\tau,2t2\tau$ &
  \shortstack{$tb\tau\nu$,\\$\tau\nu\tau\nu$} & $4b,2t2b$ & $tbtb$ \\
  \hline\hline
  & LHC300 & $-$ & $-$ & $-$ & $-$ & $-$ & $-$ & $-$ & $-$ \\ 
  (400~GeV, 7) & LHC3000 & $-$ & $-$ & $\tau\tau$ & 
  $tb$ & $\tau\tau,4\tau$& $-$ & $-$ & $tb$ \\[1mm] \cline{2-10} 
    & ILC1TeV & $4t$ & $tbtb$ & \shortstack{{}\\$4b,2b2\tau$,\\$2t2b,4t$}
  & $tbtb,tb\tau\nu$ & $4t,2t2\tau$ & \shortstack{$tbtb$,\\$tb\tau\nu$}
  & $4b,2t2b,4t$ & $tbtb$ \\ \hline\hline
  & LHC300 & $-$ & $tb$ & $-$ & $tb$ & $-$ & $tb$ & $-$ &
  $tb$ \\
  (400~GeV, 2) & LHC3000 & $-$ & $tb$ & $-$ & $tb$ & $-$ &
  $tb$ & $-$ & $tb$ \\[1mm] \cline{2-10}
    & ILC1TeV & $4t$ & $tbtb$ & $4t,2t2b$ & $tbtb$ & $4t$ & $tbtb$ &
  $4t,2t2b$ & $tbtb$ \\ \hline\hline
 \end{tabular}
\caption{The similar table as Table~\ref{tab:Benchmark1}, but for
 $m_\phi=400$~GeV.
 ILC1TeV represents the ILC run of 1~TeV.
}
 \label{tab:Benchmark2}
\end{table}

In Table~\ref{tab:Benchmark2}, the expected signals are summarized for 
each benchmark scenario with a relatively heavy mass, $m_\phi=400$~GeV. 
First, we discuss the scenario of $(m_\phi, \tan\beta) = (400~{\rm
GeV}, 20)$.
At the LHC with 300~fb$^{-1}$, while for Type-I no signature can be
observed, $\tau\tau$ and $tb$ signatures can be observed for Type-II,
and a $4\tau$ ($tb$) signature can be observed for Type-X (Type-Y). 
Thus, at least the three types (Type-II, Type-X and Type-Y) can be
discovered and discriminated by checking the pattern of the observed
signatures at the LHC with 300~fb$^{-1}$. 
With the 3000~fb$^{-1}$ data at the LHC, the situation is not 
improved, but for Type-X, one additional signature $\tau\tau$ would be
observed. 
Therefore, at the LHC with 3000~fb$^{-1}$ all types of Yukawa
interaction except Type-I can be separated basically. 
At the ILC with $\sqrt{s}=1$~TeV, signatures in various modes can be
observed for both the neutral and charged Higgs bosons 
depending on the type of Yukawa interaction. 
Signatures for Type-I are expected in $4t$ and $tbtb$ modes. 
Since the signatures are all different among the four types of
Yukawa interaction, all the types can also be discriminated at the ILC. 
This benchmark scenario demonstrates necessity of the ILC (1~TeV)
to completely separate the all four types of Yukawa interaction. 

Next, we discuss the scenario of $(m_\phi, \tan\beta) = (400~{\rm
GeV}, 7)$. 
At the LHC with 300~fb$^{-1}$, no signature is discovered for all types
of Yukawa interaction at all. 
At the LHC 3000~fb$^{-1}$, the signals of Type-II, Type-X and Type-Y can
be discovered with different signatures, while Type-I cannot be seen. 
At the ILC, all types are observed with different signatures.
Therefore, the complete discrimination or exclusion needs the ILC in
this scenario too. 

Finally, we discuss the scenario of $(m_\phi, \tan\beta) = (400~{\rm
GeV}, 2)$. 
At the LHC with 300~fb$^{-1}$, only the $H^\pm\to tb$ signature is
predicted for all types of Yukawa interaction. 
The situation does not change even with 3000~fb$^{-1}$. 
Therefore, the signals for all types of Yukawa interaction can be
discovered, but the type cannot be discriminated at the LHC. 
At the ILC, $tbtb$ signature is observed for the pair and single
production of $H^\pm$ for all types of Yukawa interaction. 
For the neutral Higgs bosons, for Type-I and Type-X only the $4t$
signature is observed, while $4t$ and $2t2b$ signatures are observed for
Type-II and Type-Y. 
Therefore, at the ILC, we are able to discriminate the type of
Yukawa interaction as either Type-I or Type-X, or either Type-II or
Type-Y. 
However, precision measurements of the number of signal events at the
ILC could be used for further discrimination.

To summarize, the additional Higgs bosons can be discovered for all the
benchmark scenarios by the combination of searches at the LHC and ILC. 
Furthermore, the type of Yukawa interaction can be separated by
looking at the pattern of the observed signatures.
For the scenarios with ($m_\phi, \tan\beta$) = (220~GeV, 20), (400~GeV,
20) and (400~GeV, 7), the ILC is necessary for the complete separation
of the type of Yukawa interaction. 
For the scenario with ($m_\phi, \tan\beta$) = (400~GeV, 2), 
the LHC cannot discriminate the type of Yukawa interaction, while 
at the ILC two groups of the type, Type-I or Type-X and
Type-II or Type-Y can be separated by looking at the difference of
signatures, and further discrimination may be possible by precision
measurements of the number of signal events. 
Therefore, the LHC and the ILC are complementary for additional Higgs
boson searches and also for discrimination the type of Yukawa
interaction in the 2HDM. 
Furthermore, the determination of $\tan\beta$ can be performed through
the observation of the branching ratio or the total decay widths of
additional Higgs bosons~\cite{Feng:1996xv,Barger:2000fi,Gunion:2002ip,%
Kanemura:2013eja}. 

We briefly give a comment for the cases with $m_\phi<200$~GeV and
$m_\phi>500$~GeV.
For $m_{H,A}<200$~GeV, the current LHC data already have excluded
regions of $\tan\beta\gtrsim5$ to 9 for Type-II in the
$H/A\to\tau^+\tau^-$ search~\cite{CMS_neutral_new} and
$\tan\beta\gtrsim15$ for Type-Y in the $H/A\to b\bar{b}$
search~\cite{CMS_neutral_old}. 
Furthermore, wide parameter regions of $\tan\beta$ with
$m_{H^\pm}<140$~GeV have been excluded for Type-II via the
$H^\pm\to\tau\nu$ search in the decay of top
quarks~\cite{ATLAS_charged_constraints}.
 For Type-I and Type-X, the $H^\pm\to\tau\nu$ signals may be searched in
the pair production process $pp\to H^+H^-$.
For Type-Y with large $\tan\beta$, $H^\pm\to cb$ decays can be
searched in the top quark decay $t\to bH^\pm$.
For $m_\phi>500$~GeV, the LHC searches can be extended into relatively
small and/or large $\tan\beta$ regions. 
On the other hand, the ILC with $\sqrt{s}\leq 1$~TeV cannot produce
additional Higgs bosons in pair. 
Single production processes of additional Higgs bosons can enhance the
number of the signal to some extent for small or large $\tan\beta$
values. 
 
In our discussion above, the SM-like limit, $\sin(\beta-\alpha)=1$, has
been commonly assumed in the benchmark scenarios in
Tables~\ref{tab:Benchmark1} and \ref{tab:Benchmark2}. 
We here discuss the case in which the SM-like limit is slightly
relaxed, i.e., $\sin^2(\beta-\alpha)=0.9$ to 0.99.
The pattern of branching ratios of additional Higgs bosons drastically
changes in this case: see for example Fig.~2 in
Ref.~\cite{Aoki:2009ha} for $\sin^2(\beta-\alpha)=1$ and Fig.~3 in
Ref.~\cite{Aoki:2009ha} for $\sin^2(\beta-\alpha)=0.96$. 
In particular, for $\sin^2(\beta-\alpha)=0.96$, $H$ can decay into weak
gauge bosons, whose decay branching ratios can easily be substantially
large. 
Consequently, our discussion above can be changed. 
We may expect that the discovery signal of $H$ can be clearer in
this case because of the decay into weak gauge boson pairs.
The analysis for such a case will be separately performed in the future. 
We also note that if $\sin^2(\beta-\alpha)$ is slightly less than unity,
the coupling constants of the SM-like Higgs boson with the SM particles
differ from the SM predictions. 
The pattern of the deviations depends on the type of Yukawa
interactions. 
Therefore, by detecting the pattern by precision measurements of the
coupling constants of the SM-like Higgs boson at the ILC, we can
fingerprint the specific type of Yukawa interaction in the
2HDM~\cite{Asner:2013psa,KTYY}. 
Notice that fingerprinting of the model by using the measurement of
SM-like Higgs boson coupling constants is powerful as long as
$\sin^2(\beta-\alpha)$ is less than unity by more than 1\%.
If the deviation is much smaller, we cannot fingerprint the 2HDM by
looking at the SM-like Higgs boson coupling constants. 
In such a case, namely the SM-like limit, only the direct searches
for the additional Higgs bosons at the LHC and the ILC are useful. 

Finally, we  mention the case where our assumption of the common
mass for additional Higgs bosons is relaxed. 
In general, masses of additional Higgs bosons are given by
\begin{align}
m^2_{\phi} = M^2 + \tilde\lambda_i v^2\left[
 1+ {\mathcal O}\left(\frac{v^2}{M^2}\right)\right],
\end{align}
where $\tilde\lambda_i$ represent specific combinations of $\lambda$
coupling constants.
Our assumption is basically reasonable when additional Higgs bosons are
heavy enough, because their masses are basically given by the unique
scale $M$, the scale of soft breaking of the discrete symmetry. 
When their masses are around the electroweak scale, they can be varied
by the contribution of the term $\tilde\lambda_i v^2$ 
without contradicting the constraints from the
rho parameter and also from perturbative unitarity etc.
In this case, the signals from neutral Higgs boson processes and those
from charged Higgs boson processes are independent.
However, even in such a case, we can repeat the discussion of
discrimination of the type of Yukawa interaction by using
Tables~\ref{tab:Benchmark1} and \ref{tab:Benchmark2}, although the
situation becomes more complicated. 

\section{Conclusions}\label{sec:sum}

In this paper, we have studied the direct searches of additional Higgs
bosons in the general 2HDM with the $Z_2$ symmetry imposed to avoid FCNCs.
We have considered the possible four types of Yukawa interaction which
are determined by generic charge assignment of the $Z_2$ parity to the
SM fermions.

We have discussed the prospect of direct searches for the additional
Higgs bosons at the LHC, and stressed that the exclusion potential is
extensive but not conclusive. 
It means that by taking into account the wide parameter space of the
general 2HDM, there are possibilities that the LHC can discover only
part of the additional Higgs bosons or that even the LHC cannot discover 
any additional Higgs boson but the ILC can discover. 

We have studied the collider signatures of additional Higgs boson
production by evaluating the production cross sections as well as the
decay branching ratios of additional Higgs bosons at the ILC for the all
types of Yukawa interaction. 
We find that various signatures can be expected depending on the type
of Yukawa interaction, the masses of additional Higgs bosons and
$\tan\beta$. 
Thus, as long as the additional Higgs bosons are kinematically accessible,
their production can be detected at the ILC, and further details around
the additional Higgs bosons, i.e.\ the type of Yukawa interaction and the
model parameters can be studied.
Therefore, the searches at the ILC would be a useful complementary
survey even after the LHC results.

\acknowledgments
We would like to thank K.~Tsumura and K.~Yagyu for fruitful discussions. 
S.K.\ and Y.Z.\ are grateful for the hospitality by National Center for
Theoretical Sciences (NCTS), where this paper was finalized.
This work was supported, in part, by Grant-in-Aid for Scientific Research
from the Ministry of Education, Culture, Sports, Science, and Technology
(MEXT), Japan, Nos.\ 22244031, 23104006 and 24340046, the Sasakawa
Scientific Research Grant from the Japan Science Society, and 
NSC of ROC. 



\end{document}